%% file: AARv_rapid_rotators.tex
\documentclass[12pt,preprint]{aastex}

\usepackage{natbib}
\usepackage{epsfig}
\usepackage{lscape}
\usepackage{graphicx}
\usepackage{multirow}
\usepackage[letterpaper]{geometry}
\slugcomment{Accepted for publication in A\&ARv}

\shorttitle{Interferometric Observations of Rapid Rotators}

\shortauthors{van Belle}

\begin{document}

\title{Interferometric Observations of Rapidly Rotating Stars}

\author{Gerard T. van Belle\altaffilmark{1}}

\altaffiltext{1}{Lowell Observatory, 1400 West Mars Hill Road, Flagstaff, Arizona, USA 86001; portions of this manuscript were prepared while in residence at European Southern Observatory,
Karl-Schwarzschild-Str. 2, 85748 Garching, Germany;
gerard@lowell.edu}

\begin{abstract}

Optical interferometry provides us with a unique opportunity to improve our understanding of stellar structure and evolution.  Through {\it direct observation} of rotationally distorted photospheres at sub-milliarcsecond scales, we are now able to characterize latitude dependencies of stellar radius, temperature structure, and even energy transport.  These detailed new views of stars are leading to revised thinking in a broad array of associated topics, such as spectroscopy, stellar evolution, and exoplanet detection.  As newly advanced techniques and instrumentation mature, this topic in astronomy is poised to greatly expand in depth and influence.

\end{abstract}

%




\section{Introduction}\label{sec_introduction}

One of the most fundamental stellar characteristics that is most frequently taken for granted is the shape of a star.  This is perhaps somewhat unsurprising --- in our direct experience of seeing stellar disks, there is only the sun, which is very nearly a perfect sphere: on an average radius of $959$\farcs$28 \pm 0$\farcs$15$ \citep{Kuhn2004ApJ...613.1241K} there is only a variation of only $9.0 \pm 1.8$mas \citep{Rozelot2003SoPh..217...39R} from equator to pole, indicating an oblateness ($b/a-1$) of less than $10^{-5}$.  However, for a surprisingly non-trivial number of stars, this degree of oblateness is in excess of 20\% and, in certain cases, even 30\%.  Hints of this have been seen for decades in stellar spectroscopy, but this phenomenon has eluded direct observation until only recently.

The basis for departures from true sphericity lie in the rapid rotation of a star.  Our sun turns about its axis at a stately rate of slightly more than once per month \citep{Snodgrass1990ApJ...351..309S}, which for a gaseous body of its size and mass leads to the $\sim$percent oblateness value seen above.  However, as we shall see below, certain objects with greater mass are spinning at rates that are $10-60\times$ more rapid, leading to distortions of their shapes.

Direct observation of these oblate spheroids has been a tantalizingly close possibility for decades, but it has only been within the last 10 years that there has been a sufficient convergence of theory, capability and technique for imaging at sub-milliarcsecond levels of resolution to be possible.  The early spectroscopic results a century ago guided theoretical expectations, which then unfortunately had to lay dormant while the technology caught up.  The advent of modern long-baseline interferometry in the visible and near-infrared has opened the door to directly characterize and even image these rapidly rotating stars - and in particular, the most recent developments in this field of ultra-high-resolution astronomy have enabled rapid strides to be made in probing the surprising stellar structure of these objects.

In \S\ref{sec_history} we will explore the history of these developments, starting with the spectroscopic background (\S\ref{sec_spectro-history}) which led to the early expectations and tests of interferometry (\S\ref{sec_interf-history}).  The basic physics will be described \S\ref{sec_physics}, including the simple Roche shape of a rigidly rotating gaseous spheroid (\S\ref{sec_roche}) and the resulting latitude dependence of flux known as the von Zeipel effect (\S\ref{sec_vonZeipel}).  The specifics of interferometric observations of such targets is addressed in \S\ref{sec_interferometry}, including a general discussion of observational quantities (\S\ref{sec_obsquans}), and discussion of extensions of the technique to improved image reconstruction (\S\ref{sec_image_reconstruction}) and greater spectral resolution (\S\ref{sec_spectrointerferometry}).  A review of observational results to date is given in \S\ref{sec_stars}, followed by a discussion of the broad impact of these results to date \S\ref{sec_impact}; a summary of 191 future targets are suggested in \S\ref{sec_future_target_list}.

\section{History}\label{sec_history}

\subsection{Spectroscopic Underpinnings}\label{sec_spectro-history}

The earliest investigations into stellar rotation have their roots in Galileo's observations of sunspots \citep{drake1957}\footnote{There are, in fact, records of pre-`modern' sunspot observations taking place in ancient China.  In the 4th century BC, astronomer Gan De from the State of Qi was the first to acknowledge sunspots as a solar phenomenon \citep{temple1986}; in the occident these were inaccurately viewed as obstructing natural satellites following the observations by the Benedictine monk Adelmus in 17-24 March 807 \citep{Wilson1917PA.....25...88W,Milone2008ASSL..352.....M}.}.   The technique of spectroscopy was developed in the late 19th century, and was the necessary tool to quantitatively observe rotational effects on stars, rather than just our sun. However, prior to that development, there is clearly evidence for consideration of the effect of rotation in the intervening years upon other astronomical observables, such as photometry \citep[cf. the work of Bouillaud in 1667 on Mira, and later by Cassini, Fontenelle, and Miraldi as presented in][]{brunet1931}.  Thus, roughly a dozen generations after Galileo, \citet{Abney1877MNRAS..37..278A} was the first to suggest axial rotation of stars could be observed from spectral line broadening.  This suggestion was swiftly rebuked by \citet{Vogel1877AN.....90...71V}, who pointed out that broadened hydrogen lines suggested by Abney are frequently accompanied by other lines which themselves appear narrow - it was only later established that there are many stars in which all lines are broadened.  An actual measurement of the rotation effect on spectral lines was carried out first by \citet{Schlesinger1909, Schlesinger1911MNRAS..71..719S} on the eclipsing binaries $\lambda$ Tauri and $\delta$ Libr\ae.  In Schlesinger's observations, the less luminous companion occulted varying parts of the rapidly rotating primary, allowing measurement of variations in apparent radial velocity. This `Rossiter-McLaughlin' effect \citet{Rossiter1924ApJ....60...15R,McLaughlin1924ApJ....60...22M} is now a phenomenon commonly observed with transiting extrasolar planets, and is a useful tool for probing the alignment of the orbital plane relative to the stellar rotation axis \citep[see e.g.,][]{Winn2005ApJ...631.1215W}.

The specific line shapes expected from rotational Doppler broadening were first predicted by \citet{Shajn1929MNRAS..89..222S}. Although their work emphasized once again the observation of this effect in binaries, they included relevant predictions for single stars.  These line contours were used by \citet{Elvey1930ApJ....71..221E} in publishing the first list of rotational velocities (although there had been a few efforts on specific individual stars a few years before), and the rest of the decade saw rapid observational progress.  \citet{Struve1931MNRAS..91..663S} linked rotation rate to spectral type, finding that A-type stars were the most likely fast rotators; \citet{Westgate1933ApJ....77..141W, Westgate1933ApJ....78...46W, Westgate1934ApJ....79..357W} published extensive observational catalogs for hundreds of stars between spectral types O and F.  These efforts solidified this new cornerstone of observational astronomy. Slettebak, in a dozen papers between 1949 and 1956, discovered that the most rapid rotators were to be found among Be stars, and established a relationship between rotation and mass.

Contemporaneously, \citet{vonZeipel1924MNRAS..84..665V, vonZeipel1924MNRAS..84..684V} demonstrated that (under the assumption of rigid body rotation) the local surface brightness at any point on a star is proportional to the local effective gravity, and as such, the temperature at the poles would be greater than at the equator for a rotating star.  \citet{Slettebak1949ApJ...110..498S} went on to take the implications of this `von Zeipel effect' and computed the first modifications for expected spectral line shapes of rapidly rotating, bright stars. These implications were developed in detail in \citet{Collins1963ApJ...138.1134C,Collins1965ApJ...142..265C} for continuum emission, and \citet{Collins1966ApJ...146..152C}, who incorporated shape distortion, aspect effects, gravity \& limb darkening, and latitude variation in calculating H-$\beta$ profiles.  \citet{Harrington1968ApJ...151.1051H} went on to characterize the intrinsic polarization expected for rapidly rotating early-type stars, and \citet{Collins1977ApJS...34...41C} demonstrated that the spread of the main sequence (e.g. $M_V$ versus $b-y$ coordinates) due to rotation alone was 2 to 3 times larger than previously expected.  This latter result had significant implications for inference of both ages and distance moduli for clusters of young stars.

Overall, the apparent position of stars on the HR diagram is significantly affected (e.g. 2-3 subtypes) in considering rapidly rotating stars to their non-rotating counterparts \citep[see][and references therein]{Slettebak1980ApJ...242..171S,Collins1985MNRAS.213..519C}.
For the interested reader, the detailed history of line profile analysis is given a substantively more thorough treatment in \citet{Slettebak1985IAUS..111..163S}.
On the broad subject of rotation the multiple editions of \citet{Tassoul1978trs..book.....T,Tassoul2000stro.book.....T}
and proceedings of the two dedicated IAU meetings (Colloquium 4 in 1969 organized by Arne Slettebak, Symposium 215 in 2002 organized by Andre Maeder \& Philippe Eenens, both titled simply ``Stellar Rotation'') are indispensable resources.

\subsection{Interferometric Observations}\label{sec_interf-history}

It is unclear as to when the possibility of directly measuring stellar rotational distortion with optical interferometry\footnote{`Optical' interferometry is the term commonly used to refer to interferometry in the visible and near-infrared.  This technology family is separate from radio interferometry in its homodyne, rather than heterodyne, nature (e.g. mix-and-detect, rather than detect-and-mix).} was first considered as a plausible exercise.  The pioneering angular diameter measurements of \citet{Michelson1921ApJ....53..249M} with the 20-foot beam interferometer on the Hooker 100'' was the technology gate that opened up the possibility of such measurements.  However, it is clear that the idea of potential observations did not fully develop until the guiding spectroscopic rotational velocities were themselves surveyed between 1930 and 1960, and the implications of the extremes of those velocities then evaluated.  Furthermore, Pease's observational experience of operating the 50-foot beam interferometer, the successor to the 20-foot, was of extreme difficultly and highly limited success.  Single measurements of individual stars were few in number, none were ever published in refereed journals \citep[see the commentary in][]{Townes1999ApJ...525C.148T}, and the collection of the sufficient data density necessary to establish a detection of rapid rotation on a given object was not forthcoming.

The entire endeavor of interferometric stellar angular diameter measurements lapsed into a state of dormancy for more than 3 decades, until the innovative proposal by \citet{HanburyBrown1956Natur.178.1046H} to pursue this task with intensity interferometry began to produce results.  These efforts led to the construction and operation of the Narrabri Intensity Interferometer (NI$^2$), which produced the seminal results on single star diameters \citep{HanburyBrown1974MNRAS.167..121H} and binary star orbits \citep{Herbison-Evans1971MNRAS.151..161H}.  The success of NI$^2$ in this regard, combined with the maturity of the underlying theory of rapid rotators that predicted observable effects, led to considerations of the possibility of using NI$^2$ to observe the oblateness of Altair, as described in the PhD dissertations of \citet{Jordahl1972PhDT.........5J} and \citet{Lake1975}.  Unfortunately, the northern hemisphere location ($\delta$=+8$^o$52') of Altair is at odds with the southern location of Narrabri (latitude = $-30^o$19'), which limited NI$^2$'s ability to collect a sufficient range of baseline projections upon the object.  Simple diameter measurements from NI$^2$ were published, but no detection of oblateness was made.  (No mention of attempts to observe the more favorable declination, bright rapid rotator Acherner with NI$^2$ is obvious in the literature.) Another contributing factor in this missed opportunity may have been the novel audacity of the intensity interferometer - and the degree to which its underlying physical principles were (and still are) poorly understood by the majority of other astronomers in the field.  Gaining general acceptance of the simple single-star diameter measurements was difficult as it was, and a more esoteric result like stellar oblateness was perhaps viewed as too ambitious. The completion of NI$^2$ operations brought this second era of optical interferometry to a close with no detection of this phenomenon.

The third, and current, generation of optical interferometers has met with resounding success in probing more than just 1-dimensional parameterizations of stellar sizes.  Many of the facilities in use today are characterized by multiple apertures, some even being relocatable, delivering dense data sets that permit sophisticated image reconstructions. An early attempt was made by Robert R. Thompson and myself using the IOTA interferometer in 1998 to detect the rotationally-induced oblateness of S~Cep, spending 2 weeks of observing using multiple baselines to characterize the shape of the object.  Unfortunately, data on our check star SS~Cep indicated spurious results, and the data had to be discarded.

Stellar oblateness was first detected when a team I led observed Altair with the Palomar Testbed Interferometer (PTI) on two separate baselines \citep{vanbelle2001ApJ...559.1155V}, leading to disparate uniform-disk size measurements, particularly in comparison to the two contemporaneous size measurements on the check star Vega, which did agree.  Extraction of a multi-parameter solution though use of Monte Carlo techniques, patterned after the similar approach applied to Keplerian orbit solving for interferometric data \citep{bkv99}, indicated a $v \sin i = 210\pm13$ km s$^{-1}$, in agreement with spectroscopic values.  The data was insufficient to further constrain the additional parameters of inclination or gravity darkening, but a new sub-field in optical interferometry was opened up by that study.

Subsequent observations of rapidly rotating stars have been carried out with the more capable facilities VLTI, NOI, and (especially) the CHARA Array.  These ensuing data sets have been sufficiently rich to allow for detailed parameterizations of the observed objects, including constraints on inclination and gravity darkening.  A full discussion on the seven objects studied in-depth to date will be reviewed in \S \ref{sec_stars}, and is summarized in Table \ref{tableData_allStars}.

\section{Description of the Geometry and Basic Physics}\label{sec_physics}

\subsection{Roche Model for Stellar Shape}\label{sec_roche}

\begin{figure*}
\epsscale{0.75}
\plotone{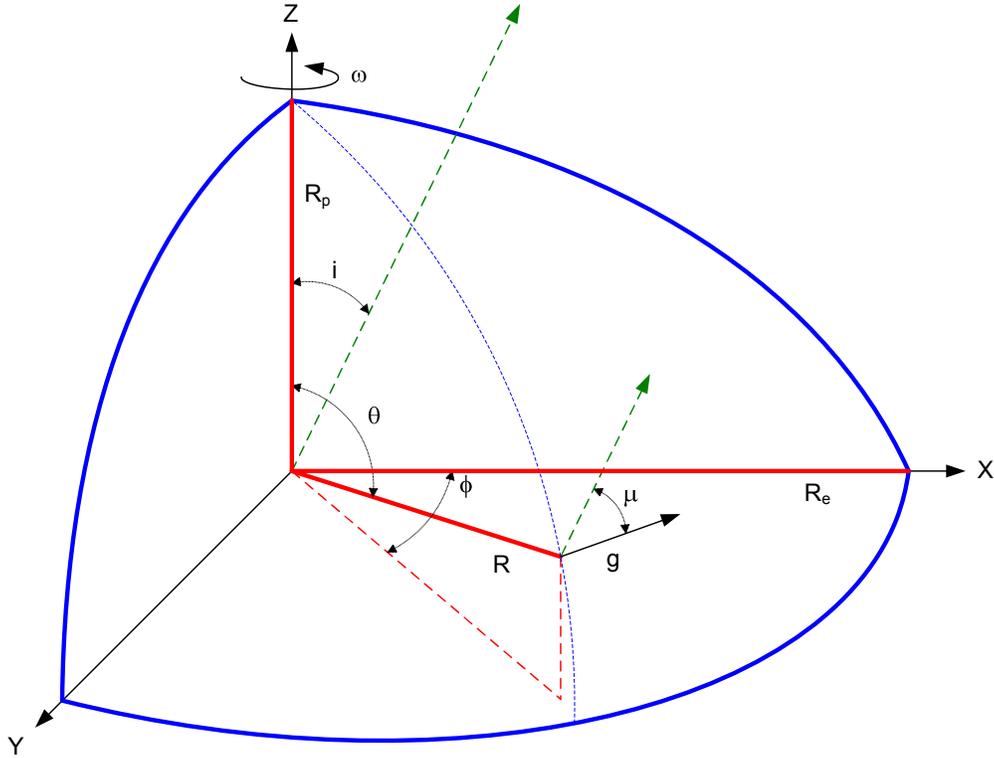}
\caption{\label{fig_geometry}  Illustration of relevant rapid rotator geometry: observer is in XZ plane, with stellar polar axis of radius $R_p$ at inclination $i$ to observer. Stellar angular rotation rate $\omega$ (and stellar mass $M$) sets equatorial radius $R_e$.  Stellar surface point at colatitude $\theta$, longitude $\phi$ has radius $R$ and local effective gravity vector $\vec g$, with $\mu$ being the cosine of the angle between $\vec g$ and the observer. Adopted to be consistent with Fig. 2 in \citet{Collins1965ApJ...142..265C}. }
\end{figure*}

For a {\it non-rotating} star of mass $M$ in radiative equlibrium, its uniformly spherical shape of radius $R$ is trivially defined in terms of an equipotential surface \citep{eddington1926}:
\begin{equation}\label{eqn_static_potl}
\Phi = {\rm constant}= {GM \over R}
\end{equation}
However, once rotation is imparted onto the star, a term must be added to account for the rotational potential:
\begin{equation}\label{eqn_rotational_potl}
\Phi = {\rm constant}= {GM \over R(\theta)}+ {1 \over 2} \Omega^2 R(\theta)^2 \sin^2 \theta = {GM \over R_{\rm pole}}
\end{equation}
where $\theta$ is the colatitude, $R_{\rm pole}$ is the polar radius, $\Omega$ is the angular velocity (since our basic model here assumes uniform rotation, $\Omega \neq \Omega(\theta)$), and $R(\theta)$ is the radius at a given colatitude; this `Roche model' \citep{Roche1837} geometry is illustrated in Figure \ref{fig_geometry}.  The last equality in Equation \ref{eqn_rotational_potl} comes from maintaining that equality across the stellar surface at the non-rotating pole.  Another important caveat of this simple treatment is that polar radius will be treated as a constant, regardless of rotation speed.  A more accurate examination of this overall phenomenon accounts for a decrease in polar radius as the rotating star approaches maximum speed - this effect is slight for most stars $(\lesssim 2\%)$, although for $\lesssim 1 M_\odot$ objects this effect is thought to increase rapidly with decreasing mass to a maximum of $\sim 8\%$ \citep{Sackmann1970A&A.....8...76S,Deupree2011ApJ...735...69D}. A important implication of advancing from Equation \ref{eqn_static_potl} to Equation \ref{eqn_rotational_potl} is that stellar models need to graduate from one-dimensional treatments to two-dimensional.

The maximum ``critical'' angular velocity, $\Omega_{\rm crit}$, where the outward centrifugal acceleration of rotation equals inward gravitational acceleration, can then be derived as
\begin{equation}\label{eqn_critical}
\Omega_{\rm crit} =  \sqrt{{8 \over 27}{ G M \over R_{\rm pole}^3}}
\end{equation}
At this rotational speed the oblateness of the object is at its greatest\footnote{This choice of symbols is consistent with that found in the more recent works by \citet{Cranmer1995ApJ...440..308C} and \citet{Aufdenberg2006ApJ...645..664A}, noting that the earlier manuscripts of \citet{Collins1963ApJ...138.1134C,Collins1965ApJ...142..265C,Jordahl1972PhDT.........5J} differ slightly (e.g. `$\omega$' is used in place of `$\Omega$', `$u$' is used instead of `$\omega$', etc.)}, with $R_{\rm eq,critical} = 3/2~R_{\rm pole}$.
The fractional angular velocity $\omega$ is then defined as
\begin{equation}\label{eqn_critical2}
\omega \equiv {\Omega \over \Omega_{\rm crit}}
\end{equation}
and solving for the cubic equipotential $\Phi(R_{\rm pole}) = \Phi(R,\theta)$ with trigonometric methods \citep[see ][]{Collins1963ApJ...138.1134C,Collins1966ApJ...146..152C,Jordahl1972PhDT.........5J}\footnote{\citet{Aufdenberg2006ApJ...645..664A} note a typographical error in Eqn. (5) of \citet{Collins1963ApJ...138.1134C}; $\omega_c^2=GM/R_e^3$}, we arrive at an expression for the stellar radius at any colatitude $\theta$ for a given fractional rotational velocity $\omega$:
\begin{equation}\label{eqn_Rtrig}
R(\omega,\theta) = {3 R_{\rm pole}\over \omega \sin \theta} \cos
\left [
{\pi + \cos^{-1}(\omega \sin \theta)\over 3}
\right ]
\end{equation}

\begin{figure*}
\epsscale{0.9}
\plotone{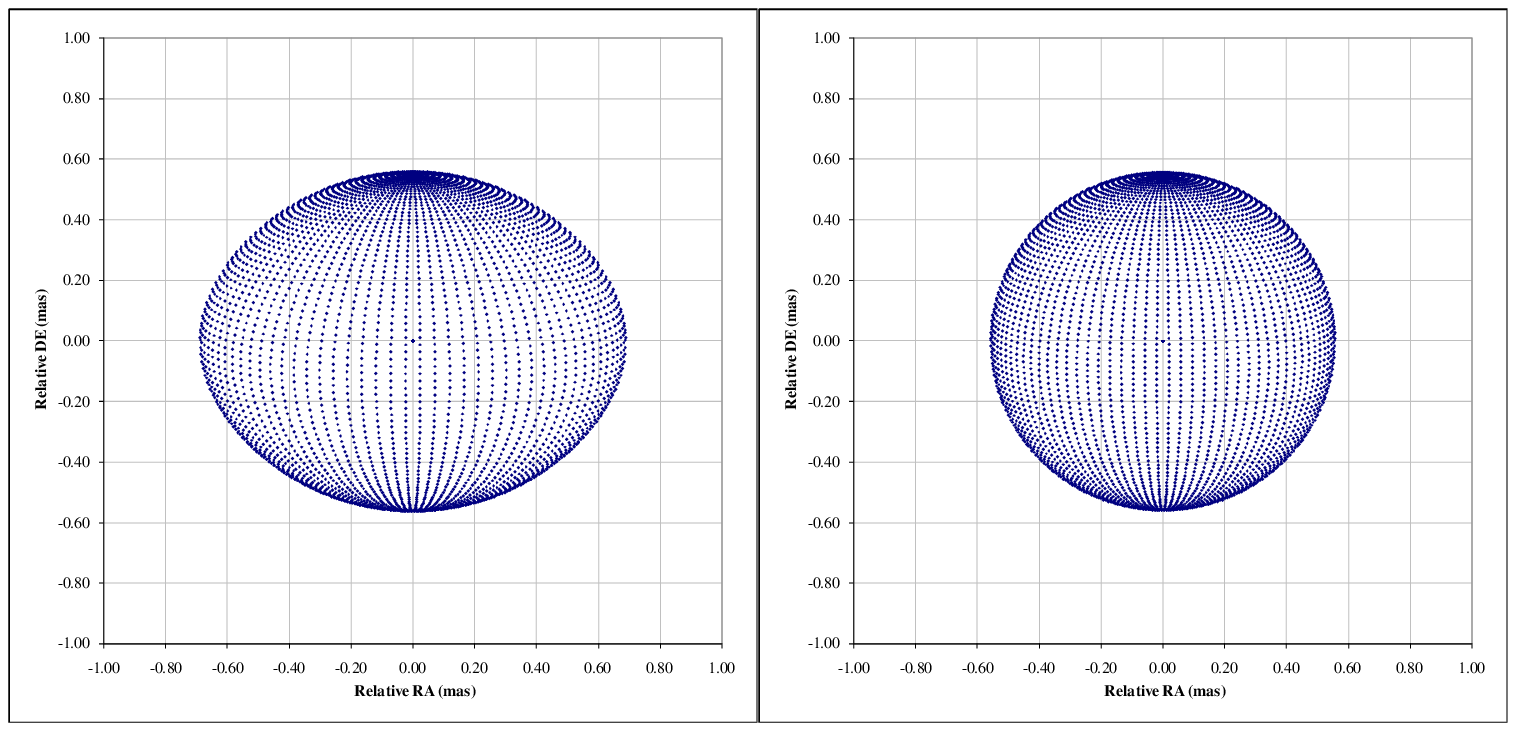}
\caption{\label{fig_relative} Illustration of the phenomenon using a simple Roche model.  Appearance is for two identical stars ($R_{\rm pole}=3 R_\odot$, $d=25$ pc, $\alpha=0^o$, $i=80^o$); the star on the left is rotating with $\omega=0.92$, while the star on the right has only $\omega=0.001$.  No contribution to pole flattening or consequences of limb and/or gravity darkening is included in this simple toy model.}
\end{figure*}

\citet{Owocki1994ApJ...424..887O} identify the convenient identities
\begin{equation}
\omega^2={27 \over 4} w_0 (1-w_0)^2, {R_{\rm pole}\over R_{\rm eq}}=1-w_0
\end{equation}
where $w_0 \equiv v_{\rm eq}^2 R_{\rm pole} / 2 G M$ and $v_{\rm eq}$ is the equatorial rotational velocity, presumably derived from spectroscopy; these identities allow use of the solution for $R(\omega,\theta)$ in Equation \ref{eqn_Rtrig}.
Figure \ref{fig_relative} illustrates the expected distortion imparted upon a (Roche-approximated) stellar surface as it progresses from a non-rotating to a rapidly rotating situation.
For this object, an increase of rotational speed from $\omega=0$ to $\omega=0.92$ induces a $\sim$15\% increase in the equatorial radius.

\citet{Owocki1994ApJ...424..887O} point out that using the point-mass form for the gravitational potential found in Equation \ref{eqn_rotational_potl} ignores high order multipole components that might arise from the rotationally distorted stellar mass distribution.  However, \citet{Orlov1961SvA.....4..845O} shows that polytropic stellar structure models using the correct potential yield a change of less than 1\% from the oblate surface radii predicted by Equation \ref{eqn_Rtrig}.

To compute the surface gravity at a given co-latitude, we compute the negative gradient of the effective potential in Equation \ref{eqn_rotational_potl}.  The two components of the local effective gravity $\vec g$ at a given stellar colatitude $\theta$ in spherical polar coordinates are
\begin{equation}\label{eqn_gravity1}
g_r(\theta) = -{\delta \Phi \over \delta r} = -{GM \over R(\theta)^2}+R(\theta)(\Omega \sin \theta)^2
\end{equation}
\begin{equation}\label{eqn_gravity1b}
g_\theta(\theta) = -{1 \over r}{\delta \Phi \over \delta \theta}
= R(\theta) \Omega^2 \sin \theta \cos \theta
\end{equation}
Thus $\vec g$, directed inwards along the local surface normal, is equal in magnitude to $\sqrt{g_r^2+g_\theta^2}$ \citep{Cranmer1995ApJ...440..308C,Aufdenberg2006ApJ...645..664A}.

The further details of the mathematics describing rotating Roche equipotential surfaces and it application to the stars can be found in
\citet{Collins1963ApJ...138.1134C,Collins1966ApJ...146..152C, Jordahl1972PhDT.........5J,Tassoul1978trs..book.....T,Cranmer1995ApJ...440..308C,vanbelle2001ApJ...559.1155V, Aufdenberg2006ApJ...645..664A}.
Interesting related discussions regarding the oblateness of a planetary objects can also be found in the appropriate literature \citep[e.g.][wherein a discussion of Uranus' oblateness is considered]{Baron1989Icar...78..119B}.

For the benefit of the reader who often encounters rotational velocities from spectroscopy expressed in terms of $v \sin i$ (rather than angular velocity expressions), it is useful to note here a few relationships.  Starting with Eqn. \ref{eqn_critical}, the critical {\it linear} velocity at the equator can be derived as
\begin{equation}\label{eqn_critical_linear}
v_{\rm crit} =  \sqrt{{2 \over 3}{ G M \over R_{\rm pole}}} = \sqrt{{ G M \over R_{\rm eq,critical}}}
\end{equation}
Also, there is a subtle difference between fractional angular velocity and fractional linear velocity, since the linear radius that relates angular velocities to linear velocities is different between actual speed and the breakup speed:
\begin{equation}\label{eqn_fractional_linear}
{v \over v_{\rm crit}} = { \omega R_{\rm eq}\over \omega_{\rm crit}R_{\rm eq,crit}} = { \omega R_{\rm eq}\over \omega_{\rm crit}{3 \over 2}R_{\rm pole}}
\end{equation}

\subsection{The von Zeipel Effect}\label{sec_vonZeipel}

\citet{vonZeipel1924MNRAS..84..665V} and later \citet{Chandrasekhar1933MNRAS..93..539C}
showed that a rotationally oblate star has a local surface flux that is proportional to the local gravity, $F \propto g$.  Thus, the Stefan-Boltzmann law, $F = \sigma_{\rm SB}T_{\rm eff}^4$ implies that $T_{\rm eff} \propto g^{0.25}$.  This particular proportionality is applied in the case of radiative equilibrium; \citet{Lucy1967ZA.....65...89L} pointed out that for stars with convective envelopes, a weaker dependency is expected, with $T_{\rm eff} \propto g^{0.08}$ given as `representative' (although the value of 0.08 is weakly dependent upon mass, radius, luminosity, and other parameters).  A closer examination of the \citet{Lucy1967ZA.....65...89L} discussion points out that progressing from slow to rapid stellar rotation should lead to pressure gradients that need to be balanced by the Coriolis force, with the consequence that the value of 0.08 quickly becoming less valid.  A general parametrization expresses the relationship as $T_{\rm eff} \propto g^{\beta}$, with $\beta$ either being a model predicted value or a parameter to fit as part of a larger overall solution for some set of observed astronomical data.

Thus, when considering the total flux from a rapidly rotating star, it is important to consider integrating over the surface accounting for the flux as it is tied to the local effective temperature.  This colatitude-dependent temperature can be written as
\begin{equation}\label{eqn_local_Teff}
T_{\rm eff}(\theta) = T_{\rm pole}
\left (
{g(\theta)\over g_{\rm pole}}
\right ) ^\beta
\end{equation}
allowing for a modification of the Stefan-Boltzmann law for characterization of the local stellar radiative flux,
\begin{equation}\label{eqn_local_flux}
F(\theta) = \sigma_{\rm SB}T_{\rm eff}(\theta)^4
\end{equation}
The von Zeipel parameter $\beta$ has been tabulated for a range of stellar models varying in mass and age by \citet{Claret1998A&AS..131..395C,Claret2000A&A...359..289C}\footnote{The gravity darkening parameter `$\beta_1$' reported in \citet{Claret1998A&AS..131..395C,Claret2000A&A...359..289C} contains a factor of 4$\times$ relative to the $\beta$ used herein.}.  It is worth noting that the validity of Equations \ref{eqn_local_Teff} and \ref{eqn_local_flux} has been questioned by \citet{ConnonSmith1974MNRAS.167..199C} when the effects of differential rotation are considered.

For the case where the concerns regarding differential rotation may be neglected, \citet{Aufdenberg2006ApJ...645..664A} use the preceding formalism to derive the temperature difference between the equator and pole:
\begin{equation}\label{eqn_Teff_diff}
\Delta T_{\rm pole-eq} = T_{\rm pole}
\left [
1-
\left (
{\omega^2 \over \eta^2} - {8 \over 27} \eta \omega
\right )^\beta
\right ]
\end{equation}
where
\begin{equation}
\eta = 3 \cos
\left [
{\pi + \cos^{-1}(\omega)\over 3}
\right ]
\end{equation}

It should be noted that use of a single `$\beta$' term in describing a star perhaps oversimplifies the true nature of the stellar surface.  As noted above, surfaces that are fully convective or fully radiative are expected to have values that are 0.08 and 0.25, respectively.  However, as the temperature ranges from hot to cool between a stellar pole and equator, the dominant mode of heat transportation may shift from radiative to convective, depending upon the value of those temperatures.  One may easily conceive of an object whose temperature profile ranges pole to equator, straddling a `sweet spot' value where such a shift occurs - the low latitudes would be convective, the high latitudes radiative. As such, a colatitude-dependent value of $\beta(\theta)$ may be appropriate to consider in constructing the most valid stellar model; the surface appearance of such an object will be most intriguing to image.

In light of the recent observational results that will be reviewed in \S \ref{sec_stars}, \citet{EspinosaLara2011A&A...533A..43E} re-examine the underlying assumptions of the von Zeipel law that lead to the $T_{\rm EFF} \propto g_{\rm EFF}^{1/4}$ formulation.  In particular, the `strong hypothesis' of barotropicity (pressure only depends on the density) is already poor for slowly rotating stars (first noticed by \citet{Eddington1925Obs....48...73E}, see review in \citet{Rieutord2006EAS....21..275R}), and breaks down in noticeable ways for rapid rotators.  These authors propose a new gravity darkening model, based on energy flux being a divergence-free vector antiparallel to the effective gravity, which appears to have good agreement with the interferometric results on Altair and Regulus.

\section{Interferometric Observations of Rapid Rotators}\label{sec_interferometry}

\subsection{Observational Quantities, Predictions}\label{sec_obsquans}

Optical interferometers are, by nature, telescopes that can rightly be considered rather `non-traditional'. Elements of an optical interferometer would be eminently familiar to the non-specialist (e.g. the individual telescopes, for example), whereas there are also elements that would be rather alien (e.g. the beam-combining back end).  This is also true of the elements of the data products - familiar concepts, such as photometry, quickly give way to unfamiliar ones, such as fringe visibility.  For the purposes of this review, we will briefly introduce the relevant observational quantities in a concise manner, directing the reader to more thorough discussions should she or he be interested.

To begin on familiar territory, consider a `normal' filled aperture telescope.  If it were to observe an unresolved point source, one would see on the telescope's image plane an Airy pattern with the characteristic width of $1.22 \lambda / D$ radians, where $\lambda$ is the wavelength of observation, and $D$ is the telescope aperture size.  For a more complex object with on-sky brightness distribution ${\rm O}(\alpha,\delta)$ being observed by a telescope with point-spread function ${\rm P}(\alpha,\delta)$, the mathematical description of this process is
\begin{equation}\label{eqn-image1}
{\rm I}(\alpha,\delta) = \int \! \int {\rm P}(\alpha-\alpha',\delta-\delta'){\rm O}(\alpha',\delta') d\alpha' d\delta'
\end{equation}
For the simple case of a point source, ${\rm O}(\alpha',\delta')$ reduces to a Dirac delta function, ${\rm P}(\alpha-\alpha',\delta-\delta')$ is the point source function associated with a filled-aperture telescope, and ${\rm I}(\alpha,\delta)$ returns the Airy pattern.

The Fourier transform of Equation \ref{eqn-image1} gives
\begin{equation}\label{eqn-image2}
I(u,v) = T(u,v) \times O(u,v)
\end{equation}
which takes us directly to the points of interest for this review.
Specifically, as a curiosity of their construction and measurement processes, optical interferometers tend to provide measurements that sample the Fourier information associated with an object, $I(u,v)$, rather than image plane information $({\rm I}(\alpha,\delta))$.  Secondly, the transfer function associated with an optical interferometer, $T(u,v)$, can be well-characterized.  Thus, since the information associated with $I(u,v)$ and $T(u,v)$ can be collected, the nature of $O(u,v)$ can be computed - and by extension, the original object brightness distribution ${\rm O}(\alpha,\delta)$.
For further expansive details on the underlying theory of optical interferometers, reference the van Cittert-Zernike theorem in a general optics textbook such as \citet{Born1980poet.book.....B}, or books specific to this topic \citep[e.g.][]{Goodman2005ifo..book.....G}.
The proceedings of recent summer schools on optical interferometry are also very instructive in considering these instruments further \citep[in particular,][]{Lawson2000plbs.conf.....L,Haniff2007NewAR..51..565H,Haniff2007NewAR..51..583H}, as is \citet{Bracewell2000fta..book.....B}'s book on the Fourier transform itself.

In practice, an optical interferometer takes light from two or more telescopes and recombines in such a way that the light from the apertures combines coherently, or interferes with itself (hence the name).  Typically the interference criteria includes maintaining wavefront quality, polarization, and path length from the point of light collection at the individual telescopes, through the system, to the point of beam recombination.
In the simplest case of two telescopes, one may consider the interference of light as it is recombined and falls on a photodetector.  Moving a mirror in the beam train that brought the light to the recombination point, the path length through the system may be swept through the point of equal path between the two arms of the interferometer, producing constructive and destructive interference.  This characteristic pattern is referred to as a `fringe', with its contrast (the intensity as measured by the photodetector between maximum constructive and maximum destructive interference) being the measured {\it visibility}:
\begin{equation}
V = {I_{\rm max} - I_{\rm min} \over I_{\rm max} + I_{\rm min}}
\end{equation}
As related to our discussion of Equation \ref{eqn-image2}, the visibility $V$ is the amplitude of a single point of complex image information $I(u,v)$.
Fringe visibility is the basic observable of an optical interferometer and is directly related to the angular size of objects being observed by the instrument.

A second observable, {\it closure phase}, is also frequently measured by optical interferometers.  Transmission of starlight through the turbulent atmosphere corrupts the phase information at optical wavelengths, so an absolute measurement of the phase information associated with $I(u,v)$ is not possible.  However, with interferometers consisting of 3 or more telescopes, these corrupted phases can be summed about triangles to recover partial, uncorrupted phase information.  If we consider a triangle of points from 3 telescopes $\{l,m,n\}$ that form baselines sampling the complex image $I(u,v)$, the measure phased for pair $(m,n)$ is $\psi_{mn}$, which consists of the source phase $\phi_{mn}$ and phase errors $(\xi_m-\xi_n)$.  Thus, summing around a triangle of $\{u,v\}$ points,
\begin{eqnarray}
CP_{lmn} &=& \psi_{lm}+\psi_{mn}+\psi_{nl}\\
 &=& \phi_{lm}+(\xi_l - \xi_m) + \phi_{mn}+(\xi_m - \xi_n) +\phi_{nl}+(\xi_n - \xi_l)\\
 &=& \phi_{lm}+\phi_{mn}+\phi_{nl},
\end{eqnarray}
we see that the measured sum $CP$ is directly related to the original uncorrupted $\phi$ values, since the $\xi$ errors cancel.  Closure phase is an observable that is sensitive to the degree of asymmetry in an image $I(u,v)$.  A good example of a striking closure phase signature is in the signal found in resolved stellar disk observations during planet transits, which can be considered `perfect' star spots \citep{vanbelle2008PASP..120..617V}.
Additionally, higher order such constructions, such as the closure amplitude ($CV_{lmno} = V_{lm}V_{no} / V_{lo}V_{mn}$) are also possible but will not be considered herein.

\begin{figure*}
\epsscale{0.9}
\plottwo{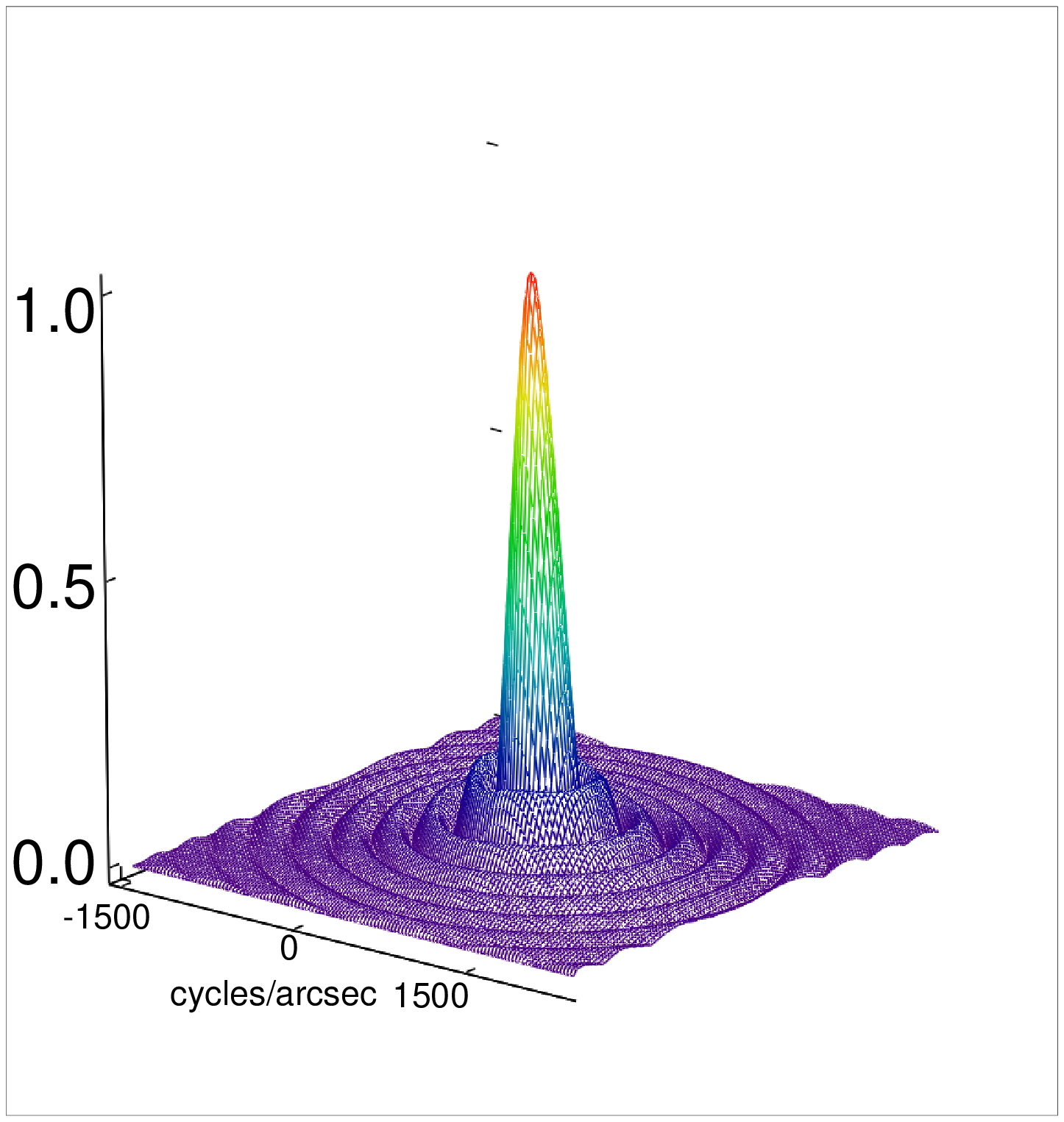}{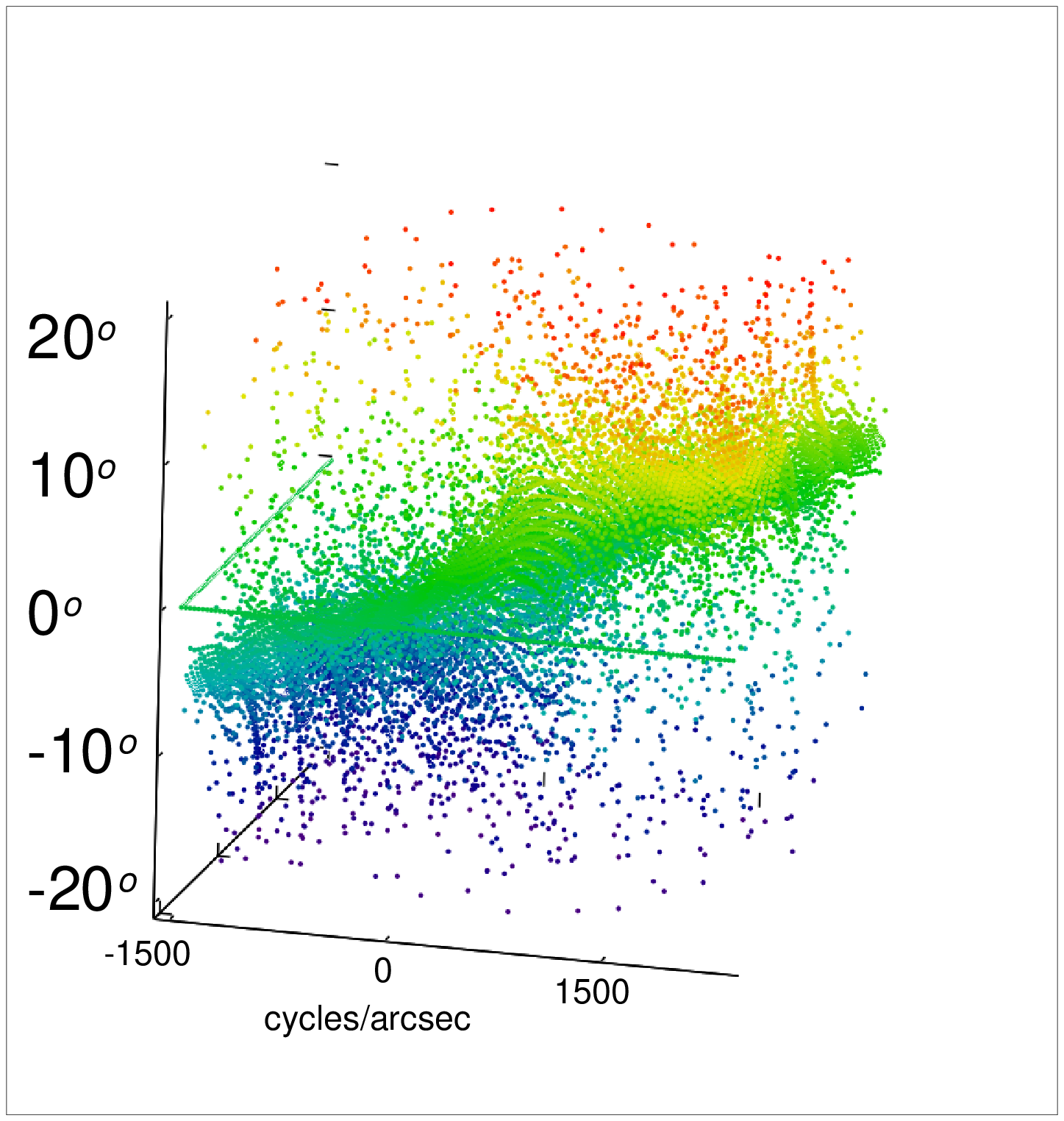}
\caption{\label{fig_altair_omega092}  The appearance of Altair in the $\{u,v\}$ plane, using parameters as reported by \citet{Monnier2007Sci...317..342M}, with $R_{\rm pole}=1.634 R_\odot$ (and $\pi=194.45$mas), $\omega=0.923$, $i=57.2^o$, $\alpha=-61.8^o$, $\beta=0.19$, $T_{\rm pole}=8450$ K; wavelength of observation is taken to be $\lambda=1.6 \mu$m.  The left panel is the modulus of the complex visibility (with visibility normalized to 1 at the center), and the right panel is the argument of the complex visibility (in units of degrees).}
\end{figure*}

\begin{figure*}
\plottwo{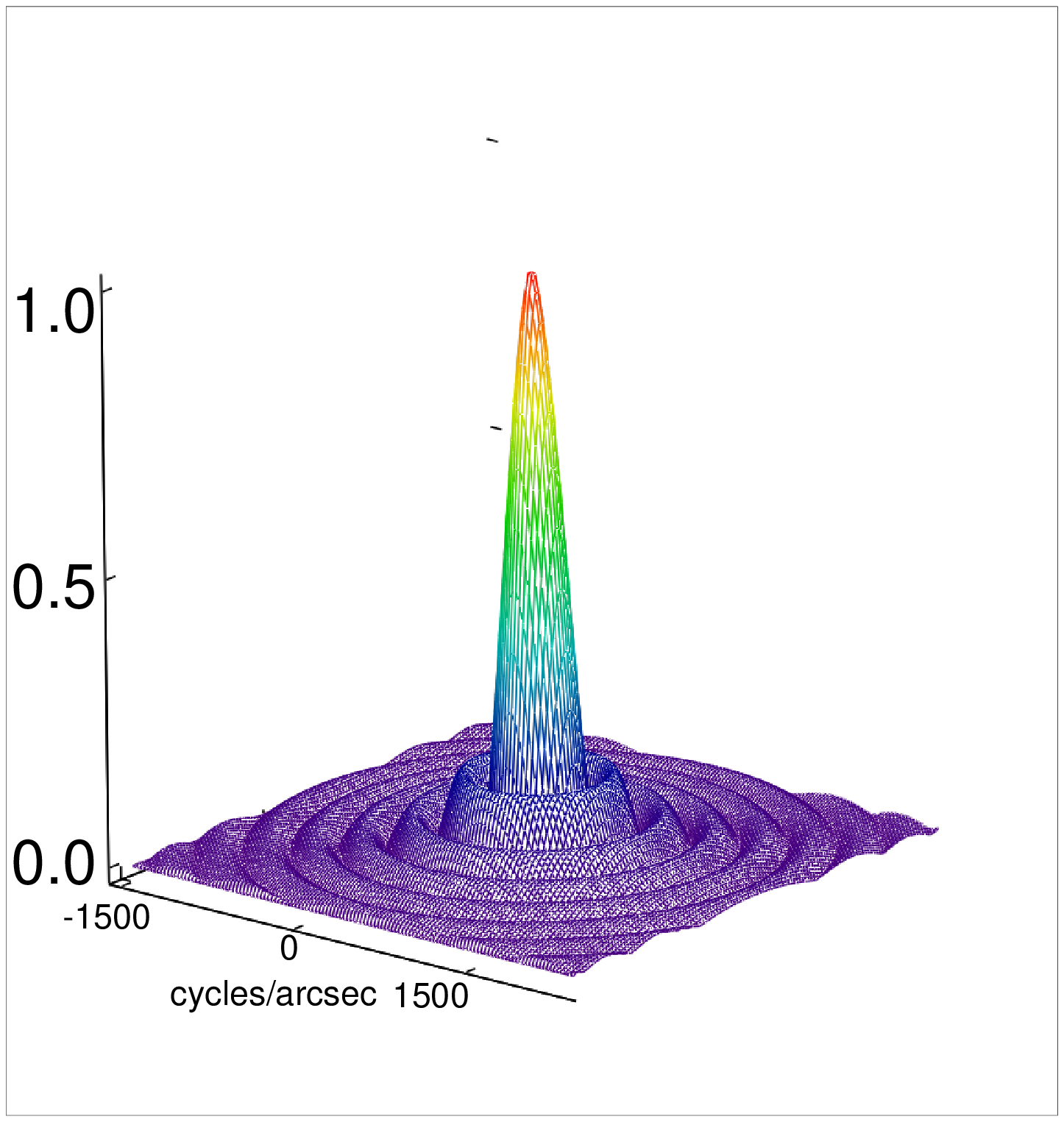}{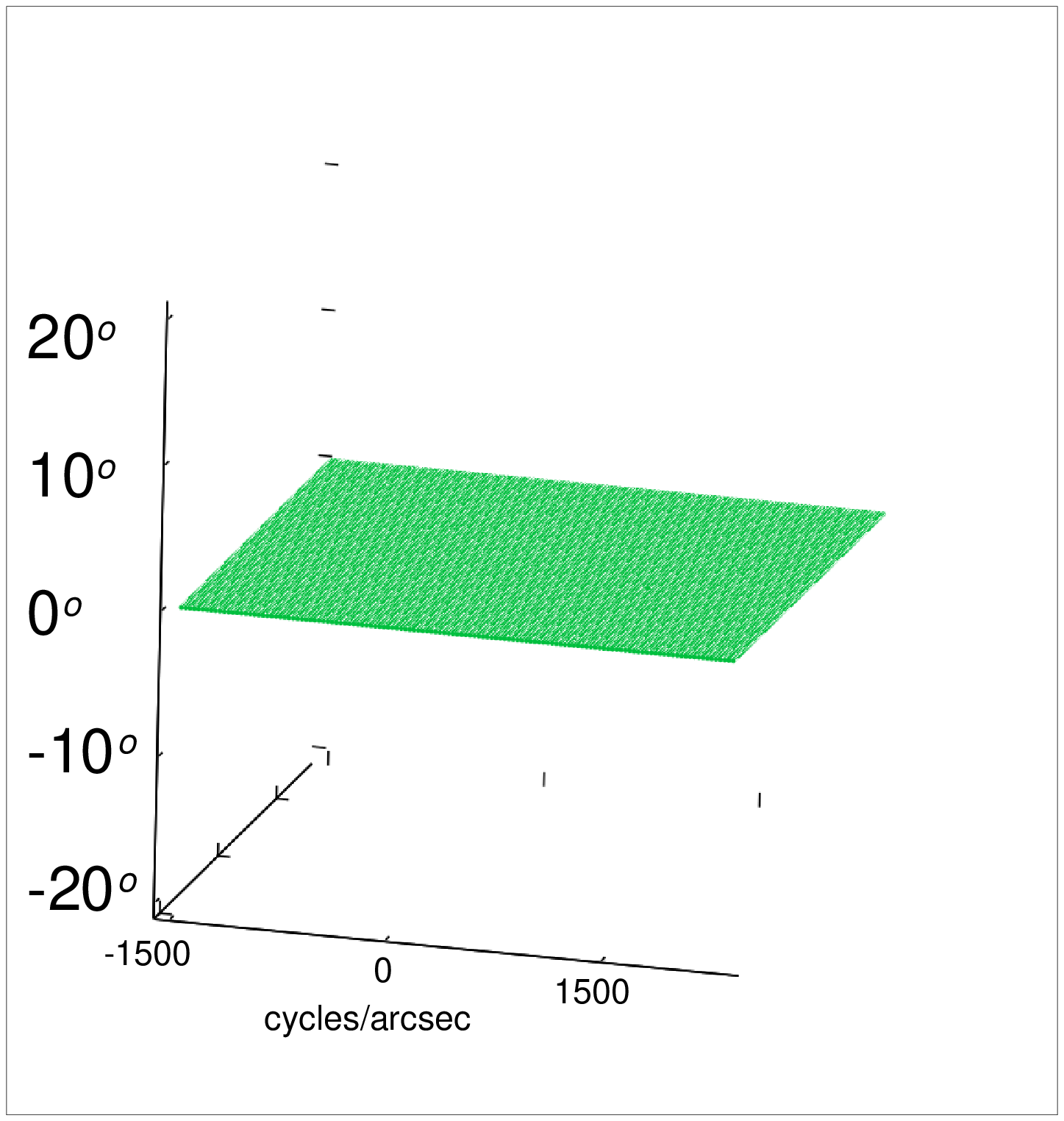}
\caption{\label{fig_altair_omega000}  As Figure \ref{fig_altair_omega092}, but with $\omega=0.01$.}
\end{figure*}

To take this discussion of interferometric observables and place it in a context of actual measurements, we shall consider the representative, pioneering case of observing the rapid rotator Altair, which we shall see in \S \ref{sec_stars} is a perennial favorite for demonstrating the prowess of an interferometer or instrument.  In Figure \ref{fig_altair_omega092} we can see the basic appearance of Altair in the $\{u,v\}$ plane, ; the left panel is a representation of the visibility amplitude, and the right panel illustrates the visibility phase.  These points were generated by construction of a Roche model based on the rotation parameters reported in \citet{Monnier2007Sci...317..342M} ($R_{\rm pole}=1.634 R_\odot$ and $\pi=194.45$mas, $\omega=0.923$, $i=57.2^o$, $\alpha=-61.8^o$, $\beta=0.19$, $T_{\rm pole}=8450$ K), and using the prescription found in \S 4 of \citet{Aufdenberg2006ApJ...645..664A}  to generate co-latitude dependent values for radius $R(\theta)$, local surface gravity $g(\theta)$, and local effective temperature $T_{\rm EFF}(\theta)$.  (Limb darkening was ignored in this example, which has a small $\sim$1\% effect on visibility amplitude measurements.)
$T_{\rm EFF}$ was then converted to 1.6 $\mu$m flux under the simplying assumption of blackbody radiation (BBR).
For a qualitative comparison of the effect rapid rotation has on these observables, a second nearly-identical set of $\{u,v\}$ points was generated, with the exception that the rotation rate was artificially reduced from $\omega=0.923$ to $\omega=0.01$, in Figure \ref{fig_altair_omega000}.  The general deviation of these two figures are most clearly seen in the right-hand phase plots: the uniform source of Figure \ref{fig_altair_omega000} has a clear plane associated with a $\phi=0$ value, while the asymmetric appearance of the gravity-darkened rapid rotator in Figure \ref{fig_altair_omega092} shows a continuum of interesting phase values.

\begin{figure*}
\plottwo{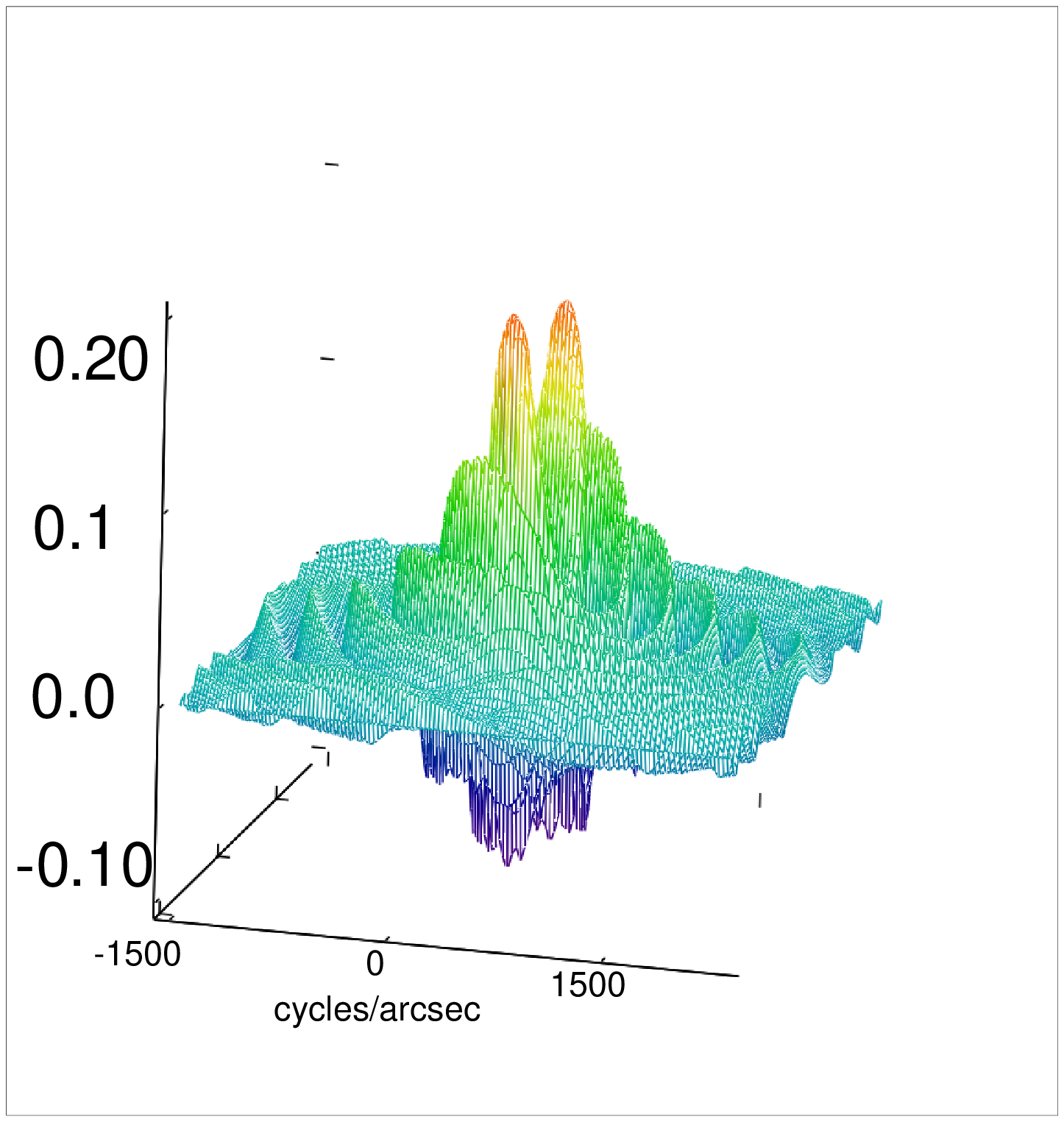}{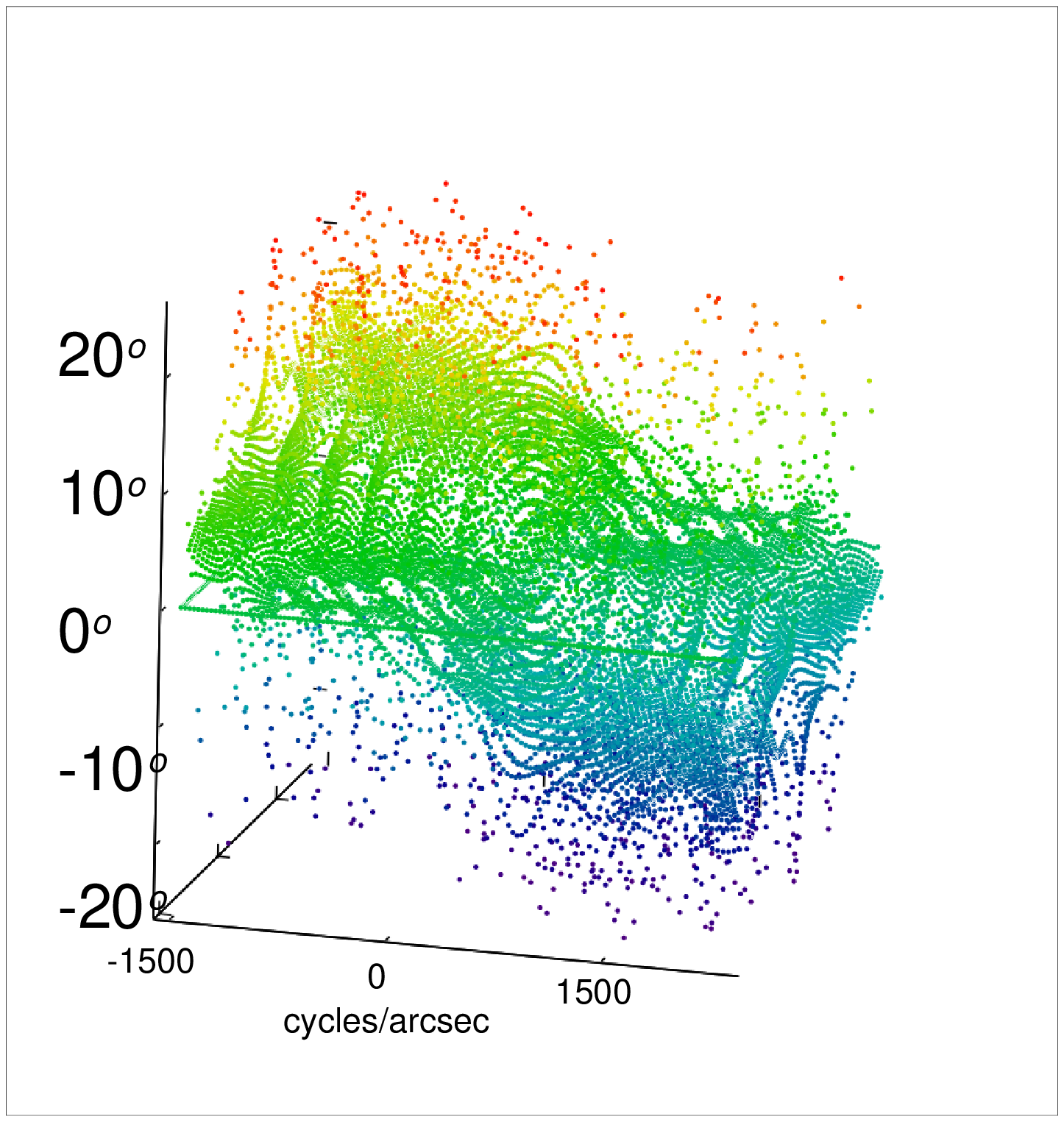}
\caption{\label{fig_altair_diff}  The difference of the data sets seen in Figures \ref{fig_altair_omega092} and \ref{fig_altair_omega000}.  The deviations in visibility amplitude on the left a due primarily to the increased angular size on the sky in the stellar equatorial region; the closure phase deviations on the right (in degrees) are most sensitive to the asymmetric brightness distribution due to the von Zeipel effect.}
\end{figure*}

These differences, and the associated visibility amplitude excursions associated with the stellar shape produced by rapid rotation, are sharply highlighted in Figure \ref{fig_altair_diff}, which plots the difference of Figures  \ref{fig_altair_omega092} \&  \ref{fig_altair_omega000}.  Not only are the obvious phase excursions seen on the right, but previously hidden $\pm10\%$ visibility amplitude differences are made obvious.  The deviations in visibility amplitude on the left a due primarily to the increased angular size on the sky in the stellar equatorial region; the closure phase deviations on the right are most sensitive to the asymmetric brightness distribution due to the von Zeipel effect.

\begin{figure*}
\epsscale{1}
\plottwo{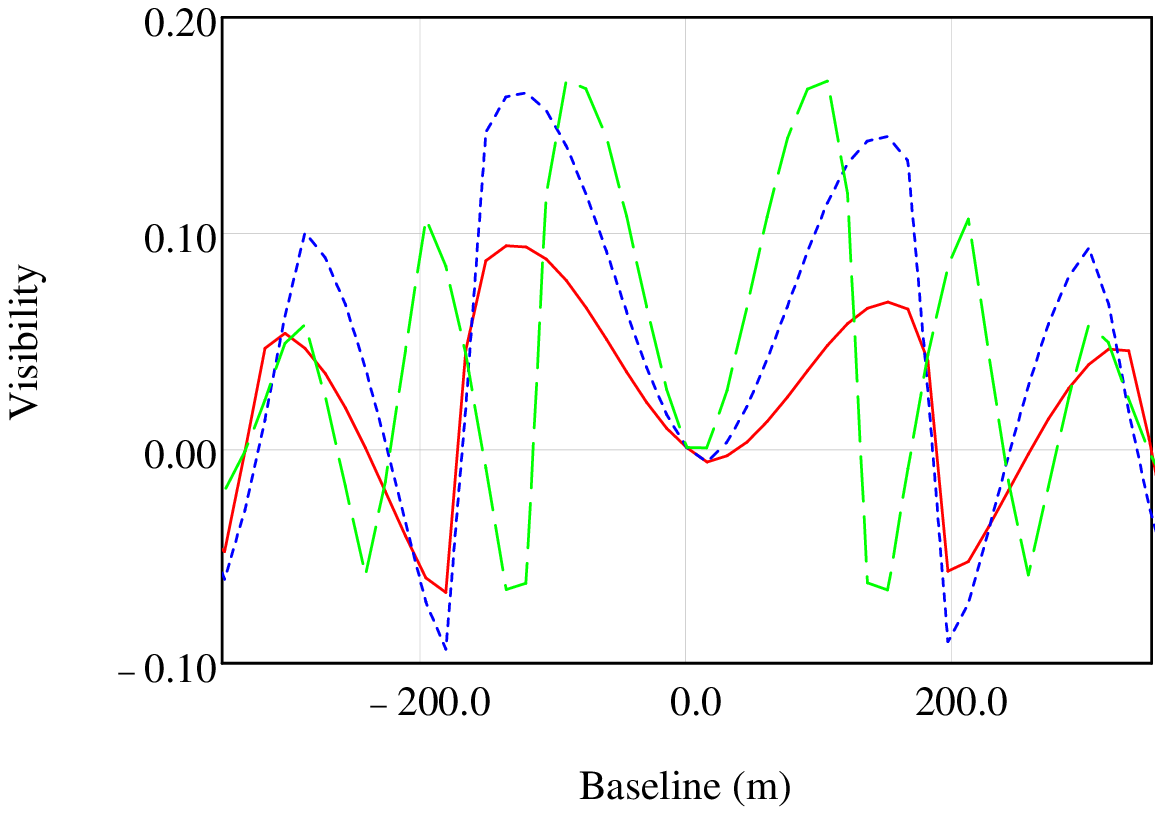}{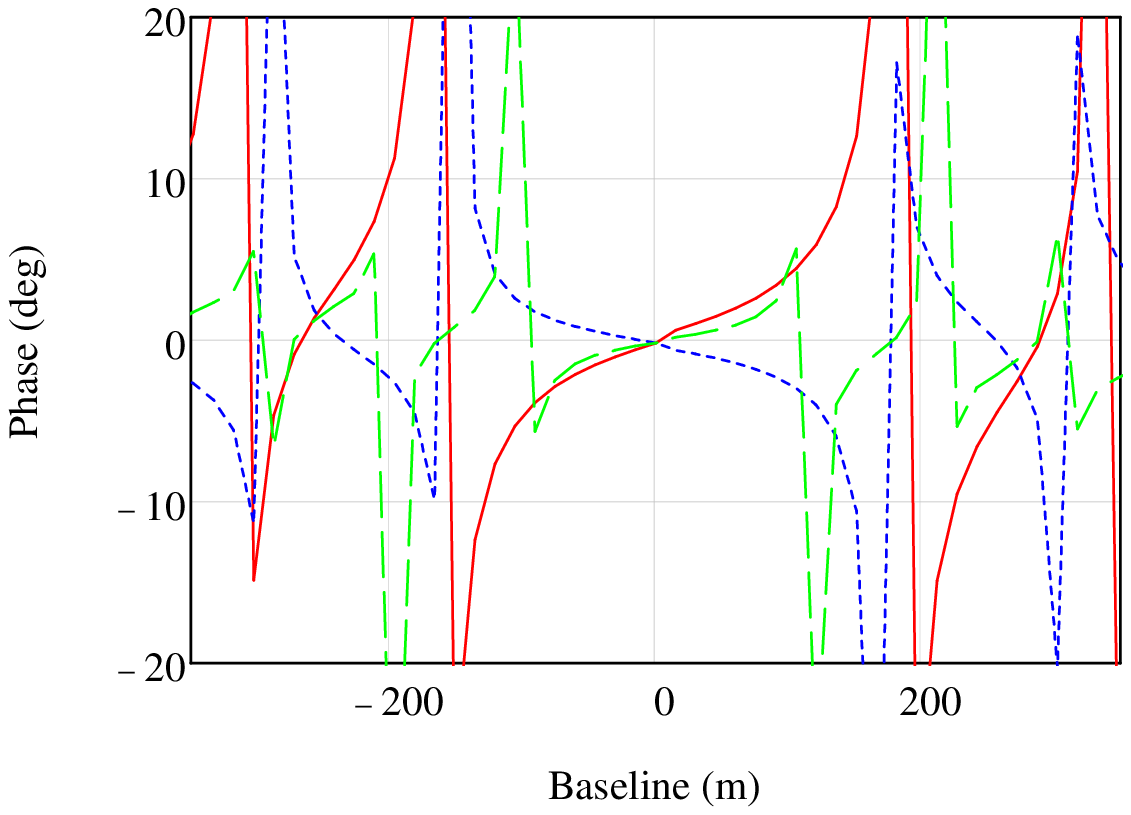}
\caption{\label{fig_altair_diff_cuts}  Cuts along the $\{u,0\}$, $\{0,v\}$ and $\{u,v\}$ lines for the left and right panels of Figure \ref{fig_altair_diff} (solid red, dotted blue, and dashed green, respectively).  Abscissal units for both panes is in meters; the lefthand ordinate is normalized visibility, while the righthand ordinate is in degrees.}
\end{figure*}

\begin{figure*}
\epsscale{1}
\plottwo{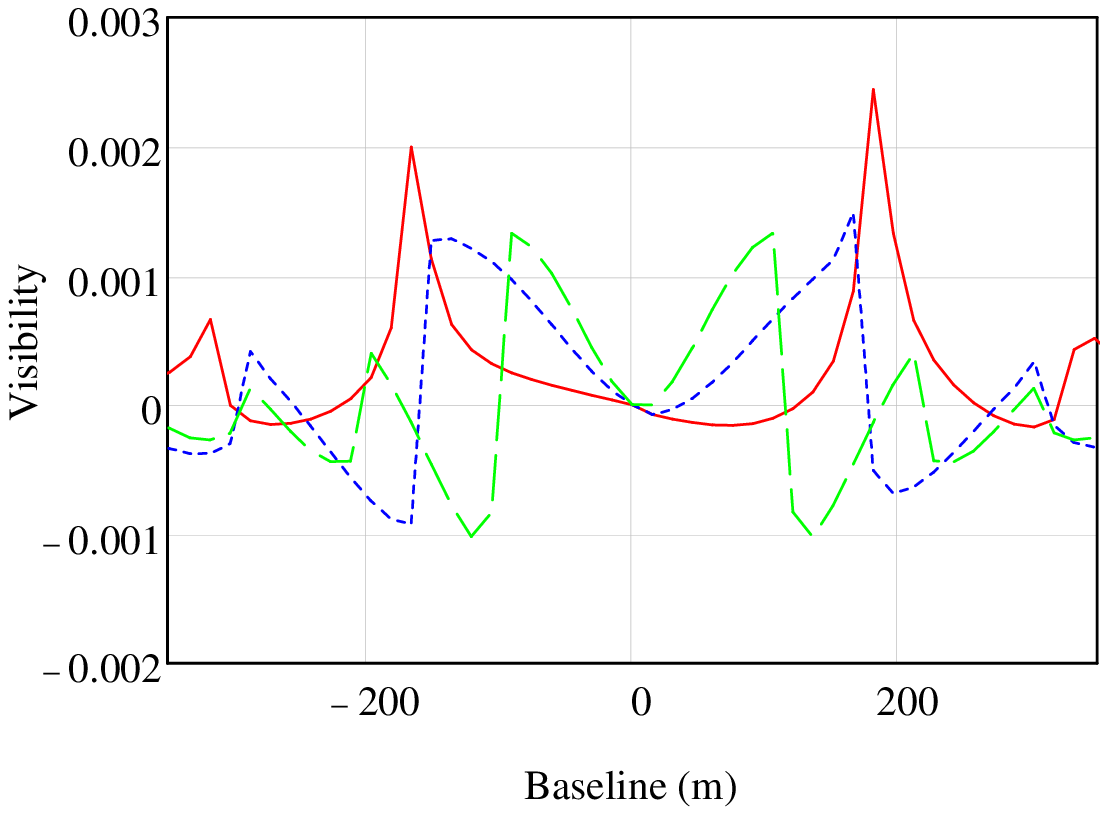}{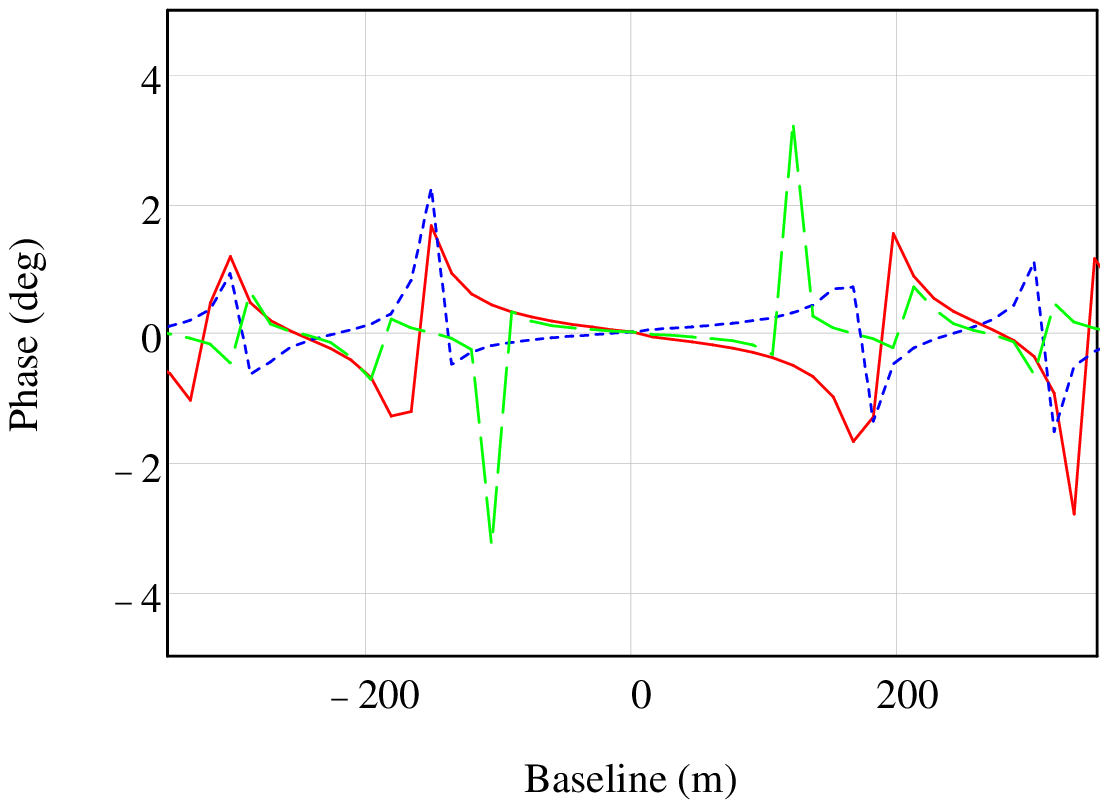}
\caption{\label{fig_altair_sigma_beta}  Changes in visibility amplitude (left) and phase (right) for 1-$\sigma$ deviations in gravity darkening $\beta$.  Lines are as in Figure \ref{fig_altair_diff_cuts}.}
\end{figure*}

\begin{figure*}
\epsscale{1}
\plottwo{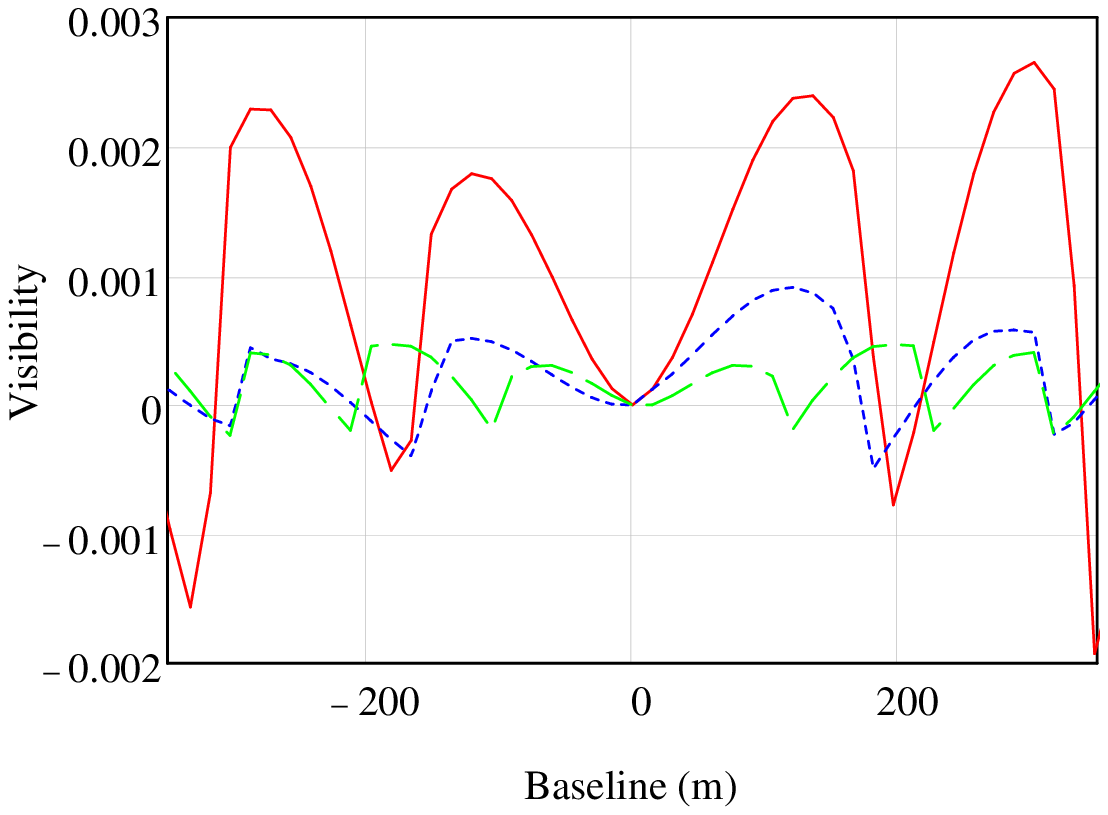}{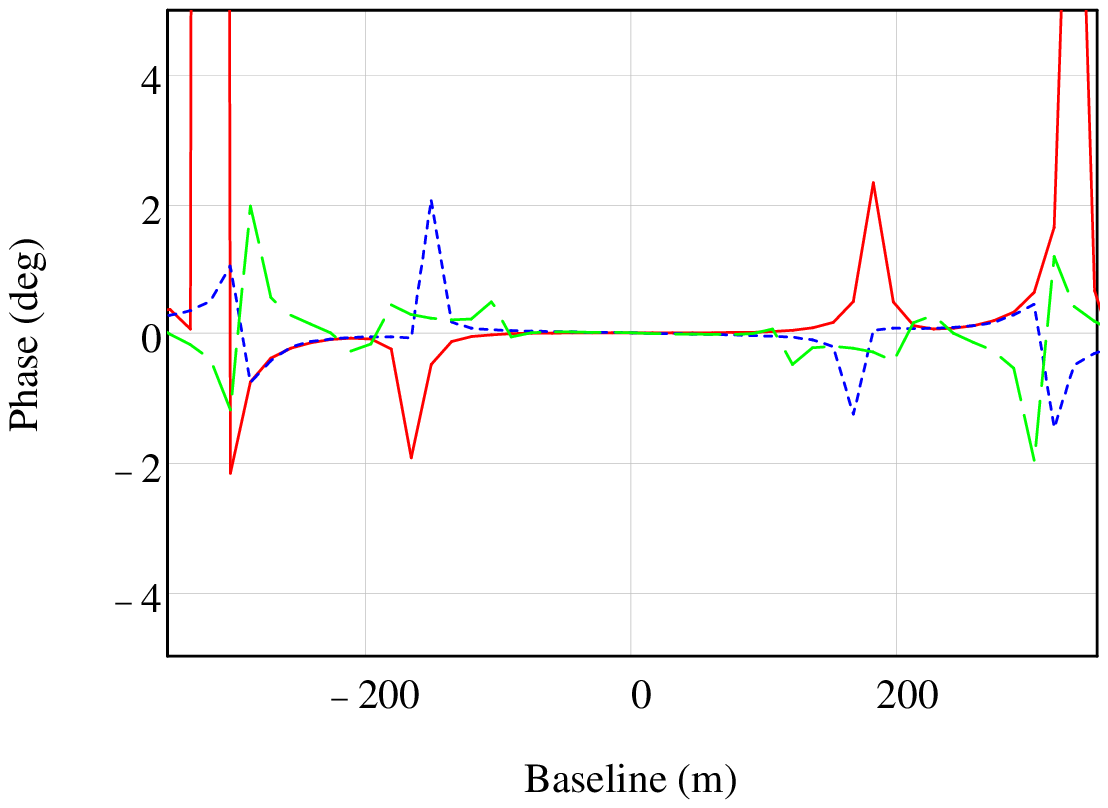}
\caption{\label{fig_altair_sigma_i}  Changes in visibility amplitude (left) and phase (right) for 1-$\sigma$ deviations in inclination $i$.  Lines are as in Figure \ref{fig_altair_diff_cuts}.}
\end{figure*}

\begin{figure*}
\epsscale{1}
\plottwo{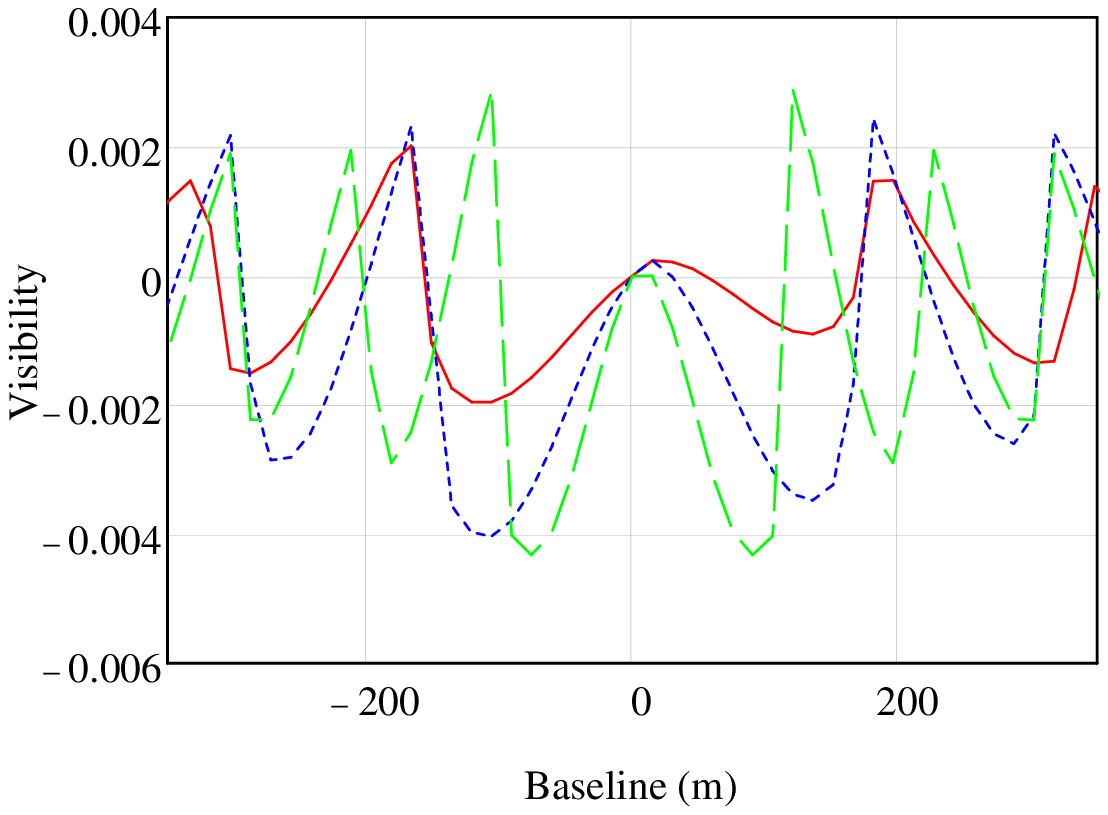}{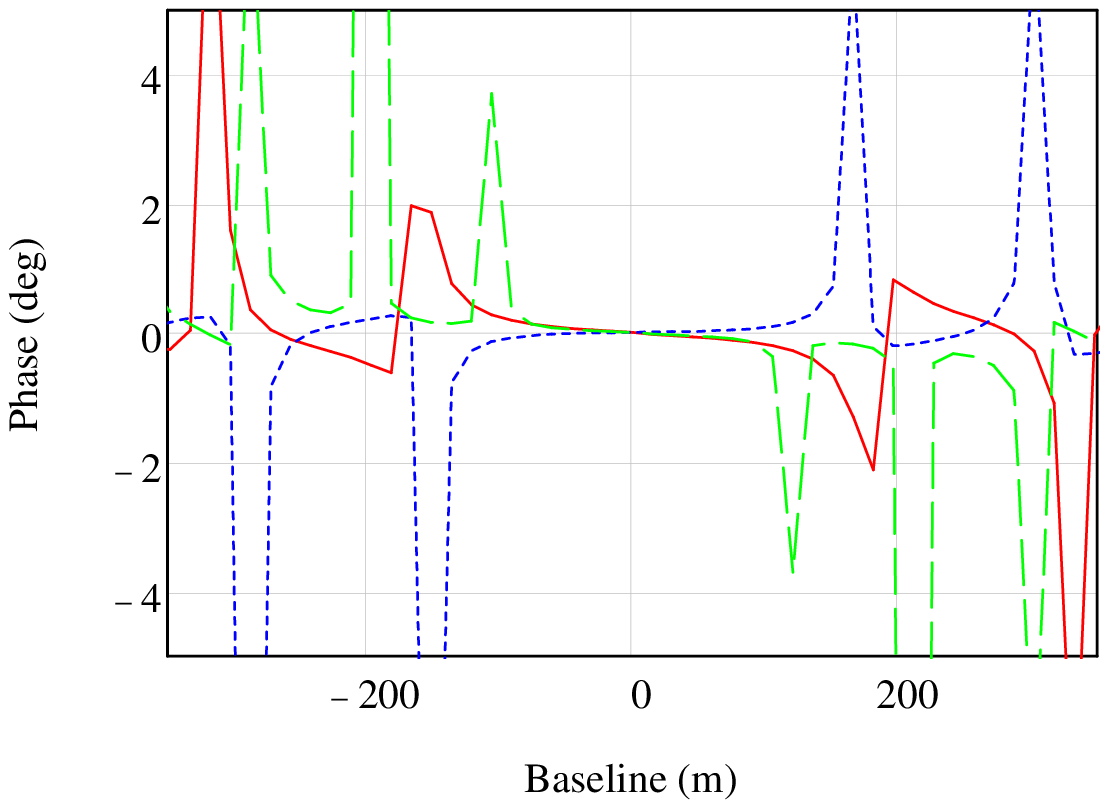}
\caption{\label{fig_altair_sigma_omega}  Changes in visibility amplitude (left) and phase (right) for 1-$\sigma$ deviations in rotation rate $\omega$.  Lines are as in Figure \ref{fig_altair_diff_cuts}.}
\end{figure*}

\begin{figure*}
\epsscale{1}
\plottwo{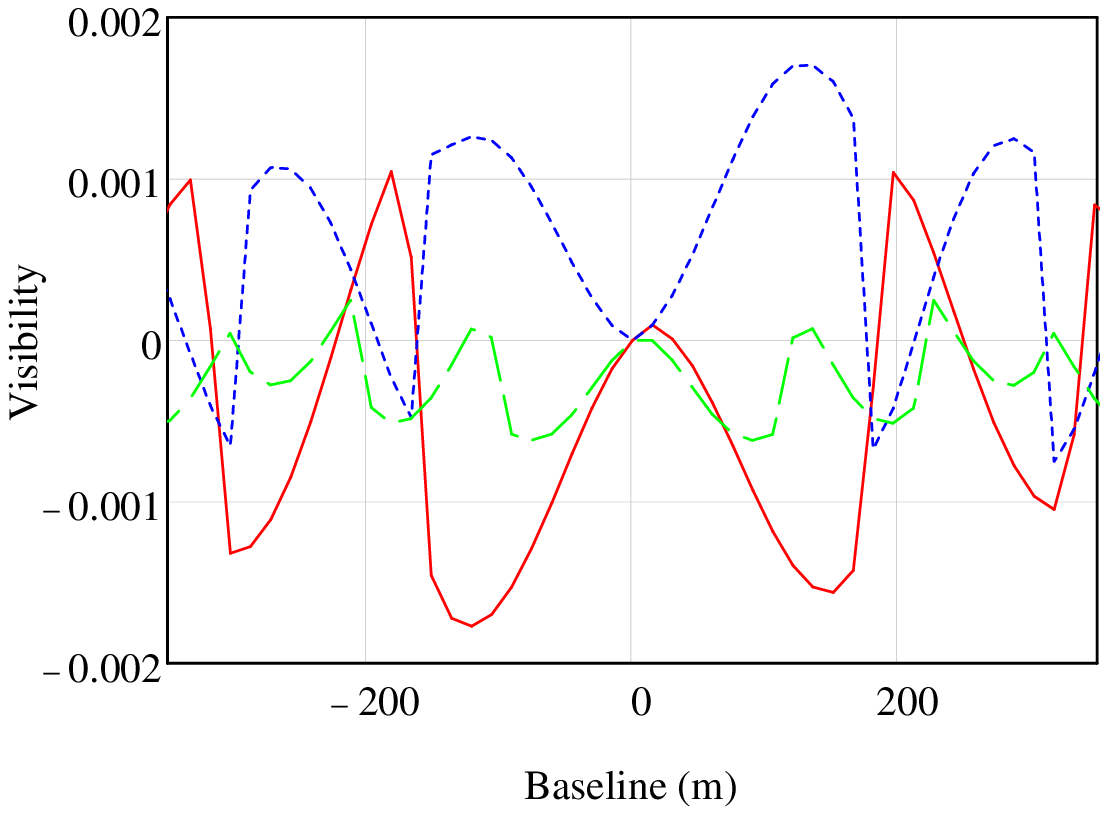}{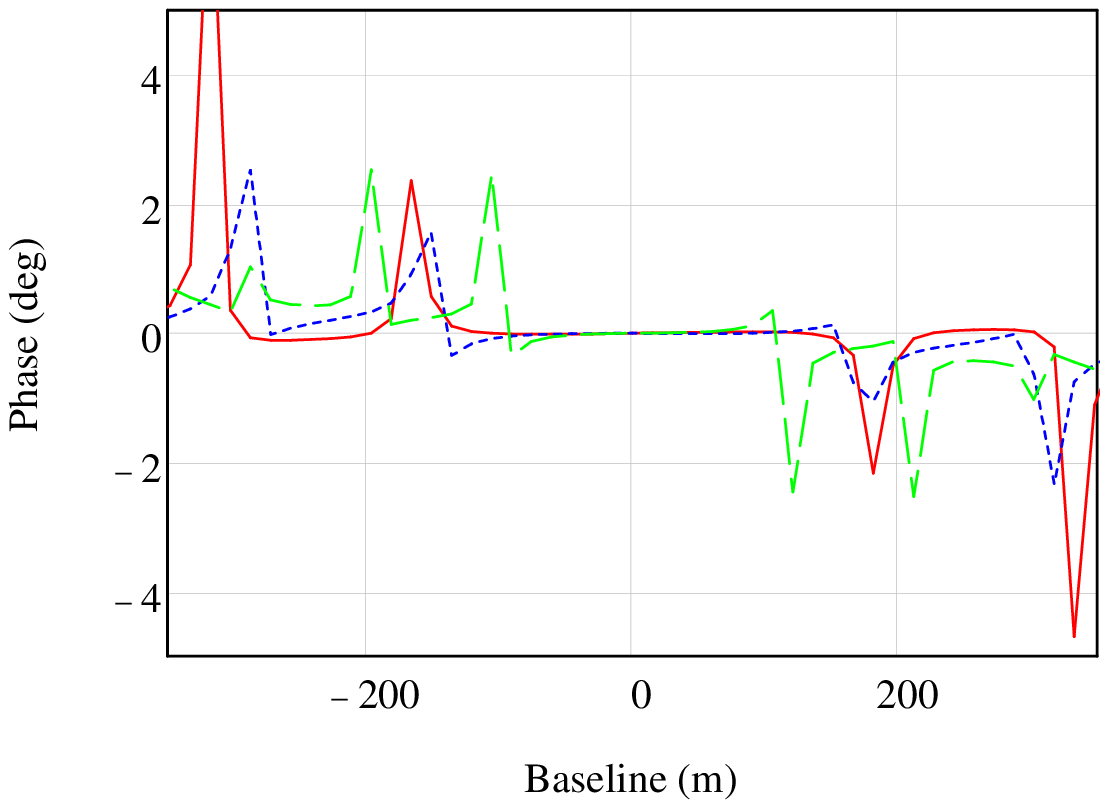}
\caption{\label{fig_altair_sigma_alpha}  Changes in visibility amplitude (left) and phase (right) for 1-$\sigma$ deviations in orientation $\alpha$.  Lines are as in Figure \ref{fig_altair_diff_cuts}.}
\end{figure*}

\begin{figure*}
\epsscale{1}
\plottwo{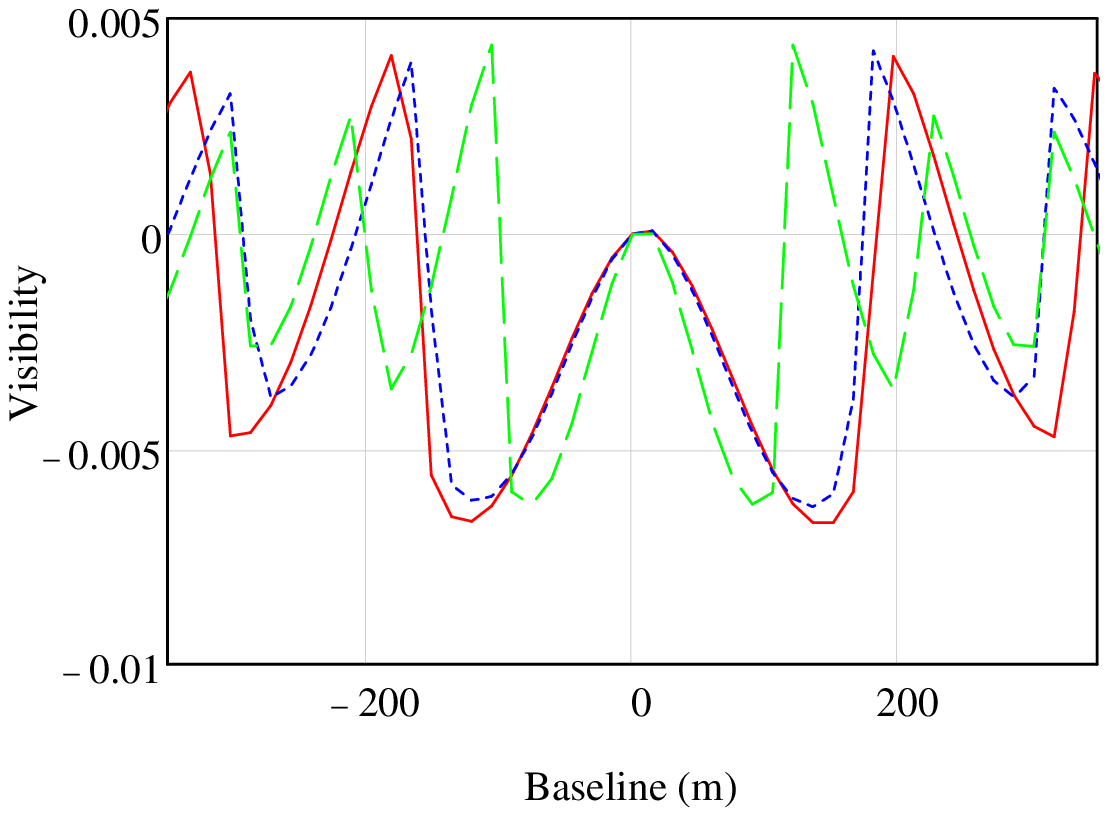}{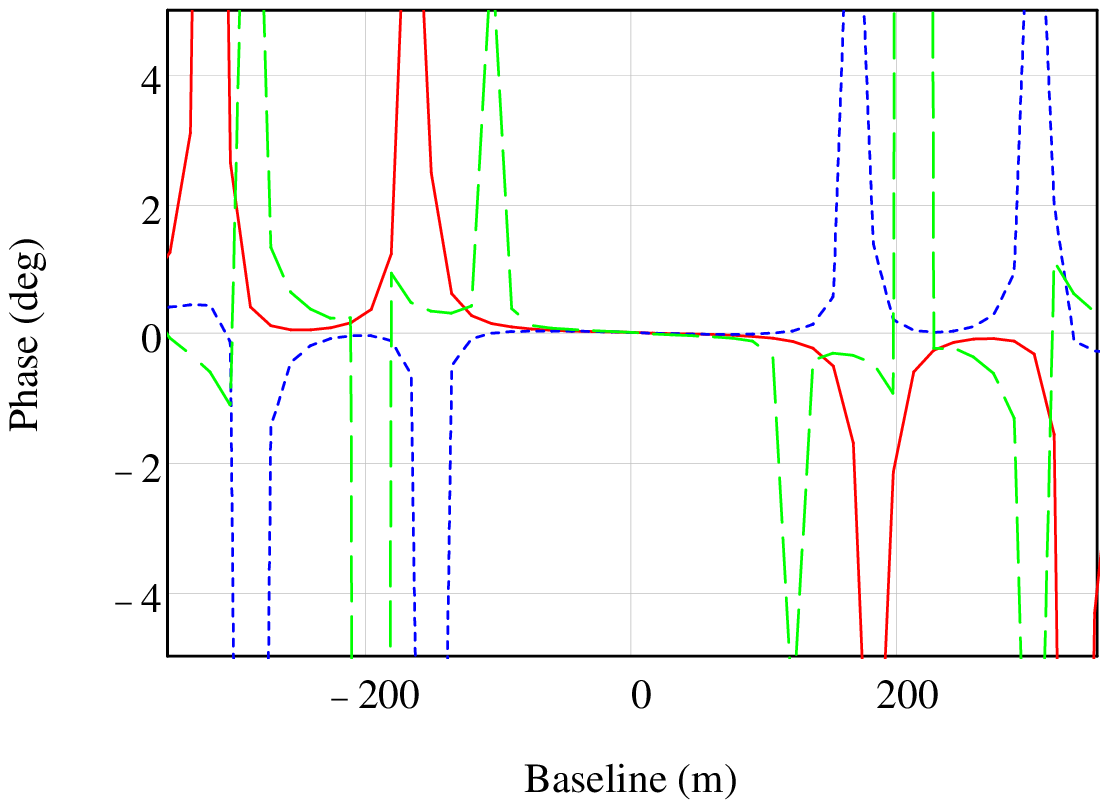}
\caption{\label{fig_altair_sigma_Rp}  Changes in visibility amplitude (left) and phase (right) for 1-$\sigma$ deviations in polar radius $R_{\rm pole}$.  Lines are as in Figure \ref{fig_altair_diff_cuts}.}
\end{figure*}

Finally, to connect the qualitative representations in these figures to some sense of what is detected by an optical interferometer, we plot in Figure \ref{fig_altair_diff_cuts} cuts along the two planes of Figure \ref{fig_altair_diff} along 3 lines ($\{u,0\}$, $\{0,v\}$ and $\{u,v\}$).  We see that the observed visibility deviates for a rapid rotator from the values expected for a non-rotating object at the $\pm10\%$ level; the phase values vary by many tens, if not hundreds, of degrees - and as such, the resultant closure phase values would deviate on similar scales.  Instruments such as CHARA Array MIRC and VLTI AMBER are able to measure visibility amplitude and closure phase; for the former, visibility amplitude precision is at the $\sim \sigma_V=5\%$ level, and closure phase precision is as good as $\sim \sigma_{CP}=1^o$ for single-measurement accuracy with substantial $\sqrt N$ improvements possible \citep{Zhao2011PASP..123..964Z}.
Figures \ref{fig_altair_sigma_beta} through \ref{fig_altair_sigma_Rp} illustrate differences in visibility amplitude and phase (e.g. Figure \ref{fig_altair_omega092} - Figure \ref{fig_altair_omega092}') as the solution parameters $\{\omega,R_{\rm pole},i,\alpha,\beta\}$ found in \citet{Monnier2007Sci...317..342M} are given $1-\sigma$ excursions from their best-fit values.

\begin{figure*}
\epsscale{1}
\plotone{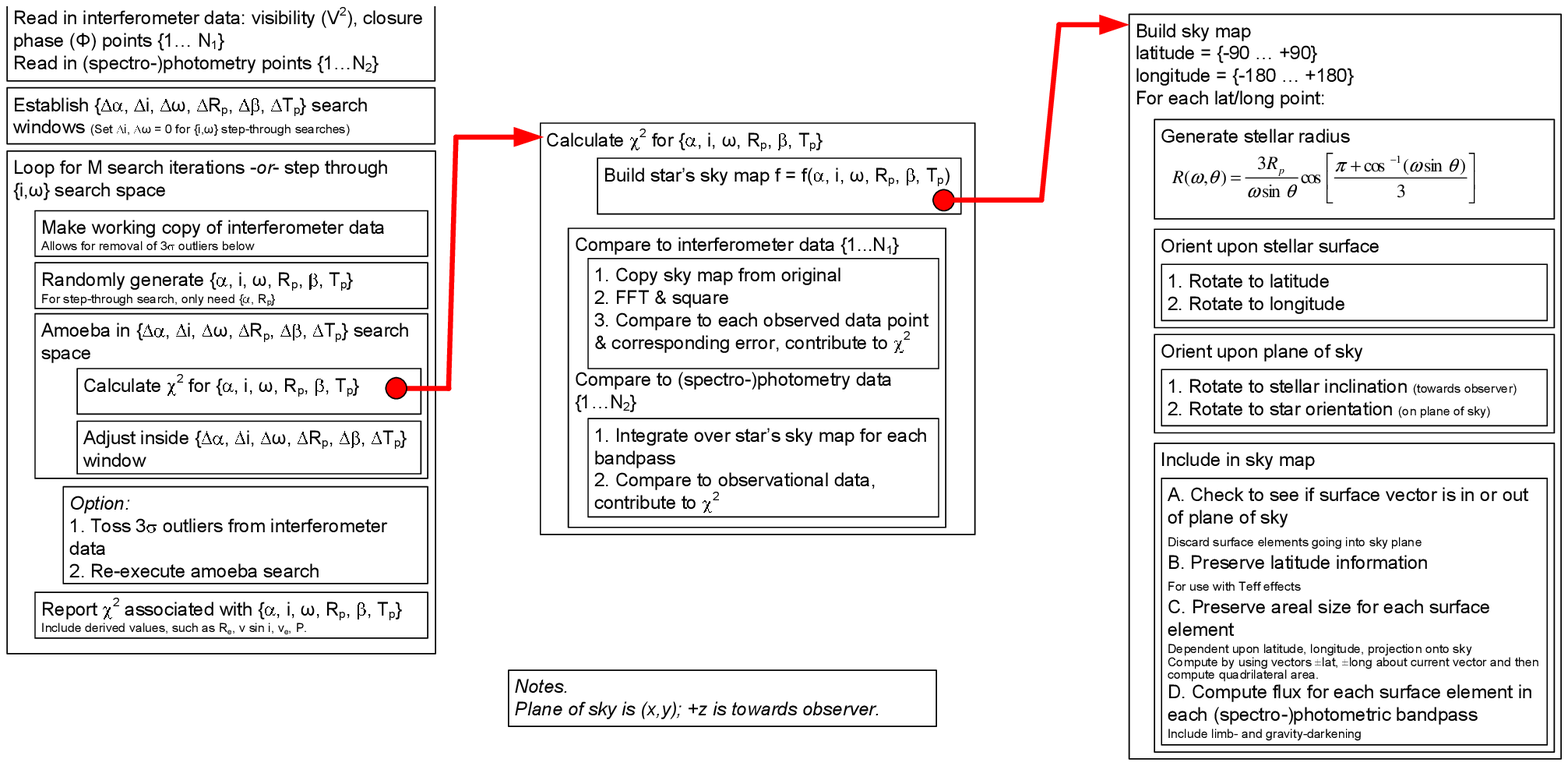}
\caption{\label{fig_flowchart} Detailed flowchart of the $\chi^2_\nu$ minimization process designed to produce a stellar $\{\omega,R_{\rm pole},i,\alpha,\beta,T_{\rm pole}\}$ solution set used for \citet{vanbelle2006ApJ...637..494V}.}
\end{figure*}

The practicalities of $\chi^2$ minimization of interferometric data are highlighted in \S6 of \citet{vanbelle2006ApJ...637..494V}  (henceforth referred to as VB06), which is illustrated pictorially in Figure \ref{fig_flowchart}.  VB06 used a Monte Carlo approach which begans by constructing models of a rapidly rotating star based upon rotation $\omega$ and polar radius $R_{\rm pole}$. Model surfaces were constructed for the star at some sufficiently dense interval (for VB06, the spacing was $0.8^o$) in both colatitude and longitude across the whole volume. Flux for a given surface area can then computed using the appropriate influence of gravity darkening as described in \S \ref{sec_vonZeipel}, with the $\beta$ parameter quantifying the effect, anchored to pole effective temperature $T_{\rm pole}$.  In addition to the gravity darkening, limb-darkening appropriate for these model stars as indicated by the quadratic laws can be generated by \citet{ClaretHauschildt2003A&A...412..241C}.  These models are then mapped onto the sky, through the use of two additional free parameters describing inclination $i$ and on-sky rotational orientation $\alpha$.

Thus, for a given set of six randomized free parameters $\{\omega,R_{\rm pole},i,\alpha,\beta,T_{\rm pole}\}$, a $\sim$100,000 element volume surface is generated, projected upon the sky, rotated and the resultant image Fourier transformed for comparison to each of the observed visibility and closure phase data points, from which a $\chi^2/$DOF is calculated.  As cleverly noted in \citet{Aufdenberg2006ApJ...645..664A}, the full Fourier transform of the model star's image does not need to be performed - only visibilities and closure phases associated with a small, specific set of $\{ u,v\}$ points corresponding to the interferometer sampling need be generated.  The resulting computational load is substantially lighter and allows for faster and more complete Monte Carlo searches.

A detailed review of optical interferometry in astronomy can be found in \citet{monnier2003RPPh...66..789M}. A detailed discussion of phases in particular is continued in \citet{Monnier2007NewAR..51..604M}, including expansive discussions on closure phase, and more esoteric observables possible with optical interferometers, such as differential closure phase, closure differential phase, and closure amplitudes.

\subsection{Image Reconstruction}\label{sec_image_reconstruction}

Relative to the VB06 study, there are two specific extensions that can be made to this technique.  First, inclusion of closure phase data can significantly expand the capability of the Monte Carlo routine to quickly recover the six characterizing parameters of the original star.  The sensitivity of closure phase to asymmetries in brightness distribution speeds and improves the fidelity of the parametric fitting approach. The code written for VB06 in fact already includes generation of the closure phase data, but no comparison was made in that investigation since the CHARA Array's Classic pairwise beam combiner (from which the VB06 data were obtained) produces no measures of closure phase.  However, both the CHARA Array's MIRC combiner and the VLTI AMBER instruments provide measurements of closure phase, and data from these instruments could be utilized for the parameter fitting; an example of this can be found in \S 4 of \citet{Zhao2009}.

Second, and most instructively, optical interferometry instrumentation and techniques is advancing beyond simple pairwise telescope recombination to now include 3-, 4-, and even 6-way telescope recombination, which allows model-independent maps to be recovered from the data.  These `imaging' techniques are advancing rapidly \citep{Cotton2008SPIE.7013E..48C,Baron2010SPIE.7734E..78B}, and their great strength lies in the fact that they use little or no {\it a priori} assumptions, making them powerful, independent checks on the model-dependent approach to interpretation of interferometric data.

An excellent example of image reconstruction can be found in \S 3 of \citet{Zhao2009}, who use the Maximum Entropy Method of radio synthesis imaging \citep{Narayan1986ARA&A..24..127N} as developed by \citet{Ireland2006SPIE.6268E..58I} into the ``Markov-Chain Imager for Optical Interferometry'' (MACIM) code.  Blind tests of the results from MACIM \citep{Lawson2006SPIE.6268E..59L} have shown it to be a robust tool for image reconstruction with optical interferometric data.  For the application to rapid rotators, \citet{Zhao2009} incorporated an ellipse prior, justified by the reasonable expectation that the stellar photosphere has a sharp cutoff at the edge.  This prior was generated through MACIM imaging of a grid of uniform surface brightness ellipses.  Additionally, a Gaussian beam deweighting (as is common in radio synthesis imaging) was also applied, for image smoothness.  Following these considerations, the final image was the one with the global maximum entropy.
As closure phase precision progresses to levels much better than a fraction of a degree \citep{Zhao2011PASP..123..964Z}, the fidelity of these image reconstructions will progress as well.

\subsection{Spectroscopy and Spectro-interferometry}\label{sec_spectrointerferometry}

Another ongoing, extremely powerful extension of these interferometric techniques is the combination of them with unresolved spectroscopy, and extension of the interferometric observables through spectral dispersion as well.  For example, the rapidly rotating model generated to fit the interferometrically observed stellar surface brightness distribution of Vega \citep{Aufdenberg2006ApJ...645..664A} was upgraded to generate high dispersion spectra \citep{Aufdenberg2007IAUS..240..271A} and found that the high rotational velocity - necessary to reproduce the surface brightness distribution  - was incompatible with the lower rotational velocity preferred by the line fits.  These observations were taken to suggest that a simple von Zeipel gravity darkening law and uniform rotation are insufficient to describe the stellar photosphere.  \citet{Yoon2010ApJ...708...71Y} demonstrate a further extension of the approach that attempts to unify high-resolution spectroscopy with interferometry, deducing that Vega's metallicity and mass are both lower than previously expected.

Additionally, the interferometric beam combiner may itself spectrally disperse the recombined light, permitting a wavelength-dependent analysis of the stellar surface brightness distribution.  Sufficiently high resolution dispersion, such as the $R \sim 12,000$ available with the VLTI AMBER instrument \citep{Petrov2007A&A...464....1P}, make it possible to examine spectral line profiles in a spatially resolved or even imaged sense.  AMBER has already been used for spectro-interferometric observations of the Herbig Be star MWC 297 as this high spectral resolution \citep{Weigelt2011A&A...527A.103W}.


\section{Stars Studied to Date}\label{sec_stars}

An increasing number of facilities have been successfully employed to make a measurement of rotationally-induced oblateness, starting in 2001 with the Palomar Testbed Interferometer \citep[PTI,][]{Colavita1999ApJ...510..505C}, the Very Large Telescope Interferometer \citep[VLTI,][]{Glindemann2003Ap&SS.286...35G}, the Navy Prototype Optical Interferometer \citep[NOI,][]{Armstrong1998ApJ...496..550A}, and by 2005 the Center for High Angular Resolution Astronomy (CHARA) Array \citep{tenbrummelaar2005ApJ...628..453T} was demonstrating the technique as well.  These interferometric observations were interpreted with parametric fits similar to the example described in \S \ref{sec_obsquans}; recently, interferometric imaging (\S \ref{sec_image_reconstruction}) has been demonstrated as a powerful tool in investigating these objects in a significantly more model-independent approach.  Below we will review the studies to date.

\include{tables/tab_results_summary}

\subsection{Altair }\label{sec_stars-Altair}
Of the rapidly rotating stars, Altair ($\alpha$ Aql, HR7557, HD187642; $\alpha$=19$^h$50$^m$47.00$^s$, $\delta$=+08$^o$52$^m$05.96$^s$) is one of the nearest, brightest prototypes, and located close enough to the celestial equator to be observed by both northern- and southern-hemisphere facilities \citep[$d=5.143\pm0.024$ pc, $m_V$=0.77, $m_K$=0.10, $\alpha$=19$^h$50', $\delta$=+08$^o$52';][]{Perryman1997A&A...323L..49P, Cutri2003tmc..book.....C}. Altair is an A7IV-V star \citep{Johnson1953ApJ...117..313J}, the 12th brightest of all stars, and has been known to be a rapid rotator for over half a century \citep[240 km s$^{-1}$ from][]{Slettebak1955ApJ...121..653S}. These characteristics made it an attractive target for the first interferometric observations of rapid rotation, and continue to make it appealing for instrument commissioning.

The interferometric observational implications were first considered theoretically in the PhD theses of \citet{Jordahl1972PhDT.........5J} and \citet{Lake1975}, who investigated the effects that rapid rotation would have upon the shape, brightness distribution, and spectral features of that particular star.  The observations of the object's size by the Narrabri Intensity Interferometer in 1965 were published \citep{HanburyBrown1974MNRAS.167..121H}, but with only a uniform disk fit to the data of $2.78\pm0.13$ mas.

More than 25 years would pass between the publication of \citet{HanburyBrown1974MNRAS.167..121H} and the first direct observation of rotational oblateness by \citet{vanbelle2001ApJ...559.1155V}, with Altair as the target.  During that time, the possibility of making such a measurement was consciously known by investigators in the field, and attempts were even made at such a measurement on various targets, without success.  However, using the Palomar Testbed Interferometer (PTI), we were able to take advantage of a number of aspects of the instrument to establish departures from spherical symmetry.  First, PTI provided at the time of the observations two baselines of sufficient length ($>80$ m) to resolve the $\sim 3.2$ mas disk of Altair in the K-band.  The baselines also provided sufficient $\{u,v\}$ coverage on the sky, such that the angle between those baselines when projected onto the sky was on order $\sim 50^o$.
Second, the visibility data quality was such that the night-to-night repeatability was (under appropriate weather conditions) on order $\sigma_{V^2} \simeq 0.018$, which was necessary to compare the differing baseline's data with sufficient precision to indicate departures from spherical symmetry.  PTI's highly automated operations also allowed for rapid collection of a large body of data points with each baseline as well.
Finally, the nature of the experimental design showed the apparent change in interferometric visibility with sky projection angle was {\it not} seen with data collected contemporaneously on the check star, Vega, as expected from that star's small $v \sin i$ value \citep[$21.9 \pm 0.1$ km s$^{-1}$;][]{Hill2004IAUS..224...35H}.

Collection of the observational data was only the first necessary step in achieving the confirmed detection of rotational oblateness.  The proper direction for interpretation of the data was, at first, unclear, given the novel nature of the investigation at the time.  Many of the publications of the time that discussed stellar rotation concentrated upon spectral line profile shapes but not the apparent shapes of the stellar disks.  Indeed, some of the best early guidance was found in solar system literature discussing the shapes of the rapidly rotating gas giants \citep[e.g.,][]{Baron1989Icar...78..119B}.  However, as recommended to us by George Collins (private communication), we found guidance in using the dissertation of \citet{Jordahl1972PhDT.........5J} set against a computational methodology similar to that employed for reducing interferometric binary star data  \citep[e.g.,][]{bkv99}.  This framework made it possible to to extend the result from a mere toy model for a star with an elliptical on-sky appearance, to a physically meaningful family of dynamic Roche models of various inclinations with a single characteristic value of $v \sin i=210\pm 13$ km s$^{-1}$, which was then compared, successfully, against the spectroscopic values.

This 2001 investigation contains one notable error\footnote{Well, at {\it least} one - only one that we know about currently.}, which fortunately does not alter the physics or the projected rotational velocity result: the $\{u,v\}$ coordinates were inadvertently swapped, resulting in an incorrect on-sky orientation angle (for example, Figure 6 in that manuscript should be rotated about the $\alpha=\delta$ diagonal).  The reported value in the manuscript of $-21.6\pm6.2 \deg$ corrects to $-68.4\pm 6.2 \deg$ when this is taken into account.

This result on Altair quickly set the stage for further investigations of the phenomenon, with a variety of instruments and technique coming to bear on this particular object, along with other additional rapid rotators, beginning with Achernar the following year (as will be discussed in the next section, \S \ref{sec_stars-Achernar}).  This result was noted in the general astrophysical review literature as being a significant development \citep{Trimble2003PASP..115..514T}.
With Altair, \citet{Ohishi2004ApJ...612..463O} followed the initial result with observations of the star with the Navy Optical Interferometer (NOI).  These investigators not only confirmed the oblate nature of the stellar photosphere, but reported a detection of a non-zero closure phase, consistent with a bright polar region produced by gravity darkening.  This result was expanded upon with the addition of observations from the Very Large Telescope Interferometer (VLTI) by
\citet{domicianodesouza2005A&A...442..567D}, who synthesized together the PTI, NOI, and VLTI data sets into a unified solution.
The following NOI observations and analysis by \citet{Peterson2006ApJ...636.1087P} was able to unambiguously establish a stellar inclination for Altair.

Ultimately, these studies collectively set the stage for the ground-breaking study by \citet{Monnier2007Sci...317..342M}, which was the first to directly image a main sequence star.  The data from the MIRC instrument on the CHARA Array directly confirmed the presence of gravity darkening using an image reconstruction independent of the previous parametric modeling approaches, while at the same time providing evidence that departures from the standard formulation of that effect were insufficient to explain the observations.  Using a value of $\beta=0.25$, the investigators were able to fit their data in a reasonably satisfactory way, but their goodness-of-fit criteria significantly improved when this was allowed to be a free parameter, which converged on a value of $\beta=0.190 \pm 0.012$.  Differential rotation, opacity, and convection-related phenomena are all cited as possible reasons for this value of $\beta$; detailed spectro-interferometric line profile analysis (as noted in \S \ref{sec_spectrointerferometry}) is suggested as the preferred tool to disentangle which of these effects is the dominant one.


Given its size and brightness, one obvious further application of interferometric observations of Altair would be an attempt to directly observe stellar rotation with time-resolved imaging.  The $\sim10$ hour rotational period of star would have to be accounted for in such observations - no one exposure could be overly long, lest it risk smearing the image; `snapshot' imaging would be a must.  However, this could in fact be an advantage: during a single evening of observing, the full face of the star would be seen over time, allowing surface morphology maps to be built on a night-to-night basis.  Such imaging would require a non-homogenous surface, but that may be possible (even likely) given the known magnetic activity of Altair \citep{Robrade2009A&A...497..511R}.


\include{tables/tab_timeline}

\subsection{Achernar ($\alpha$ Eri)}\label{sec_stars-Achernar}

Following on the heels of the Altair observations, Domiciano de Souza and collaborators presented theoretical details on modelling rapid rotators \citep{DomicianodeSouza2002A&A...393..345D} which were soon applied to VLTI observations of the rapidly rotating Be star, Achernar \citep[$\alpha$ Eri, HR472, HD10144; $\alpha$=01$^h$37$^m$42.85$^s$, $\delta$=-57$^o$14$^m$12.33$^s$;][]{DomicianodeSouza2003AA...407L..47D}.
This object is the southern hemisphere complement to the northern hemisphere object Altair: it is bright and presents an remarkably large variation of angular size as a function of sky orientation.  In fact, in evaluating their VLTI data as a solitary, rapidly rotating photosphere, the investigators found an extreme oblateness ratio of  $a/b = 1.56 \pm 0.05$ - a degree of photospheric distortion that could not be explained by Roche approximation.  The authors claimed that the lack of H$\gamma$ line emission during their observations was indicative of no disk contamination of the interferometric visibility data, and that ``the classical assumption of the Roche approximation becomes questionable''.  Investigations of more exotic photospheric models were spurred onwards by this claim \citep[e.g.,][]{Jackson2004ApJ...606.1196J}.

However, ensuing investigations of this object \citep{Kervella2006A&A...453.1059K,Carciofi2008ApJ...676L..41C} branched out to add contributions from a circumstellar environment (CSE) to model, which lessen the derived oblateness ratios; \citet{Kervella2006A&A...453.1059K} thereby derive the photospheric oblateness ratio dropped to $a/b = 1.41 \pm 0.053$ with a photosphere-to-CSE relative flux ratio of $4.7\pm0.3\%$
Additionally, there has been detection of a close-in companion by \citet{Kervella2008A&A...484L..13K}; however, inclusion of the companion and a rotationally distorted star alone in the visibility fits cannot explain the interferometry data, and a CSE contribution is still required \citep{Kanaan2008A&A...486..785K}.
Companion thermal-IR interferometric observations with VLTI-MIDI has provided data consistent with extended emission associated with a fast wind ejected along the hot polar caps of star \citep{Kervella2009A&A...493L..53K}.

Clearly this particular object has found a vigorous existence in the realm of interferometric study.  However, in contrast to Altair, the complications of disks and winds have presented additional challenges (and rewards) for investigators attempting to fully fit models for Achernar.  As the brightest Be star prototype, however, this object is uniquely situated to allow in-depth interferometric investigations to connect the effects of stellar rotation to the overall Be star phenomenon.

\subsection{Regulus ($\alpha$ Leo)}\label{sec_stars-Regulus}

As with Altair, Regulus ($\alpha$ Leo, HR 3982, HD 87901, ADS 7654A; $\alpha$=10$^h$08$^m$22.31$^s$, $\delta$=+11$^o$54$^m$01.95$^s$) was observed by the Narrabri Intensity Interferometer \citep{HanburyBrown1974MNRAS.167..121H}, with similar result: insufficient $\{u,v\}$ coverage prevented an in-depth characterization of this rapidly rotating star, although a gross size characterization was reported.  Regulus had been recognized by \citet{Slettebak1954ApJ...119..146S} as a B7V rapid rotator with $v \sin i$= 350 km s$^{-1}$, making this object an attractive target for such observing.

Observations of Regulus in 2004 were given the auspicious place of being the inaugural science investigation for the CHARA Array \citep{McAlister2005ApJ...628..439M}, operated by the Georgia State University \citep{tenbrummelaar2005ApJ...628..453T}.  In addition to demonstrating the capabilities of the new array, this investigation was notable in the specific context of rapid rotator study for the following reasons:  First, it marked the first time a combined interferometric and spectroscopic approach was applied to the data reduction methodology. Second, it was the first such investigation that resulted in a reasonably specific constraint on the inclination of the rotation axis for such an object.
As of the publication of the Regulus results, previous investigations had painted broad suggestions on what various objects' true inclinations would be, but no specific values had be offered.  The Regulus error was still quite generous ($i = 90^{+0}_{-15}$ deg) and yet specific, representing one of the ultimate benefits of this manner of investigation.

The resulting solution set for Regulus is a rather rich characterization of the object, including values for inclination $(i)$, on-sky orientation $(\alpha)$, gravity darkening $(\beta)$, rotational velocity $(v)$, fractional rotational velocity $(v_e / v_c)$, polar and equatorial radii $(R_p, R_e)$, polar and equatorial temperatures $(T_p, T_e)$, mass $(M)$, luminosity $(L)$, and interstellar extinction $A_V$.  Such a rich data set allows for in-depth evaluation of the object on a variety of fronts.  For example, comparison of the on-sky orientation with its known proper motion shows that Regulus's space motion is roughly along the axis of its rotational spin.  This observation is thought-provoking from the context of considering the formation history of the star and the dynamics of its protostellar cloud.

More specifically, providing a the true inclination solution had far-reaching consequences.  From the inclination, the true rotation velocity and fractional rotational velocity could be established.  The degree to which such objects are rotating close to their breakup velocity has significant implications regarding the nature of mass loss in Be stars \citep[e.g.,][]{Porter2003PASP..115.1153P}.  The range in temperature from the pole to the equator has cast doubt on the `young' age estimate of 150 Myr \citep{Gerbaldi2001A&A...379..162G} and a significantly older estimate of 1 Gyr has been suggested \citep{Rappaport2009ApJ...698..666R}.  The recent discovery of close companion by
\citet{Gies2008ApJ...682L.117G} to Regulus\footnote{In addition to its close companion, Regulus has a wide companion $\alpha$ Leo B at a separation of $\sim$175'', also a binary.  However, at such a distance, it will not have ever interacted with Regulus \citep{McAlister2005ApJ...628..439M}.} has led to consideration of the object's `huge' quadrupole mass moment $Q$ \citep{Iorio2008Ap&SS.318...51I}; estimates of $Q$ indicate that detection of the correction to the Keplerian period due to $Q$ would be possible with period measurements precise to $\sim 20$ seconds; however, the current state of the art for the period measurement is about two orders of magnitude beyond that at $P=40.11 \pm 0.02$ days \citep{Gies2008ApJ...682L.117G}.

Also as with Altair, Regulus has now been imaged by CHARA-MIRC \citep{Che2011ApJ...732...68C}.  The broad details of the earlier \citet{McAlister2005ApJ...628..439M} study have been validated by \citet{Che2011ApJ...732...68C} (Table \ref{tableData_allStars}), but further rich detail has been added from the imaging.  In particular, a much tighter constraint on inclination is now available ($i=86.3^{+1.0}_{-1.6}$), along with tighter constraints on pole and equator temperature, and further strong evidence of departures from the `standard' von Zeipel gravity darkening, with $\beta=0.188^{+0.012}_{-0.029}$.

\subsection{Alderamin ($\alpha$ Cep)}\label{sec_stars-Alderamin}

A second CHARA Array commissioning target, Alderamin ($\alpha$ Cep, HR 8162, HD 203280; $\alpha$=21$^h$18$^m$34.77$^s$, $\delta$=+62$^o$35$^m$08.06$^s$), was also observed with the array, also benefiting from the large amount of $\{ u,v \}$ coverage afforded by the facility's multiple baselines \citep{vanbelle2006ApJ...637..494V}. Alderamin is an A7IV star \citep{Johnson1953ApJ...117..313J}, making it an attractive observing candidate given its similarities to Altair.  Alderamin is 3$\times$ further than Altair but, strangely, at roughly the same X-ray luminosity \citep{Robrade2009A&A...497..511R}; this suggests that coronal X-ray emission sets in at about A7 for main sequence stars (at least at the sensitivity levels of current instrumentation).
Alderamin has been known to be a rapid rotator for many decades \citep{Slettebak1955ApJ...121..653S}, with spectroscopic estimates rotational velocity that range from $v \sin i$=180-200 km s$^{-1}$ \citep{Gray1980PASP...92..771G,Abt1995ApJS...99..135A} to 245-265 km s$^{-1}$ \citep{Abt1973ApJ...182..809A,Bernacca1970CoAsi.239....1B}. This object was recently re-observed \citep{Zhao2009} with the array using the newer multi-way MIRC combiner, which not only provided visibility amplitude data on baseline pairs but closure phase data from baseline triplets.

Both the \citet{vanbelle2006ApJ...637..494V} and \citet{Zhao2009} investigations attempted to fully characterize the fundamental parameters associated with the rapidly rotating stellar photosphere, including inclination $i$, sky position angle $\alpha$, polar and equatorial radii and temperatures, and most significantly, gravity darkening $\beta$.
As noted in \S \ref{sec_vonZeipel}, a single value of $\beta$ is perhaps an incomplete characterization of the true physical nature of the star's surface, particularly given the range of temperatures, from $T_{\rm pole}=8600 \pm 300$K to $T_{\rm equator}=6600 \pm 200$K; following the canonical temperatures of static models found in \citet{cox00}, we see this ranges from spectral types $\sim$A3V to $\sim$F5V, respectively.  This range of temperatures brackets the radiative-to-convective transition temperature of
$\sim 8300$K \citep{Christensen-Dalsgaard2000ASPC..210..187C,Neff2008ApJ...685..478N},
allowing for equatorial convective activity but no similar polar activity \citep{Rachford2009ApJ...698..786R}.

Contrasting the results from the earlier \citet{vanbelle2006ApJ...637..494V} study with that of \citet{Zhao2009} is instructive in illustrating the power of various interferometric data sets.  The former, having only visibility amplitude data available (only the 2-way CHARA Classic combiner was online at the time of the study), came up with an inclination much larger than the latter ($88.2^{+1.8}_{-13.3}$ versus $55.70 \pm 6.23$ degrees).  Simply put, the former was rather insensitive to the asymmetric stellar surface brightness distribution, which manifests itself in the poor value for $i$ and (probably unreasonably) low value for $\beta$; the latter, using CHARA-MIRC data which produces data sets not just with size-sensitive visibility amplitude data, but also with asymmetry-sensitive closure phase data, was able to produce a significantly more precise - and presumably accurate - values for $i$ and $\beta$.  This in turn affected the entire ensemble of the values in the solution set; it is interesting that the best value for gravity darkening still deviates from the `von Zeipel ideal' with $\beta=0.216 \pm 0.021$.

It is also noteworthy that the similar visibility-only solution for Regulus in \citet{McAlister2005ApJ...628..439M} also found a lower value for $\beta$ than the spectroscopy/visibility solution found in the same paper ($\beta = 0.13\pm 0.05$ versus $0.25 \pm 0.11$).  Investigators may wish to consider if such data sets are insufficently rich to providing constraints upon the stellar surface morphology to characterize gravity darkening: an excellent example of this can be seen in Figure 12 of \citet{Zhao2009}, where for certain on-sky orientations, degeneracies may form between various parameters (in this case, $\beta$ and $i$, when $i$ is large).  In particular, for those investigators wishing to establish interferometry-only data sets to determine absolute rotation velocities for comparison with spectroscopic $v \sin i$ values, closure phase characterizations of asymmetric surface bright distributions will be highly sensitive to rotation axis inclination values.

Extending the implications of the observations, in \citet{vanbelle2006ApJ...637..494V}, a simple analysis of the rotation history of the star via the simple application of conservation of angular momentum was considered, and implied that its present fractional breakup velocity ($v_e / v_c=0.83 \pm 0.05$ - similar to the \citet{Zhao2009} value, so this line of reasoning is unaffected by the poor values of $i$ and $\beta$ in that study) would have been even higher in its zero-age main sequence history, with $v_e / v_c \simeq 0.92-0.98$.  Evidence of such an large earlier rotation rate has significant implications for the star's formation history.

In their study of the onset of convective zones in main sequence stars, both Altair and Alderamin are cited by \citet{Neff2008ApJ...685..478N} as cases that demonstrate the lack of fidelity in categorizing stars solely by spectral type or mass.  These same investigators found no {\it FUSE} UV evidence for deepening of convective zones due to the rapid rotation of these objects, as suggested not only by the interferometry data but also by the models of \citet{MacGregor2007ApJ...663..560M}.

\subsection{Vega ($\alpha$ Lyr)}\label{sec_stars-Vega}

Vega ($\alpha$ Lyr, HR7001, HD172167; $\alpha$=18$^h$36$^m$56.34$^s$, $\delta$=+38$^o$47$^m$01.29$^s$) is a particularly interesting special case of an observed rapid rotator in that it is viewed nearly pole-on relative to the other examples cited thus far.
The long-standing use of Vega as a spectrophotometric standard \citep[e.g., ][]{Hayes1967PhDT.........7H,Straizys1976A&A....50..413S} makes understanding the surface brightness distribution of this object at the highest levels of detail more than an abstract scientific exercise, and one of fundamental utility.
The discovery of Vega's debris disk \citep{Aumann1984ApJ...278L..23A} due to its mid-infrared excess flux called into question its validity as a fundamental standard.
This situation has been further complicated by the fact that the star is a rapid rotator, as first suggested by \citet{Gray1988JRASC..82..336G}.  This result, based upon detailed spectroscopic line modelling, has been confirmed by interferometry, as we shall see below.

As with Altair and Regulus, \citet{HanburyBrown1974MNRAS.167..121H} observed Vega and obtained a limb-darkened angular size of $3.24 \pm 0.07$ mas (and a uniform disk size of $3.08\pm0.07$ mas) through a $443\pm5$nm filter.  As with all of the objects in this review, the intervening generation of optical interferometer prototypes after NII - GI2T, IRMA, IOTA (to name a few) - did not observe Vega as a scientific target, due to limited spatial resolution.  Further scientifically significant data from interferometer on this object came with the longer baselines of PTI when \citet{Ciardi2001ApJ...559.1147C} observed it and found puzzling signs of residuals in the $K$-band fits, consistent with a debris disk signal contaminating the stellar photospheric signal \citep[as discussed in more detail in][]{Barnes2009ApJ...705..683B,Lawler2009ApJ...705...89L,Akeson2009ApJ...691.1896A}.  This finding was consistent with the further investigations with the CHARA Array \citep{Absil2006A&A...452..237A} and IOTA \citep{Defrere2011arXiv1108.3698D}.  These PTI and CHARA Array interferometric studies set the stage for two further, more detailed, studies of the star's photosphere itself.

Both NOI \citep{Peterson2006Natur.440..896P} and the CHARA Array \citep{Aufdenberg2006ApJ...645..664A} published studies of Vega early in 2006, in the visible ($\sim$500-800nm) and $K'$-band, respectively.  Both studies found strong evidence for rapid rotation in a nearly pole-on star, arriving at remarkable agreement in inclination ($4.54\pm0.33^o$ and $4.7\pm0.3$), equatorial rotational velocity ($274\pm 14$ and $270\pm15$), and other parameters (see Table \ref{table_vega}) despite completely independent development of their methodologies and differing observational wavelengths.  Agreement between these two disparate approaches is reassuring; however, further investigations by the NOI group \citep{Yoon2008ApJ...681..570Y,Yoon2010ApJ...708...71Y} indicate a lower rotation speed ($v_{\rm eq} = 175 \pm 33$) coupled with lower than previously expected mass and sub-solar metallicity ($2.14 \pm 0.08 M_\odot$, $Z=0.0080 \pm 0.0033$).

\include{tables/tab_vega}

Interferometric results indicating a `fast' rotation speed of $\approx 270$ km s$^{-1}$ are disputed in the spectroscopic analysis by \citet{Takeda2008ApJ...678..446T}, who favor a more `moderate' speed of $\approx 175$ km s$^{-1}$.  However, using the parameters from \citet{Peterson2006Natur.440..896P}, \citet{Yoon2008ApJ...681..570Y} evaluate ELODIE archival spectra of Vega and invoke turbulence on large scales (`cyclones', even, to use their language) in their analysis to achieve a match between the data and their calculated spectral lines.  The implications of such a result are intriguing, to say the least: these results are themselves rather dramatic, but the underlying approach of interferometric observations to guide spectroscopic analysis could be a far-reaching in application to more than just studies of rapidly rotating stars.

It is interesting to note that there has been no detection of X-rays by CHANDRA even after 29ks of observing \citep{Pease2006ApJ...636..426P}, even though (as noted above) they have been detected for Altair and Alderamin \citep{Robrade2009A&A...497..511R}.
The overall set of `problems' with Vega are reviewed in the article by  \citet{Gray2007ASPC..364..305G}, of which the interferometrically detected phenomena of IR excess and rapid rotation play a large role.  Ultimately, all of these various pathologies may lead astronomers to use stars other than Vega for photometric standards, as already proposed by \citet{Engelke2010AJ....140.1919E}.

\subsection{Rasalhague ($\alpha$ Oph)}\label{sec_stars-Rasalhague}

Rasalhague ($\alpha$ Oph A, HR6556, HD 159561; $\alpha$=17$^h$34$^m$56.07$^s$, $\delta$=+12$^o$33$^m$36.13$^s$), due to a number of unique circumstances, is a particularly interesting rapid rotator and is quickly becoming a fundamental laboratory for exploring physics of rapid rotation. As with many of these objects, it was originally observed interferometrically by \citet{HanburyBrown1974MNRAS.167..121H}.

An A5IV \citep{Gray2001AJ....121.2148G} $\delta$ Scuti variable star, $\alpha$ Oph A has been observed to be a nearly edge-on rapid rotator with $\omega / \omega_c \sim 0.88$ with the imaging study of \citet{Zhao2009}.  As noted above in the discussion for Alderamin (\S \ref{sec_stars-Alderamin}), the edge-on orientation makes solving for $\beta$ problematic, so \citet{Zhao2009} proceeded with a fixed value of $\beta=0.25$.   \citet{Monnier2010ApJ...725.1192M} observed $\alpha$ Oph A with the {\it MOST} astroseismology satellite, noting that this inclination is particularly favorable (odd-parity $l-|m|$ modes are suppressed, simplifying mode identifications) and detected rotationally modulated $g$-modes in the object.  The (unexpected) linear relationships between $g$-mode spacings were provisionally explained as dispersion-free Kelvin waves.
\citet{Hinkley2011ApJ...726..104H} present an improved astrometric orbit of $\alpha$ Oph A and B from 8 years of AO imaging data, and point out that a significantly improved determination of the 7.9-year astrometric orbit will be possible during the $\sim$50 mas April 2012 periastron, using optical interferometry.  This in turn will permit mass determination of $\alpha$ Oph A at the few percent level.  Collectively, these studies allow Rasalhague to be a prototype for probing the physics of stellar interiors, and understanding the effects of rapid rotation in that regime.

\subsection{$\beta$ Cas (Caph)}\label{sec_stars-betaCas}

The most recent addition to the stable of rapid rotators observed with optical interferometry is $\beta$ Cas \citep{Che2011ApJ...732...68C}.  Its spectral type of F2III-IV \citep{Rhee2007ApJ...660.1556R} indicates it is a `retired' A-type star, evolving off the main sequence, and makes this observation a noteworthy one, in that it is the first object that is not (currently) `A' or `B' spectral type.  As a cooler object, it represents an expansion of the tests derived from this technique into new areas of discovery space.  The rotational velocity of $\sim 70$ km s$^-1$ \citep{Rachford2009ApJ...698..786R} did not, on its face, give an {\it a priori} suggestion that it was an rapid rotator, although the small value of $P / \sin i$ reported therein (2.48$^d$) certainly hinted at the possibility.

Observationally, $\beta$ Cas presents itself nearly pole-on, with $i=19.9 \pm 1.9 ^o$, indicating a significant rotation rate ($v=206$ km s$^{-1}$, $P=1.12^{+0.03}_{-0.04}$ days).  As with Regulus, \citet{Che2011ApJ...732...68C} also fit for $\beta$, finding a value of $0.146^{+0.013}_{-0.007}$.  This lower value is consistent with trends toward a lower expected value for a star with lower surface temperature \citep[see Figure 9 of \citet{Che2011ApJ...732...68C}, and][]{Claret2000A&A...359..289C}, although it is not nearly as small as would be expected for a fully radiative photosphere ($\beta$=0.08).


\section{Broad Impact in Astronomy}\label{sec_impact}

Although there is only a small number of results thus far, there are already broad and significant impacts throughout astronomy as the implications of such detailed new views of stars diffuse through the field.  For example, the particulars of the analysis of the spectral energy distribution of Vega already noted above \citep{Aufdenberg2006ApJ...645..664A} significantly revises our view of this fundamental standard \citep{Rieke2008AJ....135.2245R}.

The interaction between spectroscopy and rapid rotator analysis has a long history, notably from the work by \citet{Collins1963ApJ...138.1134C,Collins1965ApJ...142..265C,Collins1966ApJ...146..152C}.  The interferometric observations have begun to guide spectroscopy - for example, discrepancies in $T_{\rm EFF}$ indicated in H$\alpha$ and H$\beta$ Balmer line profiles in Altair found in \citet{Smalley2002A&A...395..601S} and \citet{Smalley2003IAUS..210P.C10S} are resolved when the lower effective temperatures indicated by interferometry are used.  Similarly, as predicted in \citet{Smalley2002A&A...395..601S}, subsequent observations of Rasalhague also indicate a lower average $T_{\rm EFF}$, particularly given the edge-on orientation of this object, which resolve similar temperature anomalies. Such agreements are contrasted with the example of $v \sin i$  estimates for Vega, which even when account for pole-on rapid rotation \citep{Hill2004IAUS..224...35H} fall short of the true magnitude of the star's extreme rotational speed \citep{Aufdenberg2007IAUS..240..271A}.  Similarly, a line profile analysis of Altair incorporating not just the Balmer lines but a larger 650-line list \citep{Reiners2004A&A...428..199R} does not quite fall into agreement is the most recent values for inclination and rotation speed \citep{Monnier2007Sci...317..342M}.  Analyses are now being forced to cast a wider net and consider such things as revised metallicities \citep{Yoon2010ApJ...708...71Y} in order to reconcile spectroscopy and interferometry, a move which perhaps mirrors other upheavals in astronomy related to metallicity \citep{Asplund2009ARA&A..47..481A}.

Folding details of rapid rotation from these observations into spectroscopic modeling has not been prevented by these challenges, however.  The line profiles of \citet{Huang2006ApJ...648..591H} are built using hemispheric averages for $\{T_{\rm EFF}, \log g, v \sin i \}$ in their analysis using line profiles to investigate evolution of stellar rotation in young clusters.
\citet{Neff2008ApJ...685..478N} examined \ion{O}{6} emission lines in detail in the context of the rapid rotation reported for Altair and Alderamin, to explore the onset of convection zones as made possible by rotational equatorial cooling - their detection of such emission in a variety of stars indicates that magnetic activity without substantial convective zones needs to be possible in current models.
\citet{Khalack2005A&A...429..677K} also point out that magnetic dipole strength can be overestimated if stellar oblateness is not taken into account.
Combining interferometric and spectroscopic observations is suggested as a way to probe the properties of the rotation law of the external layers of rapidly rotating stars \citep{Zorec2011A&A...526A..87Z}.


Underlying spectroscopy, observational characterization of rapid rotation has also started to make significant inroads with stellar modeling \citep{DomicianodeSouza2002A&A...393..345D}.
Comparison of the results of observed rapid rotators to stellar models is often difficult, in that typical models are constructed for non-rotating stars.  \citet{Che2011ApJ...732...68C} take the approach of applying corrections from predictions such as \citet{Sackmann1970A&A.....8...76S} to facilitate comparison to $Y^2$ models \citep{Yi2001ApJS..136..417Y}, with some success, although they note further work is warranted.

From an examination of collective results to date, \citet{Che2011ApJ...732...68C} suggests that for general use in modeling rapidly rotating stars, a gravity darkening value of $\beta=0.19$ is more appropriate.  The interferometric results seem to validate the suggestion in \citet{Tassoul2000stro.book.....T} that solid-body rotation is impossible for a rapid rotator: temperature and pressure consistency across the stellar surface is disrupted by such rotation.  This disruption leads to temperature and pressure gradients between the poles and equator, which in turn induces meridional circulation and breaks down strict radiative equilibrium. The differential angular momentum of the meridional circulation's matter flow - the higher latitude material carries less angular momentum - will then cause differential rotation.

Stellar rotation affects abundance, mass loss rates, and overall evolution \citep[for a discussion of the implications for the most massive stars, see the 13-part series that begins with][]{Meynet1997A&A...321..465M}.
New grids are in the process of being produced  which incorporate the observational interferometric results, including smoother transition values for gravity darkening  \citep{Claret2005A&A...440..647C}.  Alternatively, the basic parameters being discovered through interferometry are being used as anchor points in stellar evolution models to `run the clock backwards', and explore the history of parameters such as angular momentum \citep{vanbelle2006ApJ...637..494V}.


Beyond of spectroscopy and modeling, there are a broad number of areas where these observational results are being considered as noteworthy.  The naturally complementary techniques of asteroseismology and interferometry are seen as deeply connected on the point of rotation by \citet{Cunha2007A&ARv..14..217C} and \citet{Aerts2009A&A...508..409A}; observation extension of these two techniques in a conjoined fashion has already been demonstrated in \citet{Monnier2010ApJ...725.1192M}

In considering transiting extrasolar planets, \citet{Seager2002ApJ...574.1004S} noted the stellar oblateness measured by \citet{vanbelle2001ApJ...559.1155V}, and uses that as a starting point to discuss constraining the rotation rate of those planets through measurements of the {\it planetary} oblateness.  Such measurements are expected to be possible with the high-precision data being produced by the Kepler mission.  \citet{Barnes2009ApJ...705..683B} notes that the Rossiter-McLaughlin measurements that have been employed to date to probe the spin-orbit alignments of transiting systems are difficult about rapid rotators, but suggests that the unusual and distinctive transiting lightcurves associated with such systems could instead be utilized to explore those alignments.

Evaluation of rapid rotation is seen as a necessary part of understanding microlensing signals  \citep{Han2006ApJ...645..271H}, with shapes of caustics being affected by source oblateness; \citet{Rattenbury2005A&A...439..645R} attempt to derive from microlensing event photometry the shape of the source star.  Space climate is seen as directly connected to the irradiance properties of our sun, which in turn are connected to its shape \citep{Lefebvre2007AdSpR..40.1000L,Lefebvre2005MmSAI..76..994L}.
On a fundamental level, measurement of stellar oblateness is now acknowledged as one of the basic techniques for measurement of stellar rotation \citep{Royer2005MSAIS...8..124R}.


\section{Future target list}\label{sec_future_target_list}

The seminal work of \citet{HanburyBrown1974MNRAS.167..121H} providing angular diameter sizes on 32 bright, hot, nearby stars may be reviewed in considering the importance of expanding our knowledge base of rapidly rotating stars.  The error estimates on the limb darkened angular diameters presented in this paper ranged from $2\%-14\%$, with a median value of $6.3\%$.  Cross-referencing these results against the rotational velocities readily found in the catalog of \citet{Glebocki2005yCat.3244....0G}, we find 13 objects in common.  We can estimate average linear radii $R$ and mass $M$ from the appropriate sections in \citet{cox00}, and thereby estimate oblateness using the prescription found in the Appendix of \citet{vanbelle2006ApJ...637..494V}.

\begin{figure*}
\epsscale{1}
\plotone{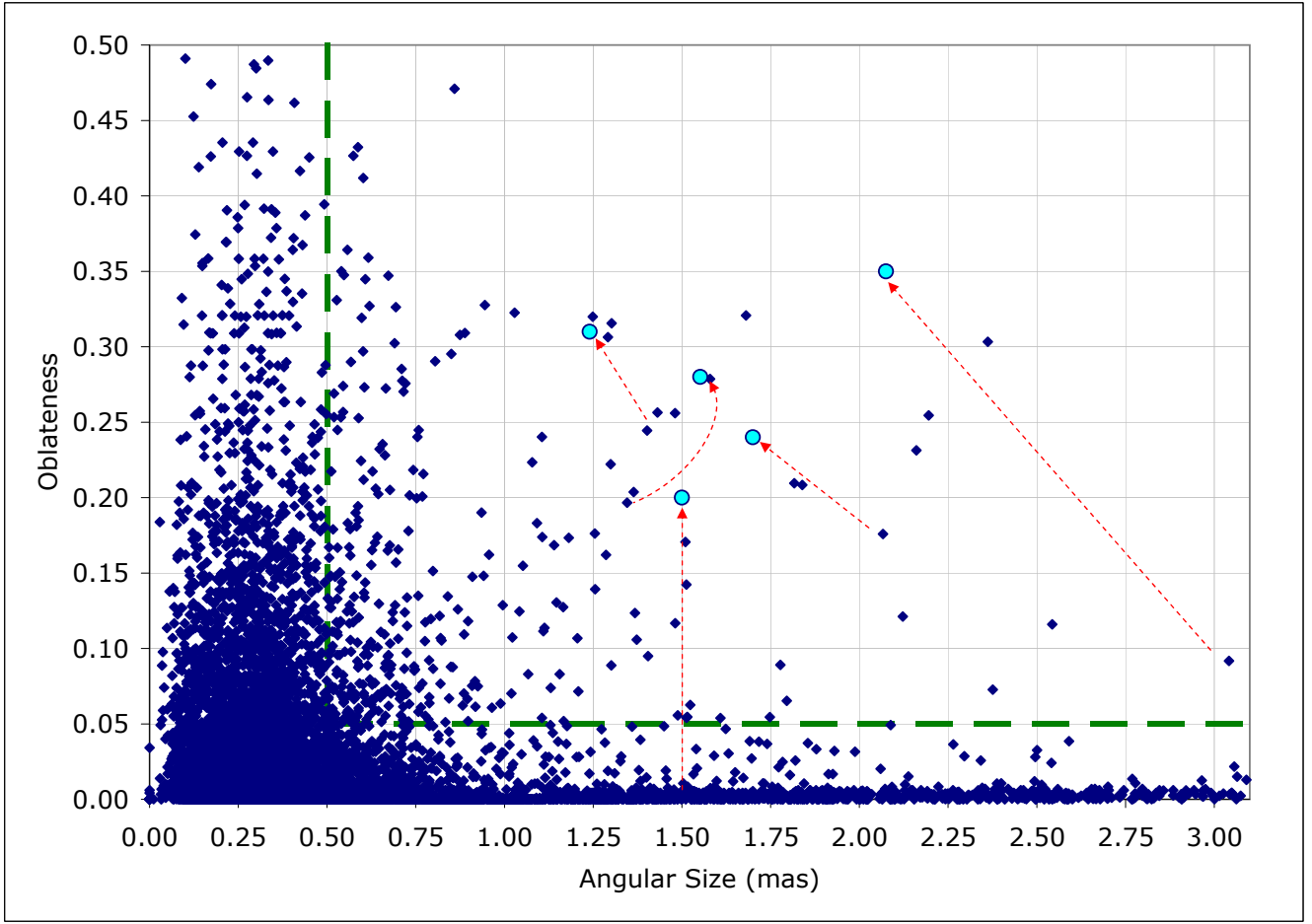}
\caption{\label{fig_future-targets} Illustration of prospective targets as outlined in \S \ref{sec_future_target_list}. Targets with {\it estimates} of $\theta>0.50$mas and oblateness $R_b / R_a-1>0.05$ are in the `allowed' region to the upper right of the dashed line.  {\it Actual} values for observed targets are shown in relationship to their proxy values via a dotted line.  Of particular interest is the clear indication that actual oblateness values are always well in excess of the simple predictions from $v \sin i$ values.}
\end{figure*}

What is instructive is to not consider the overall degree of oblateness predicted for these 13 objects, but to compare that degree of non-sphericity with the error estimates of {\it spherical} size given in \citet{HanburyBrown1974MNRAS.167..121H}.  In two cases ($\alpha$ Aql, $\alpha$ Eri), the predicted oblateness is significantly in excess ($\sim 2$-$3\times$ greater) than the spherical size quoted error; for three objects ($\beta$ Car, $\zeta$ Pup, $\epsilon$ Car) oblateness is comparable ($\sim0.9$-$1.0\times$) to the quoted error, with a fourth also being reasonably large in comparison ($\alpha$ Vir, at $0.6\times$).  Better than $45\%$ of the objects available for such a line of inquiry show strong motivation for developing a deeper understanding of their true size, {\it particularly} since the \citet{HanburyBrown1974MNRAS.167..121H} values have propagated widely in the literature.

To date, only 7 targets have been investigated with optical/near-infrared interferometry, which may leave the impression that a substantially larger sample is beyond the capabilities of current facilities.  This is simply not true.  To probe the possibilities of this manner of study, we created a list of targets that are appealing to present-day facilities.  Our assumptions were modest: angular sizes in excess of 0.50mas (roughly the current limiting angular size of modern optical interferometers), and size ratios in excess of $R_b / R_a-1>0.05$ (as dictated by the visibility precisions achievable with those facilities.) Starting with the rotational velocity catalog of \citet{Glebocki2005yCat.3244....0G}, we cross-referenced it against Hipparcos and 2MASS \citep{Perryman1997A&A...323L..49P, Cutri2003tmc..book.....C} to establish $V$ and $K$ magnitudes.  From these values, using the rough angular size predictor found in \citet{vanbelle1999PASP..111.1515V}, we estimated angular sizes and cut those stars with $<0.50$mas.  Using the quoted spectral types of Hipparcos to estimate average linear radius $R$ and mass $M$ from the nominal values found in \citet{cox00}, we then were able to estimate oblateness, cutting for $R_b / R_a-1<0.05$.

The resultant catalog of 354 objects is illustrated in Figure \ref{fig_future-targets}, of which almost a third are `Altair-class' rotators with $R_b / R_a-1>0.15$; the complete catalog is provided in an Appendix.  This catalog is, of course, incomplete, due to the imperfection of the assumptions involved, and it further reflects the incompleteness of \citet{Glebocki2005yCat.3244....0G} (although the rather uniform sky distribution suggests this degree of incompleteness is low - e.g. no bias of northern hemisphere targets over those in the south).  This approach also completely misses low-inclination rapid rotators with low values of $v \sin i$, as illustrated in the case of Caph (\S \ref{sec_stars-betaCas}).  The limitations on use of $v \sin i$ as a predictive parameter for oblateness is clearly seen in Figure \ref{fig_future-targets}: each of the targets that has been observed to date has an actual value in excess (in some cases, {\it well} in excess) of the prediction from $v \sin i$.  This of course illustrates the impact of the $\sin i$ term - and portends many fruitful future observations of oblateness with interferometry!

\section{Conclusion}\label{sec_conclusion}

\begin{figure*}
\epsscale{0.80}
\plotone{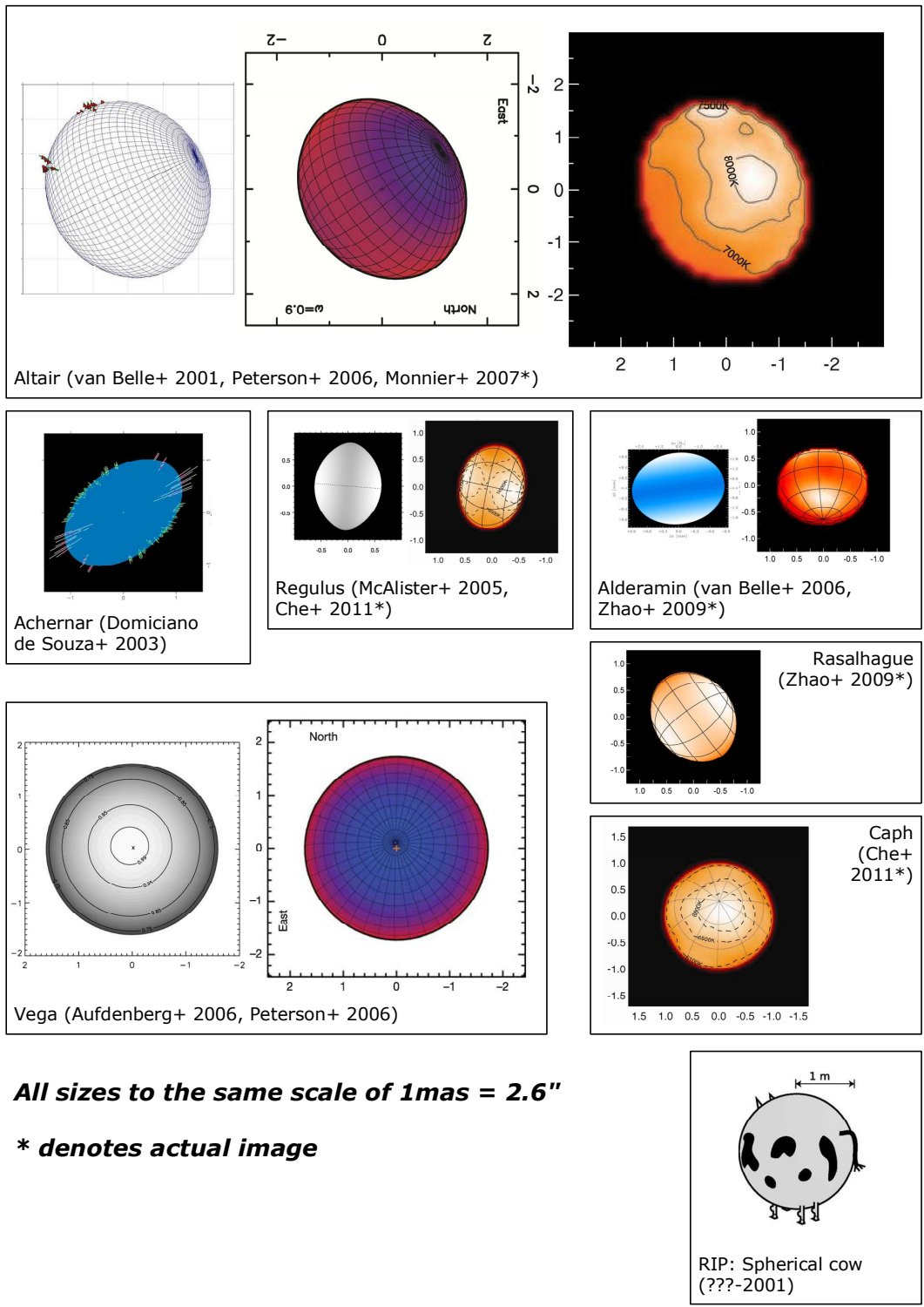}
\caption{\label{fig_stars}Illustration of figures/images of rapid rotators produced over the past 10 years by PTI, NOI, the CHARA Array, and VLTI.  All illustrations are normalized to the same page scale - relative apparent sizes are correct.}
\end{figure*}

Direct observations of rapid rotators by long-baseline interferometers have matured rapidly over the past 10 years.  Simple parametric models have given way to detailed models that probe the underlying stellar structure, and stunning images of photospheres that confirm the input physics in a dramatic, model-independent fashion (Figure \ref{fig_stars}).  The lessons learned and confidence collectively gained by the specialists carrying out these investigations is now being applied to a greater variety of objects.
As the observed sample is expanded to objects of lower temperatures, some surprises are expected and have already been found \citep{Che2011ApJ...732...68C}.

Matching this expansion of target selection has been a growth of instrumental capabilities.  The 4-telescope CHARA-MIRC instrumentation that provided the first images \citep{Monnier2007Sci...317..342M} is being expanded to full 6-way combination with sensitivity improvements through external fringe tracking \citep{Monnier2010SPIE.7734E..13M}; the first results from 4-telescope beam combination at VLTI with the PIONIER visitor instrument are very impressive \citep{Berger2010SPIE.7734E..99B};
a new NOI combiner (`VISION') will add true 6-way combination to this facility as well and is slated to be on-sky in mid-2012.  All of these developments bode well for further investigation into rapid rotators, possibly with complentary data sets spanning from the 500nm through 2.4$\mu$m, along with to increased `snapshot' capability from the richer data sets.

What has been especially gratifying to see has been the rapid ingestion of these results into a wide variety of other endeavors in the field (\S \ref{sec_impact}).  This has taken this activity from being just a `cottage industry' to being one of deep implications for many of the foundational ideas in astronomy.

\acknowledgements

The author would like to gratefully acknowledge useful discussions with Doug Gies during the preparation of this manuscript. Portions of this manuscript were prepared while the author was in residence at the European Southern Observatory (ESO). This work made use of the SIMBAD reference database, the VizieR catalogue access tool, and
the NASA Astrophysics Data System, all of which are indispensable resources for astronomy.
Funding for this research has been provided in part by Lowell Observatory.


\bibliography{interferometry-references}

\appendix

{\bf Appendix.  Prospective future targets.}

\include{tables/tab_future_targets}

\end{document}

%% file: tables/tab_results_summary.tex
\begin{landscape}
\begin{deluxetable}{llccccccccccccc}
\addtolength{\tabcolsep}{-5pt}
\tiny
\tabletypesize{\scriptsize}
\tablecolumns{13}
\tablewidth{0pc}
\tablecaption{Summary of observational results to date on rapid rotators.\label{tableData_allStars}}
\tablehead{
\colhead{} &
\colhead{Spectral} & &
\colhead{Velocity} &
\colhead{Inclination} &
\colhead{$v / v_{\rm crit}$} &
\colhead{$\omega / \omega_{\rm crit}$} &
\colhead{Orientation} &
\colhead{Gravity} &
\colhead{$T_{\rm pole}$} &
\colhead{$T_{\rm eq}$} &
\colhead{$R_{\rm pole}$} &
\colhead{$R_{\rm eq}$} &
\colhead{Oblateness} &
\colhead{} \\
\colhead{Star} &
\colhead{Type} & &
\colhead{$v$ (km/s)} &
\colhead{$i$ (deg)} &
\colhead{} &
\colhead{} &
\colhead{$\alpha$ (deg)} &
\colhead{darkening $\beta$} &
\colhead{(K)} &
\colhead{(K)} &
\colhead{($R_\odot$)} &
\colhead{($R_\odot$)} &
\colhead{} &
\colhead{Ref.}
}
\startdata
Achernar ($\alpha$ Eri) &      B3Vpe && 225\tablenotemark{a} &      $>$50 &  0.79-0.96 && $39 \pm 1$ & 0.25 (fixed) & 20000 (fixed) & 9500-14800 &    8.3-9.5 & $12.0 \pm 0.4$ & $ 0.348 \pm 0.101 $ &(5)\\
\\
\multirow{2}{*}{Regulus ($\alpha$ Leo)} &        \multirow{2}{*}{B8IVn} &
 \multirow{2}{*}{$\left\{
\begin{array}{l}
\\
\end{array}
\right.$}
& $317 \pm 3$ & $90^{+0}_{-15}$ & $0.86\pm0.03$ & $0.974\pm0.043$ & $85.5\pm2.8$ & $0.25 \pm 0.11$ & $15400 \pm 1000$ & $10300 \pm 1000$ & $3.14 \pm 0.06$ & $4.16 \pm 0.08$ & $ 0.325 \pm 0.036 $ &(8) \\
&&& $336^{+16}_{-24}$ & $86.3^{+1.0}_{-1.6}$ & $0.839 \pm 0.030$ & $0.962^{+0.014}_{-0.026}$ & $258^{+2}_{-1}$ & $0.188^{+0.012}_{-0.029}$\tablenotemark{c} & $14520^{+550}_{-690}$ & $11010^{+420}_{-520}$ & $3.22^{+0.05}_{-0.04}$ & $4.21^{+0.07}_{-0.06}$ & $ 0.307 \pm 0.030 $ &(11)\\
\\
\multirow{2}{*}{Vega ($\alpha$ Lyr)} &        \multirow{2}{*}{A0V} &
 \multirow{2}{*}{$\left\{
\begin{array}{l}
\\
\end{array}
\right.$}
& $270 \pm 15$ & $4.7 \pm 0.3$ & $0.746\pm0.034$ & $0.91\pm0.03$ &  not cited & 0.25 (fixed) & $10500 \pm 100$ & $8250^{+415}_{-315}$ & $2.26 \pm 0.07$ & $2.78 \pm 0.02$ & $ 0.230 \pm 0.039 $ &(6) \\

           &&            & $274 \pm 14$ & $4.54 \pm 0.33$ & $0.769 \pm 0.021$ & $0.921 \pm 0.021$ & $8.6 \pm 2.7$ & 0.25 (?, fixed) & $9988 \pm 61$ & $7557 \pm 261$ & $2.306 \pm 0.031$ & $2.873 \pm 0.026$ & $ 0.246 \pm 0.020 $ & (7) \\

\\
Rasalhague ($\alpha$ Oph) &       A5IV &&      $237$ & $87.70 \pm 0.43$ & $0.709 \pm 0.011$ & $0.885 \pm 0.011$ & $-53.88 \pm 1.23$ & 0.25 (fixed) & $9300 \pm 150$ & $7460 \pm 100$ & $2.390 \pm 0.014$ & $2.871 \pm 0.020$ & $ 0.201 \pm 0.011 $ &(9) \\
\\
\multirow{4}{*}{Altair ($\alpha$ Aql) } &     \multirow{4}{*}{A7IV-V}&
 \multirow{4}{*}{$\left\{
\begin{array}{l}
\\
\\
\\
\end{array}
\right.$}
& \multicolumn{ 2}{c}{$v \sin i = 210\pm13$, $i>30$} &            && $-68.4 \pm 6.2$\tablenotemark{b} & none applied & \multicolumn{ 2}{c}{$<7680 \pm 90>$} &            & $1.8868 \pm 0.0066$ & &(1)\\

           &&            &   &          &   && $35 \pm 18$ &  See note\tablenotemark{d} &  \multicolumn{ 2}{c}{$7750$ (fixed)}   &   &   & &(2) \\

           &&            & $273 \pm 13$ &        $\sim$64 & $0.729 \pm 0.019$ & $0.90 \pm 0.02$ & $123.2 \pm 2.8$ & 0.25 (fixed) & $8740 \pm 140$ & $6890 \pm 60$ & $1.636 \pm 0.022$ & $1.988 \pm 0.009$ & $ 0.215 \pm 0.017 $ & (3) \\

           &&            & $285 \pm 10$ & $57.2 \pm 1.9$ & $0.764 \pm 0.008$ & $0.923 \pm 0.006$ & $-61.8 \pm 0.8$ & $0.19 \pm 0.012$\tablenotemark{c} & $8450 \pm 140$ & $6860 \pm 150$ & $1.634 \pm 0.011$ & $2.029 \pm 0.007$ & $ 0.242 \pm 0.009 $ & (4)\\
\\
\multirow{2}{*}{Alderamin ($\alpha$ Cep)} &     \multirow{2}{*}{A7IV-V} &
 \multirow{2}{*}{$\left\{
\begin{array}{l}
\\
\end{array}
\right.$}
& $283 \pm 19$ & $88.2^{+1.8}_{-13.3}$ & $0.8287^{+0.0482}_{-0.0232}$ & $0.958 \pm 0.068$ & $3 \pm 10$ & $0.084^{+0.026}_{-0.049}$ & $8440^{+430}_{-700}$ & ``$\sim$7600'' & $2.175 \pm 0.046$ & $2.82 \pm 0.10$ & $ 0.297 \pm 0.054 $ &(10) \\

           &            &&      $225$ & $55.70 \pm 6.23$ & $0.795\pm0.025$ & $0.941\pm0.020$ & $-178.84 \pm 4.28$ & $0.216 \pm 0.021$\tablenotemark{c} & $8588 \pm 300$ & $6574 \pm 200$ & $2.162 \pm 0.036$ & $2.74 \pm 0.044$ & $ 0.267 \pm 0.029 $ &(9) \\
\\
Caph ($\beta$ Cas) &       F2III-IV &&      $72.4_{-3.5}^{+1.5}$ & $19.9_{-1.9}^{+1.9}$ & $0.760 \pm 0.040$ & $0.920^{+0.024}_{-0.034}$ & $-7.09_{-2.40}^{+2.24}$ & $0.146_{-0.007}^{+0.013}$\tablenotemark{c} & $7208_{-24}^{+42}$ & $6167_{-21}^{+36}$ & $3.06_{-0.07}^{+0.08}$ & $3.79_{-0.09}^{+0.10}$ & $ 0.239 \pm 0.046 $ &(11) \\

\enddata
\tablenotetext{a}{Fixed from \citet{Slettebak1982ApJS...50...55S}.}
\tablenotetext{b}{Reflects a correction for $\{u,v\}$ coordinates swap from manuscript value of $-21.6\pm6.2$; physics of investigation not affected.}
\tablenotetext{c}{Second solution with $\beta=0.25$ (fixed) also presented in manuscript.}
\tablenotetext{d}{The authors applied asymmetric brightness distribution at the $\sim$5\% level to simulate gravity darkening.}
\tablerefs{(1) \citet{vanbelle2001ApJ...559.1155V},
(2) \citet{Ohishi2004ApJ...612..463O},
(3) \citet{Peterson2006ApJ...636.1087P},
(4) \citet{Monnier2007Sci...317..342M},
(5) \citet{DomicianodeSouza2003AA...407L..47D},
(6) \citet{Aufdenberg2006ApJ...645..664A},
(7) \citet{Peterson2006Natur.440..896P},
(8) \citet{McAlister2005ApJ...628..439M},
(9) \citet{Zhao2009},
(10) \citet{vanbelle2006ApJ...637..494V},
(11) \citet{Che2011ApJ...732...68C}}
\tablecomments{`Oblateness' is defined as $R_{\rm eq} / R_{\rm pole}-1$. In certain cases, Eqn. \ref{eqn_fractional_linear} has been invoked to convert between $\omega / \omega_{\rm crit}$ and $v / v_{\rm crit}$.}
\end{deluxetable}
\end{landscape}

%% file: tables/tab_timeline.tex

\begin{landscape}
\begin{deluxetable}{lll}
\tablecolumns{3}
\tablewidth{0pc}
\tablecaption{Timeline of papers.\label{table_timeline}}
\tablehead{
\colhead{Target(s)} &
\colhead{Reference} &
\colhead{Notes}
}
\startdata
Altair & \citet{vanbelle2001ApJ...559.1155V}  & First direct detection of stellar oblateness \\
Achernar & \citet{DomicianodeSouza2003AA...407L..47D} & Extreme oblateness ratio of $1.56 \pm 0.05$ claimed\\
Altair & \citet{Ohishi2004ApJ...612..463O} & Indications of bright pole from closure phase\\
Regulus & \citet{McAlister2005ApJ...628..439M} & First unique solution for inclination,\\
 &  & combination of spectroscopy with interferometry\\
Vega & \citet{Peterson2006Natur.440..896P} & Rapid rotation of Vega confirmed interferometrically\\ 
Alderamin & \citet{vanbelle2006ApJ...637..494V} & First attempt to fit gravity darkening as a free parameter\\ 
Altair & \citet{Peterson2006ApJ...636.1087P} & Unique inclination for Altair \\ 
Vega & \citet{Aufdenberg2006ApJ...645..664A} & Calibration of Vega's SED\\ 
Altair & \citet{Monnier2007Sci...317..342M} & First image of stellar oblateness \\
Alderamin \& Rasalhague & \citet{Zhao2009} & Additional stellar images\\
Caph \& Regulus& \citet{Che2011ApJ...732...68C} & Unexpected rapid rotation for Caph\\
\enddata
\end{deluxetable}
\end{landscape}


%% file: tables/tab_vega.tex

\begin{landscape}
\begin{deluxetable}{lcccc}
\tablecolumns{5}
\tablewidth{0pc}
\tablecaption{A comparison of Vega results.\label{table_vega}}
\tablehead{
\colhead{Parameter} &
\colhead{Symbol} &
\colhead{P06} &
\colhead{A06} &
\colhead{Y10}
}
\startdata
Inclination (deg) & $i$ & $4.54 \pm 0.33$ & $4.7 \pm 0.3$ & $4.98 \pm 0.08$\\
Polar effective temperature ($K$) & $T_{\rm EFF}^{\rm pole}$ & $9988 \pm 61$ & $10150 \pm 100$ &$10060 \pm 10$ \\
Polar surface gravity (cm s$^-2$) & $\log g_{\rm pole}$ & $4.074 \pm 0.012$ & $4.1 \pm 0.1$ &$4.021 \pm 0.014$\\
Polar radius ($R_\odot$) & $R_{\rm pole}$ & $2.306 \pm 0.031$ & $2.26 \pm 0.07$ & $2.362 \pm 0.012$\\
Equatorial effective temperature ($K$) & $T_{\rm EFF}^{\rm eq}$ & $7557 \pm 261$ & $7900 \pm 400$ &$8152 \pm 42$\\
Equatorial radius ($R_\odot$) & $R_{\rm eq}$ & $2.873 \pm 0.026$ & $2.78 \pm 0.02$ & $2.818 \pm 0.013$\\
Equatorial rotational velocity (km s$^-1$) & $v_{\rm eq}$ & $274 \pm 14$ & $270 \pm 15$ & $236.2 \pm 3.7$\\
Projected rotational velocity (km s$^-1$) & $v \sin i$ & $21.7 \pm 1.1$ & $21.9 \pm 0.2$ & $20.48 \pm 0.11$\\
Fraction of angular break-up rate & $\omega$ & $0.926 \pm 0.021$ & $0.91 \pm 0.03$ & $0.8760 \pm 0.0057$\\
Mass ($M_\odot$) & $M$ & $2.303 \pm 0.024$ & $2.3 \pm 0.2$ & $2.135 \pm 0.074$\\
Luminosity ($L_\odot$) & $L$ & $35.0 \pm 1.5$ & $37 \pm 3$ & $40.12 \pm 0.45$\\
\enddata
\tablerefs{P06 - \citet{Peterson2006Natur.440..896P}, A06 - \citet{Aufdenberg2006ApJ...645..664A}, Y10 - \citet{Yoon2010ApJ...708...71Y}}
\end{deluxetable}
\end{landscape}


%% file: tables/tab_future_targets.tex

\begin{deluxetable}{rrrlccccc}
\tabletypesize{\scriptsize}
\tablecolumns{9}
\tablewidth{0pc}
\tablecaption{Summary of possible future targets for interferometric studies of rapid rotators as described in \S \ref{sec_future_target_list}; angular size and oblateness values here are {\it estimates}.\label{table_futureTargets}}
\tablehead{
\colhead{HD} &
\colhead{RA} &
\colhead{DE} &
\colhead{Spectral} &
\colhead{$V$} &
\colhead{$K$} &
\colhead{$v \sin i$} &
\colhead{$\theta_{\rm EST}$} &
\colhead{Oblateness} \\
\colhead{} &
\colhead{hh:mm:ss} &
\colhead{dd:mm:ss} &
\colhead{Type} &
\colhead{(mag)} &
\colhead{(mag)} &
\colhead{(km/s)} &
\colhead{(mas)} &
\colhead{$R_b / R_a-1$}
}
\startdata
225003 & 00 02 29.76 & +08 29 08.1 & F0V          & 5.70 & 4.91 & 165 & 0.51 & 0.06 \\
225132 & 00 03 44.37 & -17 20 09.5 & B9IVn        & 4.55 & 4.56 & 190 & 0.57 & 0.12 \\
360 & 00 08 17.55 & -08 49 26.5 & G8III:       & 5.99 & 3.75 & 240 & 0.94 & 0.19 \\
493 & 00 09 21.02 & -27 59 16.5 & F3V          & 5.42 & 4.38 & 170 & 0.65 & 0.07 \\
2262 & 00 26 12.12 & -43 40 47.7 & A7V          & 3.93 & 3.59 & 245 & 0.91 & 0.15 \\
2696 & 00 30 22.67 & -23 47 15.8 & A3V          & 5.17 & 4.83 & 150 & 0.51 & 0.05 \\
2884 & 00 31 32.56 & -62 57 29.1 & B9V          & 4.36 & 4.48 & 170 & 0.59 & 0.06 \\
3283 & 00 36 27.34 & +60 19 34.4 & A4III        & 5.78 & 4.79 & 100 & 0.54 & 0.07 \\
4180 & 00 44 43.50 & +48 17 03.8 & B5III        & 4.48 & 4.40 & 255 & 0.62 & 0.36 \\
5394 & 00 56 42.50 & +60 43 00.3 & B0IV:evar    & 2.15 & 1.76 & 260 & 2.12 & 0.12 \\
6903 & 01 09 49.20 & +19 39 30.2 & G0III        & 5.57 & 4.18 & 95 & 0.73 & 0.06 \\
7344 & 01 13 43.80 & +07 34 31.8 & A7IV         & 5.21 & 4.57 & 265 & 0.59 & 0.25 \\
7788 & 01 15 45.50 & -68 52 34.5 & F6IV         & 4.25 & 3.88 & 135 & 0.80 & 0.09 \\
9352 & 01 33 25.71 & +58 19 38.4 & K0Ib+...     & 5.69 & 1.77 & 50 & 2.54 & 0.12 \\
10144 & 01 37 42.75 & -57 14 12.0 & B3Vp         & 0.45 & 0.88 & 250 & 3.04 & 0.09 \\
10148 & 01 38 51.71 & -21 16 31.7 & F0V          & 5.58 & 4.75 & 170 & 0.55 & 0.06 \\
10516 & 01 43 39.62 & +50 41 19.6 & B2Vpe        & 4.01 & 3.71 & 505 & 0.86 & 0.47 \\
10830 & 01 45 38.65 & -25 03 08.8 & F2IV         & 5.29 & 4.46 & 120 & 0.62 & 0.06 \\
13041 & 02 08 29.15 & +37 51 33.1 & A5IV-V       & 4.78 & 4.43 & 135 & 0.62 & 0.06 \\
13174 & 02 09 25.29 & +25 56 23.9 & F2III        & 4.98 & 4.06 & 160 & 0.75 & 0.24 \\
14228 & 02 16 30.50 & -51 30 43.6 & B8IV-V       & 3.56 & 4.13 & 250 & 0.68 & 0.17 \\
14055 & 02 17 18.84 & +33 50 50.4 & A1Vnn        & 4.03 & 3.96 & 240 & 0.76 & 0.13 \\
15008 & 02 21 45.02 & -68 39 33.9 & A3V          & 4.08 & 3.96 & 180 & 0.76 & 0.07 \\
14690 & 02 22 12.41 & -00 53 05.1 & F0Vn         & 5.42 & 4.58 & 185 & 0.59 & 0.08 \\
15233 & 02 24 53.99 & -60 18 41.9 & F2III        & 5.36 & 4.42 & 140 & 0.64 & 0.17 \\
15130 & 02 25 57.01 & -12 17 25.6 & A0V          & 4.88 & 4.81 & 210 & 0.51 & 0.10 \\
15257 & 02 28 09.99 & +29 40 10.3 & F0III        & 5.29 & 4.59 & 85 & 0.59 & 0.06 \\
236970 & 02 33 18.35 & +56 19 05.0 & A2Iab        & 8.84 & 4.90 & 170 & 0.60 & 0.30 \\
16555 & 02 37 24.26 & -52 32 35.1 & A6V          & 5.30 & 4.53 & 315 & 0.61 & 0.27 \\
16978 & 02 39 35.22 & -68 16 01.0 & B9III        & 4.12 & 4.25 & 100 & 0.65 & 0.05 \\
16754 & 02 39 47.92 & -42 53 29.9 & A2V          & 4.74 & 4.46 & 245 & 0.61 & 0.14 \\
16970 & 02 43 18.12 & +03 14 10.2 & A3V          & 3.47 & 3.08 & 190 & 1.16 & 0.08 \\
17566 & 02 45 32.53 & -67 37 00.2 & A2IV/V       & 4.83 & 4.60 & 135 & 0.57 & 0.05 \\
17573 & 02 49 58.99 & +27 15 38.8 & B8Vn         & 3.61 & 3.86 & 240 & 0.78 & 0.12 \\
17584 & 02 50 34.91 & +38 19 08.1 & F2III        & 4.22 & 3.24 & 160 & 1.11 & 0.24 \\
18331 & 02 56 37.45 & -03 42 44.0 & A3Vn         & 5.16 & 4.86 & 255 & 0.51 & 0.16 \\
18411 & 02 58 45.65 & +39 39 46.2 & A2Vn         & 4.68 & 4.44 & 170 & 0.61 & 0.06 \\
18866 & 02 58 47.77 & -64 04 16.7 & A5III        & 4.98 & 4.62 & 115 & 0.57 & 0.10 \\
18978 & 03 02 23.59 & -23 37 27.6 & A4V          & 4.08 & 3.57 & 180 & 0.92 & 0.07 \\
19319 & 03 03 36.90 & -59 44 15.4 & F0IV         & 5.12 & 4.28 & 130 & 0.68 & 0.07 \\
19107 & 03 04 16.48 & -07 36 03.2 & A8V          & 5.26 & 4.74 & 150 & 0.54 & 0.05 \\
20313 & 03 07 31.90 & -78 59 21.9 & F2II-III     & 5.67 & 4.94 & 75 & 0.50 & 0.05 \\
19275 & 03 11 56.24 & +74 23 37.9 & A2Vnn        & 4.85 & 4.71 & 250 & 0.54 & 0.15 \\
20121 & 03 12 25.68 & -44 25 10.8 & F3V+...      & 5.92 & 4.83 & 195 & 0.54 & 0.10 \\
22192 & 03 36 29.36 & +48 11 33.7 & B5Ve         & 4.32 & 4.11 & 375 & 0.71 & 0.28 \\
22928 & 03 42 55.48 & +47 47 15.6 & B5III SB     & 3.01 & 3.26 & 245 & 1.03 & 0.32 \\
23302 & 03 44 52.52 & +24 06 48.4 & B6III        & 3.72 & 3.92 & 215 & 0.76 & 0.24 \\
23480 & 03 46 19.56 & +23 56 54.5 & B6IV         & 4.14 & 4.22 & 285 & 0.67 & 0.27 \\
23630 & 03 47 29.06 & +24 06 18.9 & B7III        & 2.85 & 2.64 & 210 & 1.40 & 0.24 \\
23850 & 03 49 09.73 & +24 03 12.7 & B8III        & 3.62 & 3.88 & 195 & 0.77 & 0.22 \\
23401 & 03 50 21.48 & +71 19 56.5 & A2IVn        & 4.59 & 4.36 & 205 & 0.63 & 0.17 \\
24554 & 03 54 17.49 & -02 57 17.0 & G8III        & 4.46 & 2.45 & 180 & 1.68 & 0.32 \\
25945 & 04 05 37.30 & -27 39 07.3 & F0IV/V       & 5.59 & 4.73 & 135 & 0.55 & 0.05 \\
25642 & 04 06 35.06 & +50 21 04.9 & A0IVn        & 4.25 & 4.15 & 205 & 0.69 & 0.16 \\
25940 & 04 08 39.67 & +47 42 45.3 & B3Ve         & 3.96 & 3.80 & 230 & 0.82 & 0.08 \\
26612 & 04 10 50.43 & -41 59 37.5 & A9V          & 4.93 & 3.95 & 250 & 0.80 & 0.15 \\
26574 & 04 11 51.93 & -06 50 16.0 & F2II-III     & 4.04 & 3.21 & 105 & 1.11 & 0.11 \\
27901 & 04 24 57.06 & +19 02 31.5 & F4V          & 5.97 & 4.98 & 150 & 0.50 & 0.06 \\
28024 & 04 26 18.39 & +22 48 49.3 & A8Vn         & 4.28 & 3.76 & 195 & 0.85 & 0.09 \\
28052 & 04 26 20.67 & +15 37 06.0 & F0V...       & 4.48 & 4.03 & 195 & 0.75 & 0.09 \\
29992 & 04 42 03.45 & -37 08 41.2 & F3V          & 5.04 & 4.09 & 185 & 0.75 & 0.09 \\
30478 & 04 44 21.12 & -59 43 58.2 & A8/A9III/IV  & 5.28 & 4.73 & 230 & 0.55 & 0.26 \\
30211 & 04 45 30.14 & -03 15 16.6 & B5IV         & 4.01 & 4.40 & 160 & 0.60 & 0.07 \\
30739 & 04 50 36.72 & +08 54 00.9 & A1Vn         & 4.35 & 4.17 & 235 & 0.69 & 0.13 \\
30780 & 04 51 22.41 & +18 50 23.8 & A7IV-V       & 5.08 & 4.49 & 145 & 0.61 & 0.06 \\
30912 & 04 52 47.09 & +27 53 51.3 & F2IV         & 5.97 & 4.98 & 155 & 0.50 & 0.11 \\
31109 & 04 52 53.68 & -05 27 09.9 & A9IV         & 4.36 & 3.72 & 170 & 0.87 & 0.13 \\
32045 & 04 59 55.71 & -12 32 13.9 & F0V          & 4.78 & 4.01 & 185 & 0.77 & 0.08 \\
33111 & 05 07 51.03 & -05 05 10.5 & A3IIIvar     & 2.78 & 2.40 & 190 & 1.58 & 0.28 \\
32991 & 05 07 55.43 & +21 42 17.4 & B2Ve         & 5.84 & 4.78 & 220 & 0.55 & 0.06 \\
33328 & 05 09 08.78 & -08 45 14.7 & B2IVn        & 4.25 & 4.71 & 325 & 0.52 & 0.25 \\
33802 & 05 12 17.89 & -11 52 08.9 & B8V          & 4.45 & 4.65 & 195 & 0.54 & 0.07 \\
34658 & 05 19 11.23 & +02 35 45.4 & F5IIvar      & 5.34 & 4.33 & 80 & 0.67 & 0.09 \\
36705 & 05 28 44.78 & -65 26 56.2 & K1III(p)     & 6.88 & 4.69 & 100 & 0.61 & 0.13 \\
36267 & 05 30 47.05 & +05 56 53.6 & B5V          & 4.20 & 4.61 & 180 & 0.55 & 0.05 \\
36576 & 05 33 31.63 & +18 32 24.8 & B2IV-Ve      & 5.67 & 4.78 & 285 & 0.54 & 0.14 \\
37202 & 05 37 38.68 & +21 08 33.3 & B4IIIp       & 2.97 & 2.81 & 320 & 1.29 & 0.31 \\
37507 & 05 38 53.09 & -07 12 45.8 & A4V          & 4.77 & 4.42 & 185 & 0.62 & 0.08 \\
37490 & 05 39 11.15 & +04 07 17.3 & B3IIIe       & 4.50 & 4.81 & 190 & 0.50 & 0.13 \\
37795 & 05 39 38.94 & -34 04 26.6 & B7IV         & 2.65 & 2.83 & 210 & 1.26 & 0.14 \\
39014 & 05 44 46.42 & -65 44 07.9 & A7V          & 4.34 & 3.84 & 225 & 0.82 & 0.12 \\
38678 & 05 46 57.35 & -14 49 19.0 & A2Vann       & 3.55 & 3.29 & 230 & 1.04 & 0.12 \\
40248 & 05 56 20.94 & -31 22 56.8 & F2III        & 5.52 & 4.43 & 110 & 0.64 & 0.10 \\
41335 & 06 04 13.50 & -06 42 32.2 & B2Vne+       & 5.19 & 4.78 & 445 & 0.53 & 0.33 \\
41695 & 06 06 09.33 & -14 56 07.0 & A0V          & 4.67 & 4.52 & 215 & 0.59 & 0.10 \\
42818 & 06 18 50.78 & +69 19 12.1 & A0Vn         & 4.76 & 4.67 & 325 & 0.55 & 0.27 \\
44769 & 06 23 46.10 & +04 35 34.2 & A5IV         & 4.39 & 3.92 & 125 & 0.79 & 0.06 \\
45725 & 06 28 49.07 & -07 01 59.0 & B3Ve         & 3.76 & 4.08 & 325 & 0.70 & 0.17 \\
45542 & 06 28 57.79 & +20 12 43.8 & B6III        & 4.13 & 4.35 & 149 & 0.62 & 0.10 \\
46273 & 06 29 49.13 & -50 14 20.3 & F2V          & 5.28 & 4.33 & 175 & 0.67 & 0.07 \\
45910 & 06 30 32.94 & +05 52 01.2 & B2:IIIpshev  & 6.70 & 4.41 & 430 & 0.69 & 0.30 \\
46304 & 06 32 23.13 & -05 52 07.4 & F0Vnn+...    & 5.60 & 4.91 & 200 & 0.51 & 0.09 \\
46933 & 06 35 03.38 & -22 57 53.4 & A0III        & 4.54 & 4.50 & 145 & 0.59 & 0.12 \\
47670 & 06 37 45.67 & -43 11 45.3 & B8III SB     & 3.17 & 3.56 & 225 & 0.89 & 0.31 \\
50506 & 06 40 02.91 & -80 48 49.4 & A5III        & 5.61 & 4.94 & 200 & 0.50 & 0.26 \\
48917 & 06 44 28.47 & -31 04 13.9 & B2V          & 5.23 & 4.89 & 200 & 0.50 & 0.05 \\
50241 & 06 48 11.54 & -61 56 31.1 & A7IV         & 3.24 & 2.57 & 230 & 1.48 & 0.26 \\
50013 & 06 49 50.47 & -32 30 30.6 & B1.5IVne     & 3.50 & 3.55 & 200 & 0.91 & 0.08 \\
50019 & 06 52 47.34 & +33 57 40.9 & A3III        & 3.60 & 3.16 & 130 & 1.11 & 0.11 \\
51199 & 06 55 37.40 & -20 08 11.7 & F2IV/V       & 4.66 & 3.95 & 145 & 0.79 & 0.07 \\
50973 & 06 57 37.12 & +45 05 38.8 & A2Vn         & 4.90 & 4.79 & 210 & 0.52 & 0.10 \\
52690 & 07 02 06.73 & -03 45 17.4 & M1Ib comp SB & 6.58 & 2.01 & 50 & 2.36 & 0.30 \\
55057 & 07 11 23.63 & -00 18 06.9 & F2V          & 5.44 & 4.64 & 150 & 0.57 & 0.05 \\
55185 & 07 11 51.86 & -00 29 34.0 & A2V          & 4.15 & 3.90 & 155 & 0.79 & 0.05 \\
56537 & 07 18 05.61 & +16 32 25.7 & A3V...       & 3.58 & 3.54 & 165 & 0.92 & 0.06 \\
57150 & 07 18 18.40 & -36 44 02.3 & B2V+...      & 4.65 & 4.52 & 360 & 0.59 & 0.19 \\
56169 & 07 18 31.98 & +49 27 53.1 & A4IIIn       & 5.00 & 4.62 & 215 & 0.57 & 0.29 \\
57167 & 07 19 28.08 & -16 23 41.7 & F2III/IV     & 5.70 & 4.72 & 100 & 0.56 & 0.06 \\
56986 & 07 20 07.39 & +21 58 56.4 & F0IV...      & 3.50 & 2.56 & 115 & 1.51 & 0.05 \\
58954 & 07 27 07.99 & -17 51 53.5 & F2V          & 5.60 & 4.76 & 185 & 0.55 & 0.08 \\
58715 & 07 27 09.07 & +08 17 21.9 & B8Vvar       & 2.89 & 3.10 & 285 & 1.11 & 0.17 \\
59037 & 07 29 20.46 & +28 07 06.3 & A4V          & 5.07 & 4.74 & 220 & 0.53 & 0.12 \\
61715 & 07 38 18.21 & -48 36 05.2 & F4Iab        & 5.68 & 4.11 & 55 & 0.76 & 0.06 \\
61110 & 07 39 09.96 & +34 35 04.7 & F3III        & 4.89 & 3.84 & 90 & 0.84 & 0.07 \\
61497 & 07 43 00.46 & +58 42 37.8 & A3IVn        & 4.93 & 4.63 & 200 & 0.56 & 0.17 \\
63462 & 07 48 05.17 & -25 56 13.8 & B1IV:nne     & 4.40 & 4.17 & 375 & 0.69 & 0.33 \\
64760 & 07 53 18.16 & -48 06 10.6 & B0.5Ib       & 4.22 & 4.64 & 250 & 0.54 & 0.07 \\
64145 & 07 53 29.84 & +26 45 57.1 & A3V          & 4.97 & 4.66 & 160 & 0.56 & 0.06 \\
65925 & 07 59 28.43 & -39 17 48.6 & F3V          & 5.22 & 4.22 & 165 & 0.71 & 0.07 \\
65810 & 07 59 52.06 & -18 23 56.9 & A1V          & 4.61 & 4.31 & 220 & 0.65 & 0.11 \\
67006 & 08 08 27.50 & +51 30 24.0 & A2V          & 4.78 & 4.66 & 175 & 0.55 & 0.07 \\
67797 & 08 09 01.64 & -19 14 42.0 & B5V          & 4.40 & 4.77 & 185 & 0.51 & 0.06 \\
72072 & 08 25 39.14 & -71 03 02.6 & K2III        & 7.74 & 4.73 & 200 & 0.62 & 0.33 \\
71935 & 08 27 36.65 & -53 05 18.7 & A9/F0III/IV  & 5.08 & 4.44 & 160 & 0.62 & 0.16 \\
72041 & 08 31 30.57 & +24 04 52.4 & F0IIIn       & 5.71 & 4.91 & 110 & 0.51 & 0.10 \\
72779 & 08 35 19.47 & +19 35 24.3 & G0III        & 6.55 & 5.00 & 85 & 0.51 & 0.05 \\
73262 & 08 37 39.41 & +05 42 13.7 & A1Vnn        & 4.14 & 4.03 & 285 & 0.73 & 0.20 \\
75710 & 08 49 47.65 & -45 18 28.5 & A2III        & 4.94 & 4.62 & 110 & 0.57 & 0.07 \\
76143 & 08 50 34.68 & -66 47 35.6 & F5IV         & 5.34 & 4.32 & 140 & 0.67 & 0.10 \\
75486 & 08 53 22.57 & +61 57 44.0 & F2III        & 5.72 & 4.88 & 115 & 0.51 & 0.11 \\
79837 & 08 56 41.88 & -85 39 47.6 & F0III        & 5.43 & 4.67 & 115 & 0.57 & 0.11 \\
76543 & 08 57 14.91 & +15 19 21.8 & A5III        & 5.22 & 4.87 & 90 & 0.51 & 0.06 \\
76644 & 08 59 12.84 & +48 02 32.5 & A7IV         & 3.12 & 2.66 & 150 & 1.40 & 0.09 \\
77327 & 09 03 37.56 & +47 09 24.0 & A1Vn         & 3.57 & 3.38 & 235 & 0.99 & 0.13 \\
77601 & 09 05 24.11 & +48 31 49.3 & F6II-III     & 5.48 & 4.34 & 160 & 0.67 & 0.35 \\
80007 & 09 13 12.24 & -69 43 02.9 & A2IV         & 1.67 & 1.49 & 140 & 2.38 & 0.07 \\
79439 & 09 16 11.28 & +54 01 18.2 & A5V          & 4.80 & 4.29 & 155 & 0.66 & 0.06 \\
80081 & 09 18 50.67 & +36 48 10.4 & A1V          & 3.82 & 3.42 & 170 & 0.99 & 0.06 \\
81471 & 09 23 59.34 & -51 44 13.5 & A7Iab        & 6.05 & 4.27 & 65 & 0.72 & 0.05 \\
82554 & 09 24 09.73 & -80 47 13.9 & F3/F5IV      & 5.34 & 4.20 & 160 & 0.72 & 0.09 \\
82434 & 09 30 42.11 & -40 28 00.8 & F2IV         & 3.60 & 2.67 & 260 & 1.43 & 0.26 \\
81937 & 09 31 31.57 & +63 03 42.5 & F0IV         & 3.65 & 2.86 & 145 & 1.30 & 0.09 \\
82621 & 09 34 49.49 & +52 03 05.6 & A2V          & 4.47 & 4.35 & 185 & 0.64 & 0.08 \\
83446 & 09 36 49.66 & -49 21 18.5 & A5V          & 4.34 & 3.94 & 155 & 0.78 & 0.06 \\
83953 & 09 41 17.03 & -23 35 29.5 & B5V          & 4.76 & 4.54 & 315 & 0.58 & 0.18 \\
85376 & 09 51 53.02 & +24 23 44.9 & A5IV         & 5.29 & 4.66 & 130 & 0.56 & 0.07 \\
87427 & 10 04 21.02 & -24 17 08.1 & F0V          & 5.70 & 4.87 & 180 & 0.52 & 0.07 \\
87696 & 10 07 25.73 & +35 14 40.9 & A7V          & 4.49 & 4.00 & 160 & 0.76 & 0.06 \\
87901 & 10 08 22.46 & +11 58 01.9 & B7V          & 1.36 & 1.64 & 330 & 2.16 & 0.23 \\
88215 & 10 10 05.96 & -12 48 56.4 & F2/F3IV/V    & 5.30 & 4.40 & 195 & 0.65 & 0.13 \\
88824 & 10 13 22.88 & -51 13 58.6 & A7V          & 5.27 & 4.62 & 235 & 0.58 & 0.13 \\
89080 & 10 13 44.28 & -70 02 16.5 & B8III        & 3.29 & 3.45 & 230 & 0.94 & 0.33 \\
89254 & 10 17 37.90 & -08 04 08.1 & F2III        & 5.25 & 4.44 & 90 & 0.63 & 0.07 \\
90132 & 10 23 29.41 & -38 00 35.0 & A8V          & 5.34 & 4.69 & 270 & 0.56 & 0.18 \\
91465 & 10 32 01.48 & -61 41 07.3 & B4Vne        & 3.30 & 3.24 & 305 & 1.05 & 0.15 \\
91312 & 10 33 14.00 & +40 25 31.9 & A7IV         & 4.72 & 4.20 & 135 & 0.69 & 0.08 \\
94601 & 10 55 36.85 & +24 44 59.1 & A1           & 4.30 & 4.35 & 185 & 0.63 & 0.08 \\
95370 & 11 00 09.25 & -42 13 33.1 & A3IV         & 4.37 & 4.07 & 115 & 0.73 & 0.05 \\
96202 & 11 05 20.03 & -27 17 36.9 & F3IV/V       & 4.92 & 4.09 & 240 & 0.74 & 0.22 \\
97603 & 11 14 06.41 & +20 31 26.5 & A4V          & 2.56 & 2.14 & 195 & 1.78 & 0.09 \\
98058 & 11 16 39.76 & -03 39 05.5 & A7IVn        & 4.45 & 4.13 & 240 & 0.71 & 0.29 \\
98718 & 11 21 00.44 & -54 29 27.7 & B5Vn         & 3.90 & 4.31 & 340 & 0.63 & 0.22 \\
100841 & 11 35 46.93 & -63 01 11.4 & B9II:        & 3.11 & 3.07 & 165 & 1.14 & 0.17 \\
100889 & 11 36 40.95 & -09 48 08.1 & B9.5Vn       & 4.70 & 4.78 & 190 & 0.51 & 0.07 \\
101431 & 11 40 12.82 & -34 44 40.8 & B9V          & 4.70 & 4.79 & 245 & 0.51 & 0.13 \\
102124 & 11 45 17.00 & +08 15 29.4 & A4V          & 4.84 & 4.41 & 170 & 0.63 & 0.07 \\
102776 & 11 49 41.09 & -63 47 18.6 & B3V          & 4.30 & 4.68 & 270 & 0.53 & 0.11 \\
103287 & 11 53 49.74 & +53 41 41.0 & A0V SB       & 2.41 & 2.43 & 170 & 1.52 & 0.06 \\
106490 & 12 15 08.76 & -58 44 56.0 & B2IV         & 2.79 & 3.53 & 175 & 0.88 & 0.06 \\
106591 & 12 15 25.45 & +57 01 57.4 & A3Vvar       & 3.32 & 3.10 & 180 & 1.13 & 0.07 \\
106661 & 12 16 00.23 & +14 53 56.9 & A3V          & 5.09 & 4.89 & 175 & 0.50 & 0.07 \\
106911 & 12 18 20.94 & -79 18 44.2 & B5Vn         & 4.24 & 4.56 & 260 & 0.56 & 0.12 \\
108283 & 12 26 24.07 & +27 16 05.7 & F0p          & 4.92 & 4.15 & 230 & 0.72 & 0.12 \\
108483 & 12 28 02.41 & -50 13 50.2 & B3V          & 3.91 & 4.48 & 215 & 0.58 & 0.07 \\
108722 & 12 29 27.05 & +24 06 32.1 & F5III        & 5.47 & 4.55 & 90 & 0.60 & 0.07 \\
108767 & 12 29 51.98 & -16 30 54.3 & B9.5V        & 2.94 & 3.00 & 160 & 1.17 & 0.05 \\
109026 & 12 32 28.11 & -72 07 58.7 & B5V          & 3.84 & 4.25 & 205 & 0.64 & 0.07 \\
109141 & 12 32 36.09 & -13 51 32.3 & F3IV/V       & 5.74 & 4.88 & 135 & 0.52 & 0.06 \\
109387 & 12 33 29.04 & +69 47 17.6 & B6IIIp       & 3.85 & 3.82 & 230 & 0.80 & 0.29 \\
109536 & 12 35 45.61 & -41 01 19.0 & A7III        & 5.12 & 4.57 & 110 & 0.59 & 0.09 \\
109787 & 12 37 42.33 & -48 32 28.6 & A2V          & 3.85 & 3.71 & 330 & 0.85 & 0.30 \\
110411 & 12 41 53.01 & +10 14 09.0 & A0V          & 4.88 & 4.68 & 175 & 0.55 & 0.07 \\
110335 & 12 41 56.60 & -59 41 08.9 & B6IV         & 4.91 & 4.79 & 240 & 0.52 & 0.18 \\
110432 & 12 42 50.28 & -63 03 31.0 & B2pe         & 5.27 & 4.04 & 275 & 0.78 & 0.10 \\
112429 & 12 55 28.56 & +65 26 18.8 & A5n          & 5.23 & 4.43 & 160 & 0.63 & 0.06 \\
113314 & 13 03 33.35 & -49 31 38.1 & A0V          & 4.83 & 4.74 & 240 & 0.53 & 0.13 \\
114529 & 13 12 17.63 & -59 55 13.9 & B8V          & 4.58 & 4.71 & 280 & 0.53 & 0.17 \\
116842 & 13 25 13.42 & +54 59 16.8 & A5V SB       & 3.99 & 3.15 & 230 & 1.15 & 0.13 \\
117150 & 13 29 25.26 & -51 09 54.4 & A0V          & 5.04 & 4.72 & 315 & 0.54 & 0.25 \\
118232 & 13 34 27.37 & +49 00 57.3 & A5V          & 4.68 & 4.27 & 175 & 0.67 & 0.07 \\
118098 & 13 34 41.75 & -00 35 45.4 & A3V          & 3.38 & 3.22 & 190 & 1.07 & 0.08 \\
118261 & 13 37 12.20 & -61 41 29.7 & F6V          & 5.63 & 4.44 & 150 & 0.64 & 0.06 \\
118623 & 13 37 27.70 & +36 17 41.4 & A7III        & 4.82 & 4.15 & 235 & 0.72 & 0.27 \\
118716 & 13 39 53.27 & -53 27 58.9 & B1III        & 2.29 & 3.04 & 140 & 1.11 & 0.05 \\
120315 & 13 47 32.55 & +49 18 47.9 & B3V SB       & 1.85 & 2.27 & 195 & 1.61 & 0.05 \\
120324 & 13 49 37.01 & -42 28 25.3 & B2IV-Ve      & 3.47 & 4.01 & 180 & 0.72 & 0.05 \\
121263 & 13 55 32.43 & -47 17 17.8 & B2.5IV       & 2.55 & 3.22 & 225 & 1.02 & 0.11 \\
122408 & 14 01 38.78 & +01 32 40.5 & A3V          & 4.23 & 4.09 & 165 & 0.71 & 0.06 \\
123255 & 14 06 42.91 & -09 18 48.7 & F2IV         & 5.46 & 4.60 & 140 & 0.59 & 0.09 \\
124675 & 14 13 28.95 & +51 47 24.0 & A8IV         & 4.53 & 4.10 & 130 & 0.72 & 0.07 \\
124367 & 14 14 57.16 & -57 05 09.9 & B4Vne        & 5.03 & 4.77 & 270 & 0.53 & 0.12 \\
125238 & 14 19 24.23 & -46 03 29.1 & B2.5IV       & 3.55 & 4.10 & 235 & 0.69 & 0.12 \\
126248 & 14 24 11.39 & +05 49 12.4 & A5V          & 5.10 & 4.77 & 190 & 0.53 & 0.09 \\
127762 & 14 32 04.76 & +38 18 28.4 & A7IIIvar     & 3.04 & 2.51 & 145 & 1.51 & 0.17 \\
127972 & 14 35 30.45 & -42 09 27.9 & B1Vn + A     & 2.33 & 2.75 & 345 & 1.29 & 0.16 \\
128345 & 14 37 53.25 & -49 25 32.7 & B5V          & 4.05 & 4.51 & 185 & 0.57 & 0.06 \\
129246 & 14 41 08.92 & +13 43 42.0 & A3IVn        & 3.78 & 3.70 & 150 & 0.85 & 0.09 \\
129988 & 14 44 59.25 & +27 04 27.0 & A0           & 2.35 & 0.12 & 165 & 4.96 & 0.06 \\
129422 & 14 45 17.25 & -62 52 31.6 & A7Vn         & 5.36 & 4.58 & 270 & 0.59 & 0.19 \\
129926 & 14 46 00.18 & -25 26 34.5 & F0V + G/K    & 5.15 & 4.43 & 210 & 0.63 & 0.10 \\
130109 & 14 46 14.99 & +01 53 34.6 & A0V          & 3.73 & 3.65 & 340 & 0.87 & 0.31 \\
131492 & 14 56 44.00 & -62 46 51.6 & B4Vnp        & 5.08 & 4.60 & 210 & 0.58 & 0.07 \\
134481 & 15 11 56.16 & -48 44 15.7 & B9V          & 3.88 & 3.97 & 205 & 0.75 & 0.09 \\
135742 & 15 17 00.47 & -09 22 58.3 & B8V          & 2.61 & 2.91 & 230 & 1.21 & 0.11 \\
135734 & 15 18 32.05 & -47 52 30.7 & B8V          & 4.27 & 4.43 & 400 & 0.60 & 0.41 \\
135382 & 15 18 54.69 & -68 40 46.1 & A1V          & 2.87 & 2.53 & 225 & 1.48 & 0.12 \\
137422 & 15 20 43.75 & +71 50 02.3 & A3II-III     & 3.00 & 2.71 & 160 & 1.36 & 0.20 \\
136298 & 15 21 22.34 & -40 38 50.9 & B1.5IV       & 3.22 & 3.96 & 225 & 0.72 & 0.10 \\
137058 & 15 25 20.25 & -38 44 00.9 & A0V          & 4.60 & 4.48 & 300 & 0.60 & 0.22 \\
137898 & 15 28 38.29 & +01 50 31.8 & A8IV         & 5.15 & 4.59 & 130 & 0.58 & 0.07 \\
138629 & 15 31 46.99 & +40 53 57.7 & A5V          & 4.98 & 4.61 & 175 & 0.57 & 0.07 \\
138749 & 15 32 55.80 & +31 21 33.0 & B6Vnn        & 4.14 & 4.43 & 385 & 0.60 & 0.32 \\
138690 & 15 35 08.46 & -41 10 00.1 & B2IV         & 2.80 & 3.38 & 270 & 0.96 & 0.16 \\
140159 & 15 41 33.09 & +19 40 13.8 & A1V          & 4.51 & 4.31 & 175 & 0.65 & 0.07 \\
142105 & 15 44 03.46 & +77 47 40.2 & A3Vn         & 4.29 & 4.20 & 210 & 0.68 & 0.10 \\
140417 & 15 44 04.42 & -15 40 21.6 & A6IV         & 5.41 & 4.82 & 115 & 0.52 & 0.05 \\
141003 & 15 46 11.21 & +15 25 18.9 & A3V          & 3.65 & 3.55 & 185 & 0.92 & 0.08 \\
141637 & 15 50 58.75 & -25 45 04.4 & B1.5Vn       & 4.63 & 4.78 & 310 & 0.51 & 0.13 \\
141851 & 15 51 15.65 & -03 05 25.5 & A3Vn         & 5.09 & 4.70 & 185 & 0.55 & 0.08 \\
142114 & 15 53 36.73 & -25 19 37.5 & B2.5Vn       & 4.59 & 4.79 & 320 & 0.51 & 0.15 \\
141891 & 15 55 08.81 & -63 25 47.1 & F2III        & 2.83 & 2.15 & 90 & 1.80 & 0.07 \\
143466 & 15 57 47.59 & +54 44 58.2 & F0IV         & 4.96 & 4.28 & 140 & 0.68 & 0.08 \\
142983 & 15 58 11.38 & -14 16 45.5 & B8Ia/Iab     & 4.95 & 4.59 & 405 & 0.57 & 0.43 \\
143118 & 16 00 07.34 & -38 23 47.9 & B2.5IV       & 3.42 & 4.09 & 240 & 0.68 & 0.12 \\
143275 & 16 00 20.01 & -22 37 17.8 & B0.2IV       & 2.29 & 2.43 & 180 & 1.52 & 0.05 \\
143474 & 16 03 32.22 & -57 46 29.5 & A7IV         & 4.63 & 4.10 & 175 & 0.73 & 0.13 \\
144294 & 16 06 35.56 & -36 48 08.0 & B2.5Vn       & 4.22 & 4.70 & 320 & 0.52 & 0.15 \\
144668 & 16 08 34.29 & -39 06 18.1 & A8/A9        & 7.00 & 4.39 & 180 & 0.71 & 0.07 \\
145502 & 16 11 59.74 & -19 27 38.3 & B2IV         & 4.00 & 3.88 & 210 & 0.79 & 0.09 \\
147547 & 16 21 55.24 & +19 09 10.9 & A9III        & 3.74 & 2.94 & 145 & 1.25 & 0.18 \\
147933 & 16 25 35.12 & -23 26 49.6 & B2V          & 4.57 & 3.17 & 300 & 1.17 & 0.13 \\
149630 & 16 34 06.19 & +42 26 12.8 & B9Vvar       & 4.20 & 4.05 & 280 & 0.73 & 0.18 \\
149757 & 16 37 09.53 & -10 34 01.7 & O9.5V        & 2.54 & 2.68 & 385 & 1.35 & 0.20 \\
151890 & 16 51 52.24 & -38 02 50.4 & B1.5IV + B   & 3.00 & 3.70 & 235 & 0.82 & 0.11 \\
152427 & 16 54 01.02 & -18 44 36.1 & K1III        & 8.65 & 5.16 & 92 & 0.52 & 0.11 \\
154494 & 17 05 22.66 & +12 44 27.1 & A4IV         & 4.89 & 4.61 & 115 & 0.57 & 0.05 \\
155203 & 17 12 09.18 & -43 14 18.6 & F3p          & 3.32 & 2.26 & 150 & 1.75 & 0.05 \\
156164 & 17 15 01.92 & +24 50 22.5 & A3IVv SB     & 3.12 & 2.81 & 295 & 1.30 & 0.32 \\
156729 & 17 17 40.29 & +37 17 28.8 & A2V          & 4.64 & 4.44 & 155 & 0.61 & 0.05 \\
157778 & 17 23 40.97 & +37 08 45.3 & B9.5III      & 4.15 & 4.23 & 195 & 0.66 & 0.23 \\
157246 & 17 25 23.66 & -56 22 39.7 & B1Ib         & 3.31 & 3.79 & 285 & 0.79 & 0.12 \\
158352 & 17 28 49.69 & +00 19 50.1 & A8V          & 5.41 & 4.81 & 190 & 0.53 & 0.08 \\
158094 & 17 31 05.98 & -60 41 01.0 & B8V          & 3.60 & 3.71 & 255 & 0.84 & 0.13 \\
158643 & 17 31 24.95 & -23 57 45.3 & A0V          & 4.78 & 4.30 & 220 & 0.66 & 0.11 \\
158427 & 17 31 50.52 & -49 52 33.5 & B2Vne        & 2.84 & 2.49 & 315 & 1.51 & 0.14 \\
166205 & 17 32 12.90 & +86 35 10.8 & A1Vn         & 4.35 & 4.26 & 180 & 0.66 & 0.07 \\
159561 & 17 34 56.00 & +12 33 38.1 & A5III        & 2.08 & 1.68 & 230 & 2.20 & 0.25 \\
159532 & 17 37 19.13 & -42 59 52.2 & F1II         & 1.86 & 0.84 & 125 & 3.34 & 0.19 \\
159975 & 17 37 50.72 & -08 07 07.4 & B8II-IIIMNp  & 4.58 & 4.25 & 135 & 0.67 & 0.09 \\
160365 & 17 38 57.87 & +13 19 45.0 & F6III        & 6.12 & 4.70 & 90 & 0.58 & 0.07 \\
161868 & 17 47 53.57 & +02 42 26.9 & A0V          & 3.75 & 3.62 & 220 & 0.89 & 0.11 \\
163955 & 17 59 47.56 & -23 48 57.6 & B9V          & 4.74 & 4.47 & 250 & 0.60 & 0.14 \\
164577 & 18 01 45.19 & +01 18 18.4 & A2Vn         & 4.42 & 4.23 & 260 & 0.67 & 0.16 \\
166045 & 18 07 49.56 & +26 06 04.4 & A3V          & 5.83 & 5.12 & 175 & 0.46 & 0.07 \\
168914 & 18 21 01.02 & +28 52 11.4 & A7V          & 5.12 & 4.48 & 195 & 0.61 & 0.09 \\
170073 & 18 23 54.65 & +58 48 02.1 & A3V          & 4.98 & 4.78 & 180 & 0.52 & 0.07 \\
169022 & 18 24 10.35 & -34 23 03.5 & B9.5III      & 1.79 & 1.77 & 175 & 2.07 & 0.18 \\
169702 & 18 24 13.80 & +39 30 26.1 & A3IVn        & 5.11 & 4.86 & 200 & 0.51 & 0.17 \\
169985 & 18 27 12.51 & +00 11 46.1 & G0III+...    & 5.20 & 3.39 & 270 & 1.08 & 0.22 \\
170296 & 18 29 11.85 & -14 33 56.9 & A1IV/V       & 4.67 & 4.36 & 255 & 0.64 & 0.21 \\
170479 & 18 31 04.85 & -32 59 20.4 & A5V          & 5.37 & 4.87 & 160 & 0.51 & 0.06 \\
169978 & 18 31 22.43 & -62 16 41.5 & B8III        & 4.63 & 4.83 & 125 & 0.50 & 0.08 \\
172365 & 18 39 36.88 & +05 15 51.4 & F8Ib-II      & 6.36 & 4.46 & 67.3 & 0.66 & 0.09 \\
172777 & 18 43 46.94 & -38 19 23.9 & A0/A1V       & 5.11 & 4.86 & 175 & 0.50 & 0.07 \\
173582 & 18 44 20.34 & +39 40 11.9 & F1V          & 4.67 & 4.23 & 195 & 0.68 & 0.09 \\
173607 & 18 44 22.78 & +39 36 45.3 & A8Vn         & 4.59 & 4.16 & 175 & 0.70 & 0.07 \\
173649 & 18 44 48.19 & +37 35 40.4 & F0IVvar      & 5.73 & 4.96 & 240 & 0.50 & 0.29 \\
172555 & 18 45 26.86 & -64 52 15.2 & A7V          & 4.78 & 4.30 & 175 & 0.66 & 0.07 \\
174602 & 18 49 52.92 & +32 33 03.9 & A3V          & 5.22 & 4.90 & 150 & 0.50 & 0.05 \\
173948 & 18 52 13.04 & -62 11 15.2 & B2II-III     & 4.22 & 4.39 & 190 & 0.61 & 0.10 \\
175824 & 18 54 47.17 & +48 51 35.0 & F3III        & 5.84 & 4.67 & 80 & 0.58 & 0.05 \\
175191 & 18 55 15.92 & -26 17 47.7 & B2.5V        & 2.05 & 2.44 & 205 & 1.49 & 0.06 \\
175639 & 18 56 14.61 & +04 12 07.4 & A5Vn         & 4.98 & 4.46 & 200 & 0.62 & 0.10 \\
175813 & 18 58 43.47 & -37 06 25.5 & F3IV/V       & 4.83 & 3.88 & 175 & 0.82 & 0.11 \\
176723 & 19 03 17.69 & -38 15 11.5 & F2III/IV     & 5.73 & 4.87 & 265 & 0.52 & 0.27 \\
177724 & 19 05 24.61 & +13 51 49.4 & A0Vn         & 2.99 & 2.88 & 345 & 1.25 & 0.32 \\
177756 & 19 06 14.95 & -04 52 56.4 & B9Vn         & 3.43 & 3.56 & 160 & 0.90 & 0.05 \\
178233 & 19 06 37.68 & +28 37 42.2 & F0III        & 5.53 & 4.81 & 165 & 0.53 & 0.24 \\
178449 & 19 07 25.50 & +32 30 06.0 & F0V          & 5.20 & 4.19 & 155 & 0.71 & 0.05 \\
178596 & 19 08 59.92 & +06 04 24.2 & F0III-IV     & 5.23 & 4.33 & 125 & 0.66 & 0.09 \\
178253 & 19 09 28.28 & -37 54 15.3 & A0/A1V       & 4.11 & 4.05 & 225 & 0.73 & 0.11 \\
180868 & 19 17 49.00 & +11 35 43.4 & F0IV         & 5.28 & 4.71 & 115 & 0.55 & 0.05 \\
181623 & 19 23 13.06 & -44 47 58.7 & F2III        & 4.27 & 2.93 & 155 & 1.30 & 0.22 \\
184006 & 19 29 42.34 & +51 43 46.1 & A5Vn         & 3.76 & 3.60 & 220 & 0.90 & 0.12 \\
186005 & 19 42 31.09 & -16 07 26.3 & F3IV/V       & 5.06 & 4.28 & 160 & 0.68 & 0.09 \\
186882 & 19 44 58.44 & +45 07 50.5 & B9.5III      & 2.86 & 2.68 & 140 & 1.37 & 0.11 \\
187076 & 19 47 23.27 & +18 32 03.3 & M2II + B6    & 3.68 & -0.56 & 50 & 7.56 & 0.07 \\
187362 & 19 48 58.65 & +19 08 31.1 & A3V          & 5.01 & 4.73 & 240 & 0.54 & 0.14 \\
186219 & 19 49 25.29 & -72 30 12.3 & A4III        & 5.39 & 4.80 & 125 & 0.53 & 0.11 \\
187642 & 19 50 46.68 & +08 52 02.6 & A7IV-V       & 0.76 & 0.10 & 240 & 4.61 & 0.20 \\
189037 & 19 55 37.82 & +52 26 20.5 & A4Vn         & 4.91 & 4.49 & 285 & 0.60 & 0.21 \\
190004 & 20 02 01.37 & +24 56 16.3 & F2III        & 5.23 & 4.30 & 150 & 0.68 & 0.21 \\
192514 & 20 13 18.04 & +46 48 56.4 & A5IIIn       & 4.80 & 4.41 & 145 & 0.63 & 0.17 \\
192696 & 20 13 23.80 & +56 34 03.1 & A3IV-Vn      & 4.28 & 4.08 & 280 & 0.72 & 0.28 \\
192518 & 20 14 14.52 & +28 41 41.5 & A7IVn        & 5.19 & 4.65 & 190 & 0.57 & 0.16 \\
192425 & 20 14 16.59 & +15 11 50.9 & A2V          & 4.94 & 4.77 & 160 & 0.52 & 0.06 \\
195627 & 20 35 34.77 & -60 34 52.7 & F1III        & 4.75 & 4.04 & 150 & 0.75 & 0.20 \\
197157 & 20 44 02.19 & -51 55 15.0 & A6:var       & 4.51 & 3.82 & 150 & 0.83 & 0.05 \\
197937 & 20 48 29.00 & -43 59 17.8 & F1IV         & 5.11 & 4.20 & 115 & 0.71 & 0.06 \\
199611 & 20 56 25.44 & +50 43 43.1 & F0III        & 5.83 & 4.92 & 125 & 0.51 & 0.13 \\
199629 & 20 57 10.41 & +41 10 01.9 & A1Vn         & 3.94 & 3.52 & 250 & 0.94 & 0.15 \\
200120 & 20 59 49.55 & +47 31 15.4 & B1ne         & 4.74 & 4.35 & 379 & 0.64 & 0.20 \\
199532 & 21 04 43.03 & -77 01 22.3 & F4III        & 5.13 & 3.60 & 85 & 0.96 & 0.06 \\
177482 & 21 08 46.01 & -88 57 23.4 & F0III        & 5.45 & 4.67 & 145 & 0.56 & 0.18 \\
202904 & 21 17 55.07 & +34 53 48.8 & B2Vne        & 4.41 & 4.48 & 255 & 0.59 & 0.09 \\
203280 & 21 18 34.58 & +62 35 07.6 & A7IV-V       & 2.45 & 2.07 & 245 & 1.84 & 0.21 \\
202730 & 21 19 51.88 & -53 26 57.4 & A5V          & 4.39 & 4.15 & 210 & 0.70 & 0.11 \\
203803 & 21 23 58.74 & +24 16 26.7 & F1IV         & 5.70 & 4.92 & 110 & 0.51 & 0.05 \\
205114 & 21 31 27.46 & +52 37 11.5 & G2Ib+...     & 6.17 & 3.87 & 50 & 0.89 & 0.07 \\
205835 & 21 36 56.98 & +40 24 48.6 & A5V          & 5.04 & 4.51 & 180 & 0.60 & 0.08 \\
205637 & 21 37 04.82 & -19 27 57.6 & B3V:p        & 4.51 & 4.79 & 225 & 0.51 & 0.07 \\
205767 & 21 37 45.04 & -07 51 14.9 & A7V          & 4.68 & 4.25 & 165 & 0.67 & 0.06 \\
205852 & 21 37 45.37 & +19 19 06.9 & F1IV         & 5.46 & 4.61 & 180 & 0.58 & 0.15 \\
207155 & 21 47 44.17 & -30 53 53.9 & A1V          & 5.02 & 4.85 & 175 & 0.50 & 0.07 \\
208450 & 21 57 55.03 & -54 59 33.2 & F0IV         & 4.40 & 3.42 & 130 & 1.02 & 0.07 \\
209409 & 22 03 18.83 & -02 09 19.2 & B7IVe        & 4.74 & 4.66 & 305 & 0.55 & 0.35 \\
209952 & 22 08 13.88 & -46 57 38.2 & B7IV         & 1.73 & 2.02 & 250 & 1.82 & 0.21 \\
210049 & 22 08 22.95 & -32 59 18.2 & A2V          & 4.50 & 4.31 & 300 & 0.65 & 0.23 \\
210334 & 22 08 40.86 & +45 44 31.7 & K2III comp   & 6.11 & 4.27 & 70 & 0.72 & 0.07 \\
210271 & 22 09 55.71 & -34 00 54.1 & A5IV         & 5.37 & 4.75 & 260 & 0.54 & 0.35 \\
210459 & 22 09 59.25 & +33 10 41.8 & F5III        & 4.28 & 3.12 & 135 & 1.18 & 0.17 \\
210839 & 22 11 30.58 & +59 24 52.3 & O6e          & 5.05 & 4.50 & 275 & 0.60 & 0.10 \\
212581 & 22 27 19.87 & -64 57 59.0 & B8V          & 4.51 & 4.55 & 240 & 0.57 & 0.12 \\
213310 & 22 29 31.82 & +47 42 24.8 & M0II         & 4.34 & 0.27 & 50 & 5.13 & 0.10 \\
213998 & 22 35 21.33 & -00 07 02.5 & B9IV-Vn      & 4.04 & 4.24 & 280 & 0.66 & 0.24 \\
214748 & 22 40 39.33 & -27 02 37.0 & B8V          & 4.18 & 4.40 & 375 & 0.61 & 0.34 \\
214923 & 22 41 27.67 & +10 49 53.0 & B8.5V        & 3.41 & 3.57 & 185 & 0.90 & 0.07 \\
215789 & 22 48 33.20 & -51 19 00.1 & A3V          & 3.49 & 3.19 & 270 & 1.09 & 0.18 \\
217675 & 23 01 55.25 & +42 19 33.5 & B6pv SB      & 3.62 & 3.89 & 320 & 0.77 & 0.20 \\
217782 & 23 02 36.34 & +42 45 28.1 & A3Vn         & 5.09 & 4.69 & 205 & 0.55 & 0.10 \\
218045 & 23 04 45.62 & +15 12 19.3 & B9.5III      & 2.49 & 2.65 & 150 & 1.37 & 0.12 \\
217831 & 23 04 52.15 & -68 49 13.4 & F4III        & 5.53 & 4.69 & 120 & 0.56 & 0.13 \\
218918 & 23 11 44.19 & +08 43 12.5 & A5Vn         & 5.15 & 4.74 & 220 & 0.54 & 0.12 \\
219586 & 23 15 37.71 & +70 53 17.1 & F0IV         & 5.55 & 4.81 & 150 & 0.53 & 0.10 \\
219571 & 23 17 25.81 & -58 14 09.3 & F1III        & 3.99 & 3.04 & 95 & 1.21 & 0.07 \\
219688 & 23 17 54.20 & -09 10 57.0 & B5Vn         & 4.41 & 4.76 & 340 & 0.51 & 0.22 \\
220061 & 23 20 38.22 & +23 44 25.3 & A5V          & 4.58 & 4.09 & 150 & 0.73 & 0.05 \\
221565 & 23 33 16.63 & -20 54 52.3 & A0V          & 4.70 & 4.52 & 385 & 0.59 & 0.43 \\
222095 & 23 37 50.94 & -45 29 32.4 & A2V          & 4.74 & 4.52 & 165 & 0.59 & 0.06 \\
222439 & 23 40 24.44 & +44 20 02.3 & B9IVn        & 4.15 & 4.57 & 190 & 0.56 & 0.12 \\
223352 & 23 48 55.48 & -28 07 48.1 & A0V          & 4.59 & 4.53 & 280 & 0.58 & 0.19 \\
223781 & 23 52 37.12 & +10 56 50.4 & A4Vn         & 5.30 & 4.71 & 170 & 0.55 & 0.07 \\
224392 & 23 57 34.97 & -64 17 53.1 & A1V          & 5.00 & 4.82 & 250 & 0.51 & 0.15 \\
224686 & 23 59 54.91 & -65 34 37.5 & B9IV         & 4.49 & 4.60 & 295 & 0.56 & 0.36 \\
\enddata
\tablecomments{A full version of this table is available electronically.}
\end{deluxetable}
